\documentclass[10pt,a4paper]{article}
\usepackage{mathtools}
\usepackage{mathrsfs}
\usepackage{amsmath}
\usepackage[utf8x]{inputenc}
\usepackage{amssymb}
\usepackage{bbm}
\usepackage{bm}
\usepackage{color}
\usepackage[draft]{todonotes}
\usepackage[a4paper,top=3cm,bottom=3cm,left=2cm,right=2cm,bindingoffset=5mm]{geometry}
\usepackage{booktabs} 
\usepackage{tikz}
\usepackage{cancel}
\usepackage[bottom]{footmisc}
\usepackage{cite}

\newcommand*\diff{\mathop{}\!\mathrm{d}}

\usepackage{hyperref}

\hypersetup{
colorlinks=true,         
linkcolor=blue,          
citecolor=red,        
urlcolor=red            
}

\renewcommand{\vec}[1]{\mathbf{#1}}

\newcommand{\1}{\mathbbm{1}}

\newcommand{\ahat}{\hat{a}}
\newcommand{\bhat}{\hat{b}}
\newcommand{\chat}{\hat{c}}
\newcommand{\dhat}{\hat{d}}
\newcommand{\ehat}{\hat{e}}
\newcommand{\fhat}{\hat{f}}

\title{\textbf{Pauli--Jordan Function and Scalar Field Quantization in $\kappa$-Minkowski Noncommutative Spacetime}}

\author{Flavio~Mercati\footnote{\href{mailto:flavio.mercati@gmail.com}{flavio.mercati@gmail.com}} ~and Matteo~Sergola\footnote{\href{mailto:matteo.sergola@gmail.com}{matteo.sergola@gmail.com}}
\vspace{12pt} \\
\it Dipartimento di Fisica, Sapienza Universit{\`a} di Roma,\\
\it P.le A. Moro 2, 00185 Roma, Italy.}

\begin{document}

\maketitle

\begin{abstract}
We study a complex free scalar field theory on a noncommutative background spacetime called $\kappa$-Minkowski. In particular we address the problem of second quantization. We obtain the algebra of creation and annihilation operators in an explicitly covariant way. Our procedure does not use canonical/Hamiltonian formulations, which turn out to be ill-defined in our context. Instead we work in a spacetime covariant way by introducing a noncommutative Pauli--Jordan function. This function is obtained as a generalization of the ordinary, commutative, one by taking into account the constraints imposed by the symmetries of our noncommutative spacetime.
The Pauli--Jordan function is later employed to study the structure of the light cone in $\kappa$-Minkowski spacetime, and to draw conclusions on the superluminal propagation of  signals. 
\end{abstract}

\section{Introduction}

Quantum Field Theory (QFT) on Minkowski spacetime is arguably the most successful paradigm in physics, both for the precision with which some of its predictions have been tested, and for the variety of phenomena it is capable to describe. As a description of nature, flat-space QFT is clearly an effective theory, as it ignores the dynamics of the gravitational field, for which we currently have no satisfying quantum formulation. The unspoken underlying assumption is that, whatever the correct quantum theory of gravity is, it will admit a `ground state' which looks like Minkowski spacetime. Then flat-space QFT would be a good description of matter as long as graviton production can be ignored. This last condition is a safe assumption in the experimental regimes we have access to. This is because the coupling between matter and gravity is controlled by the Planck scale $E_p \sim 10^{28} eV$, which is an enormous energy scale that makes for an extremely small coupling constant.
Our current understanding of Quantum Gravity suggests that the aforementioned assumption might not be correct. The ground state of General Relativity might be something different from Minkowski space, which only looks like it in the low-energy limit. The strongest indications come from 2+1 dimensional Quantum Gravity, which, because it lacks local propagating degrees of freedom (gravitons), can be quantized with topological QFT methods. Coupling this theory to matter and integrating away the gravitational degrees of freedom, one ends up with a nonlocal effective theory~\cite{matschull,freidel}. This theory admits a description as a QFT on a noncommutative background, in the sense that the ordinary algebra of functions on spacetime [which is an abelian algebra when endowed with the pointwise product $(f\cdot g)(x)=f(x) g(x)$] is replaced with a noncommutative algebra. The Planck scale (or rather its inverse, which in $\hbar = c =1$ units is the Planck length $L_p \sim 10^{-35} m$) plays the role of noncommutativity parameter, similar to that played by $\hbar$ in ordinary quantum mechanics. In particular it appears on the right-hand side of uncertainty relations between coordinate functions $x^\mu$, and therefore there is a sense in which the noncommutative geometry described by this algebra should look like the commutative geometry of Minkowski spacetime in the large-scale/infrared limit. Let us restate this important point: the `background state' of 2+1 D Quantum Gravity coupled with matter is not QFT on Minkowski spacetime. It is rather a QFT on a noncommutative geometry which reduces to Minkowski space only in the low-energy limit.

In light of the lesson of 2+1 D Quantum Gravity, studying noncommutative geometries in 3+1 D, and the QFTs that can be built upon them, acquires a great interest. Because of our lack of understanding of 3+1 D Quantum Gravity, we presently have no way to repeat the exercise done in 2+1 D of integrating away the gravitational field to uncover the correct effective theory of matter on a quantum-gravity background. For this reason, we are compelled to study all the possible 3+1 D noncommutative geometries whose noncommutativity parameter depends on the Planck length. If we are able to develop consistent QFTs on such backgrounds, there is a chance that their phenomenological implications could be experimentally tested, giving us precious hints towards the correct quantum theory of gravity in 3+1 D.

A powerful way to describe a geometry is through its symmetries. Minkowski spacetime, for example, is completely characterized by the fact that it is a \emph{maximally symmetric space} which is  \emph{flat}. The first condition (maximal symmetry) reduces the possible choices to only three: de Sitter (for positive curvature), anti-de Sitter (negative curvature) and Minkowski (flat). Then the spacetime can be understood as the quotient of a 10-dimensional Lie group [(A)-dS or Poincar\'e, depending on the curvature] by a 6-dimensional isotropy subgroup (the Lorentz group).
Interestingly, these structures generalize to noncommutative spaces.  Lie groups are generalized to something called \emph{quantum groups}~\cite{compact,majid}: essentially, the algebra of functions on the group manifold becomes noncommutative (see next Section).

Remarkably, there is only a limited number of quantum groups that reduce to the Poincar\'e group when the noncommutativity parameter vanishes~\cite{zaz}. In particular, if we require the noncommutativity parameter to depend linearly on the Planck scale, and to admit a regular limit as the cosmological constant is sent to zero, we are left with essentially only one choice. This is the \emph{$\kappa$-Poincar\'e} quantum group~\cite{commutatori, majid22}, where $\kappa$ refers to the (inverse of the) noncommutativity parameter, which has the dimensions of an energy. Field theories that are invariant under $\kappa$-Poincar\'e symmetries  have been studied extensively in the past (see for instance  \cite{FKKKN, arzanooooo}), especially at the classical level (\emph{i.e.} in the limit $\hbar \to 0$ while keeping $\kappa$ finite). However, the understanding of $\kappa$-deformed field theories is still far from complete, especially for what regards their second quantization. 

In the present work we discuss a new strategy for the construction and quantization of a complex scalar field on $\kappa$-Minkowski. Our approach is explicitly $\kappa$-Poincar\'e covariant, and independent on the choice of basis of the spacetime symmetry algebra (an important feature which most previous approaches lacked -- see below). A key element of our analysis is the observation that, in order to preserve Lorentz invariance, the momentum space of fields on $\kappa$-Minkowski must have the topology of \emph{elliptic de Sitter space}. It has in fact been observed long ago~\cite{firstds} that the $\kappa$-deformed momentum space has a de Sitter geometry. However, at the global level the momentum space covers only half of the de Sitter hyperboloid, and that half is not closed under the action of Lorentz transformations. The Lorentz orbits become complete (\emph{i.e.} do not terminate at a finite boundary) only if we assume the elliptic topology $d S/\mathbbm{Z}^2$ for the de Sitter hyperboloid (see Sec.~\ref{Sec_k-Poincare_group_and_k-Minkowski}).

Our approach will allow us to write the algebra of creation and annihilation operators for a scalar filed, which in turn defines the Fock space of the theory. An important feature of our analysis is that we define the quantization rules of the theory in a covariant way, using the Pauli--Jordan function. This is a necessary step that was overlooked in previous analyses, which typically relied, in one form or another, upon canonical quantization and therefore Hamiltonian formulations. Such formulations are flawed, because $\kappa$-Minkowski (the noncommutative spacetime whose symmetries are described by the  $\kappa$-Poincar\'e group) does not admit the notion of constant-time slices.\footnote{The commutation relations of $\kappa$-Minkowski coordinates are such that a sharply defined time coordinate implies infinite uncertainty on the spatial coordinates (see Sec.~\ref{PauliJordanSec}).}

Our covariant approach has an additional bonus: it gives us a generalization of the notion of light cone. In fact the commutative Pauli--Jordan function vanishes outside the light cone: it measures the commutator between quantum fields at different points, and such a commutator must vanish on spacelike-separated points in order for causality to be respected. In the noncommutative case the Pauli--Jordan function turns into an element of a noncommutative algebra, and therefore some more work is necessary in order to extract a notion of light cone from it. Our proposal is to interpret the Pauli--Jordan function as an operator on a Hilbert space of `geometrical' states, \emph{i.e.} states of the background quantum geometry. By calculating the expectation value of said operator on a state that is peaked around a classical pair of points (what we call \emph{semiclassical state}) we are able to extract the dependence of the Pauli--Jordan function on the coordinates of the classical points around which our state is peaked. It turns out that this dependence is not confined inside the classical light cone, but it rather `spills out' within a region whose size is the geometric mean between the Planck length and the distance from the origin of the light cone. The derivation of these results is the subject of Sec.~\ref{MinimalUncertaintyStatesSec}.

Our calculation of the Pauli--Jordan function allows us to intervene in a debate that has been ongoing since the early days of $\kappa$-Poincar\'e. This is whether the deformations of relativistic kinematics predicted by $\kappa$-Poincar\'e imply anomalous in-vacuum dispersion and whether such dispersion can be detected with present-day technology.
At the turn of the century it was proposed that the Planck scale might enter new physics in a fashion similar to that of the speed of light. $c$ in fact is a speed constant that takes the same value for all inertial observers, and in order to accommodate the relativistic invariance of $c$ one needs to deform Galilean relativity in a $c$-dependent way. Similarly, it was suggested in~\cite{deformed}, $E_p$ could be a new energy scale that appears the same to all inertial observers, but this requires a deformation of special relativity into a new relativistic theory with two invaraint scales (this proposal was dubbed Doubly Special Relativity~\cite{GiovanniLRR}). Soon after this idea was proposed, $\kappa$-Poincar\'e was identified as a candidate model to realize such two-scale generalization of special relativity~\cite{amelinowaves}. Indeed, as we show in Eq.~\eqref{1+1DLorentzTransform} below, one finds deformed Lorentz transformations laws for energy and momentum which depend on both $c$ and the Planck scale. $\kappa$-Poincar\'e is still today the best candidate for a consistent model realizing the idea of Doubly Special Relativity. Unfortunately introducing the Planck scale as an observer-independent constant makes it harder to detect its effects. In fact, most meaningful constraints for a possible role of $E_p$ in physics come from testing some conjectured Lorentz-breaking effects, whereupon 
Michelson--Morley-like experiments putting constraints on a `quantum-gravity aether' are easy to devise and extremely powerful~\cite{82}. If the Planck scale enters as a deformation of some laws of physics in a way that does not depend on the inertial frame, it becomes much harder to detect.
One of the best proposals for such a test is in-vacuum dispersion: if the relativistic kinematics is deformed in a Planck-scale-dependent way, it is conceivable that the energy-momentum dispersion relation acquires Planck-scale corrections. Then the group velocity of particles propagating in a vacuum should get Planck-scale corrections, and it is just a matter of dimensional analysis to show which kind of dependences on energy and momentum such a deformation could acquire~\cite{21}.
One possibility is that the dispersion is such that distant point-like sources arrive on Earth with a time delay of the form:
\begin{equation}\label{LinearDispRel}
\delta t \propto L_p \, E \, L \,,
\end{equation}
where $\delta t$ might be a systematic delay of all particle or a random uncertainty on the time of arrival, different for each particle. $E$ is the energy of the particle and $L$ is the distance of the source (in units $\hbar = c =1$). Such a law can be meaningfully tested with present-day observations: in particular Gamma Ray Bursts provide bright sources of photons of energies up to $1 TeV$ at distances of the order of a billion light-years. With such numbers the law~(\ref{LinearDispRel}) predicts time delays of the order of one second. Indeed the Fermi-LAT experiment has been testing such hypothesis for years, putting constraint on the energy scale of the proposed quantum-gravity effect~\cite{Fermi2009Nature} of the order of $E_p$ or more. More recently, a proposal to test the hypothesis that the high-energy neutrinos observed by the IceCube observatory are originated in Gamma Ray Bursts, and are subject to the same proposed quantum-gravity effects gained a lot of attention~\cite{icecubbbo}.
So, soon after a law like~(\ref{LinearDispRel}) was first proposed, it was conjectured that it would be a prediction of $\kappa$-Poincar\'e, and then Gamma Ray Bursts would be the perfect arena to test the predictions of this model.
In the present paper we develop $\kappa$-Poincar\'e-invariant QFT to an unprecedented level of consistency, and this will allow us to establish with a higher level of confidence whether Gamma Ray Burst phenomenology is able to put meaningful bounds on our model.

In the next Section (\ref{Sec_k-Poincare_group_and_k-Minkowski}) we briefly revise the mathematical tools needed for our analysis, and give a brief recap of the things we already know about scalar fields and the geometry of momentum space. In Section~\ref{FreeQuantumKleinGordonFieldSection} we generalize the Klein-Gordon equation to $\kappa$-Minkowski and in Section~\ref{PauliJordanSec}  we introduce the `$\kappa$-deformed' Pauli-Jordan function. In Section~\ref{MinimalUncertaintyStatesSec} we use the Pauli--Jordan function to study  the noncommutative lightcone of our theory and in Section~\ref{ConclusionSec} we present future perspectives and conclusions.

\section{The $\kappa$-Poincar\'e quantum group and the $\kappa$-Minkowski spacetime}\label{Sec_k-Poincare_group_and_k-Minkowski}

\subsection{The standard Poincar\'e group as a Hopf algebra}

The structure of the Poincar\'e group $\mathcal G$ can be described in algebraic terms by considering the algebra of complex-valued functions on the group $\mathbbm{C}[\mathcal G]$, and by introducing three maps which are dual\footnote{In the category-theory sense of inverting all the arrows.} to the three defining axioms of Lie groups. First one needs to introduce \emph{coordinate functions} on the group $ \Lambda^\mu{}_\nu , a^\mu : \mathcal G \to \mathbbm{C}$, \emph{i.e.} elements of $\mathbbm{C}[\mathcal G]$ which associate to each group element $g \in \mathcal G$ its coordinates in the standard  representation of the group.

The group product can be described by a \emph{coproduct} map $\Delta : \mathbbm{C}[\mathcal G]  \to \mathbbm{C}[\mathcal G] \otimes \mathbbm{C}[\mathcal G]$:
\begin{equation}
\Delta [ \Lambda^\mu{}_\nu ] =  \Lambda^\mu{}_\rho  \otimes  \Lambda^\rho{}_\nu \,,
\qquad
\Delta[a^\mu] =  a^\mu \otimes \1 +  \Lambda^\mu{}_\nu \otimes a^\nu \,,
\end{equation}
$\Delta [ \Lambda^\mu{}_\nu ]$ is now a function that associates to two group elements $g,h \in \mathcal{G}$ the coordinates, in the coordinate system $\Lambda^\mu{}_\nu, a^\mu$, of the product element $g \cdot h$: $\Delta [ \Lambda^\mu{}_\nu ] (g,h) = \Lambda^\mu{}_\nu(g)  \Lambda^\nu{}_\mu (h) = \Lambda^\mu{}_\nu(g \cdot h)$,
and similarly for $\Delta[a^\mu]$: $\Delta [ a^\mu ] (g,h) = a^\mu (g) +  \Lambda^\mu{}_\nu (g)  a^\nu(h)  = a^\mu (g \cdot h)$.

The group inverse can be encoded into an \emph{antipode} map $S : \mathbbm{C}[\mathcal G] \to \mathbbm{C}[\mathcal G]$:
\begin{equation}
S [ \Lambda^\mu{}_\nu ] =  (\Lambda^{-1})^\mu{}_\nu   \,,
\qquad
S[a^\mu] = - (\Lambda^{-1})^\mu{}_\nu  a^\nu \,,
\end{equation}
which, when calculated on the coordinate functions give two functions whose value on a group element $g$ is the coordinates of the inverse element $g^{-1}$: $S [ \Lambda^\mu{}_\nu ] (g) =  (\Lambda^{-1})^\mu{}_\nu (g)  = \Lambda^\mu{}_\nu (g^{-1})$, $S[a^\mu]  (g)= - \Lambda^\mu{}_\nu  a^\nu (g) = a^\mu (g^{-1})$.

Finally, the map that stands in for the unit is called \emph{counit} $\varepsilon : \mathbbm{C}[\mathcal G] \to \mathbbm{C}$:
\begin{equation}
\varepsilon [ \Lambda^\mu{}_\nu ] =  \delta^\mu{}_\nu  \,,
\qquad
\varepsilon [a^\mu] = 0 \,,
\end{equation}
where $\varepsilon  [ \Lambda^\mu{}_\nu ] = \Lambda^\mu{}_\nu (e)$ and $\varepsilon  [ a^\mu  ] = a^\mu  (e)$ give the coordinates of the identity element $e \in \mathcal{G}$ in the coordinate system $\Lambda^\mu{}_\nu, a^\mu$.

In relation to the structures of the commutative algebra $\mathbbm{C}[\mathcal G]$ (product and linear combination), the maps $\Delta$ and $\varepsilon$ are \emph{algebra homomorphisms:} $\Delta[f g] =\Delta[f] \Delta[g]$, $\Delta[f + g] =\Delta[f] + \Delta[g]$, while $S$ is an \emph{anti-}homomorphism.

With the three maps $\Delta$, $S$ and $\varepsilon$ we can completely describe the group, provided that they satisfy a compatibility axiom:
\begin{equation}\label{HopfAxiom}
\mu \circ (S \otimes  \text{id} ) \circ \Delta = \mu \circ (\text{id} \otimes S) \circ \Delta  = \1 \, \varepsilon \,,
\end{equation}
where $\mu : \mathbbm{C}[\mathcal G] \otimes \mathbbm{C}[\mathcal G] \to \mathbbm{C}[\mathcal G]$ is the product of $\mathbbm{C}[\mathcal G]$ (the pointwise product between functions) and $\text{\textit{id}}$ is the identity map. This axiom coincides with the definition of the inverse $g^{-1} \cdot g = g \cdot g^{-1} = e$.

\subsection{Generalization to Quantum Groups}

Having achieved this unusual description of the well-known Poincar\'e group, we can disclose now the reason for going through all this trouble. The commutative algebra of functions $\mathbbm{C}[\mathcal G]$ can be replaced with a \emph{nonabelian algebra} $\mathbbm{C}_\kappa[\mathcal G]$, whose product now does not admit anymore the interpretation of pointwise product between functions, nor its elements the interpretation of functions on $\mathcal G$. We are now dealing with a \emph{quantum group}~\cite{majid, elvis}, whose algebra of functions is noncommutative. In the case of the Poincar\'e group in 3+1 dimensions, there appears to be essentially only a seven-parameter space of candidates which admit the interpretation of the flat-space limit of a Planck-scale deformation (first order in the Planck length) of Poincar\'e group~\cite{MatteoFlavioToAppear}. If three of these parameters (those associated to a `Reshetikhin  twist'~\cite{MatteoFlavioToAppear}) are set to zero, we obtain the \emph{$\kappa$-Poincar\'e group}, defined by the commutation relations~\cite{commutatori}
\begin{equation}\label{kappaPoincareGroup}
\begin{gathered}
[ a^\mu , a^\nu ] = i \left( v^\mu \, a^\nu - v^\nu \, a^\mu \right) \,, \qquad [ \Lambda^\mu{}_\nu ,  \Lambda^\rho{}_\sigma ] = 0 \,,
\\
[ \Lambda^\mu{}_\nu , a^\rho ] =-  i \left[  \left( \Lambda^\mu{}_\sigma v^\sigma - v^\mu \right) \Lambda^\rho{}_\nu + \left( \Lambda^\sigma{}_\nu v_\sigma - v_\nu \right) \eta^{\mu\rho} \right] \,,
\end{gathered}
\end{equation}
where $\eta_{\mu\nu} = \text{diag} (-,+,+,+)$ and $v^\mu = (v^0,v^1,v^2,v^3)$ are four deformation parameters, which should be of the order of the Planck length.
In a mathematical sense, all choices of $v^\mu$ with the same sign of the norm, $v_\mu v^\mu$, are equivalent (they are related by algebra automorphisms). Then the mathematically-inequivalent cases are only three: when $v^\mu$ is spacelike, lightlike or timelike~\cite{koso}. Mathematical equivalence aside, the question whether different choices of $v^\mu$ lead to different physics remains open, and we do not intend to dwell on these issues in the present paper. For now, it is sufficient to study one particular case, and we choose the most-studied one, which is $v^\mu$ timelike and of the form:
\begin{equation}\label{timelike_v}
v^\mu = \frac{1}{\kappa} \delta^\mu_0 \,,
\end{equation}
where $\kappa$ is the eponymous parameter of the $\kappa$-Poincar\'e group.  $\kappa$ has the dimensions of an energy, and it is expected to be close to the Planck energy, if the $\kappa$-Poincar\'e group is to describe deformations originated in a presently-unknown quantum theory of gravity. 

Notice that the commutation relations~(\ref{kappaPoincareGroup}) admit two subalgebras, one generated by the four translation generators $a^\mu$ and another one generated by the Lorentz matrices $\Lambda^\mu{}_\nu$ (moreover the latter algebra is commutative - therefore the Lorentz subgroup $SO(3,1)$ is classical). The fact that the translation generators close a subalgebra is a consequence of the fact that the $\kappa$-Poincar\'e algebra is \emph{coisotropic} with respect to Lorentz transformations~\cite{homo}. This is required to be able to talk about a \emph{quantum homogeneous space,} generated by quotienting $\mathbbm{C}_\kappa[\mathcal G]$ by the Lorentz subgroup $\mathcal A = \mathbbm{C}_\kappa[\mathcal G] / SO(3,1)$. This is a noncommutative algebra generated by
\begin{equation}
[ x^\mu , x^\nu ] = i \left( v^\mu \, x^\nu - v^\nu \, x^\mu \right) \,,
\end{equation}
or, in the timelike case we are concerned with in the present paper, 
 \begin{equation}\label{kappaMinkowskiCommutationRelations}
[ x^0 , x^i ] = \frac{i}{\kappa} x^i \,, \qquad [ x^i , x^j ] = 0  \,.
\end{equation}
The $x^\mu$ generators should be interpreted as \emph{coordinate functions} on a noncommutative spacetime, which we call \emph{$\kappa$-Minkowski}. The algebra $\mathcal A$ is to be interpreted as the algebra of functions on $\kappa$-Minkowski, or, which is the same thing, the algebra of scalar fields. $\mathcal A$ 
is obtained by taking all possible (finite or infinite) products and sums of $x^\mu$, \emph{i.e.}  $\sum_{n=0}^\infty c_{\mu_1 \dots \mu_n} x^{\mu_1} \dots  x^{\mu_n}$ with $c_{\mu_1 \dots \mu_n} \in \mathbbm{C}$.

Just like commutative homogeneous spaces, $\kappa$-Minkowski comes equipped with an action of the $\kappa$-Poincar\'e group which leaves it invariant. In the Hopf algebra language, as could be expected, this is expressed by means of a (left-) \emph{coaction} map, $\Delta_L: \mathcal A \to  \mathbbm{C}[\mathcal G] \otimes \mathcal A$,
\begin{equation}\label{Coaction_kappa_Poincare_group}
\Delta_L [x^\mu] = \Lambda^\mu{}_\nu \otimes x^\nu +  a^\mu \otimes \1  \,.
\end{equation}
The map above is an $\mathcal A$-homomorphism, as can be explicitly seen by calculating the commutator between two transformed coordinates:
\begin{equation}
[\Delta_L [x^\mu] , \Delta_L [x^\nu] ] = \Delta_L [ i \left( v^\mu \, x^\nu - v^\nu \, x^\mu \right)] =  i \left( v^\mu \, \Delta_L [x^\nu] - v^\nu \, \Delta_L [x^\mu] \right) \,,
\end{equation}
the above relation is left invariant only if $\Lambda^\mu{}_\nu$ and $a^\mu$ satisfy the commutation relations~\eqref{kappaPoincareGroup}.

\subsection{Scalar fields on $\kappa$-Minkowski and the $\kappa$-Poincar\'e algebra}

As we remarked above, scalar fields on $\kappa$-Minkowski are simply generic elements of $\mathcal A$ (spacetime coordinates are scalar fields too: in the commutative case they are four scalar fields whose values at a spacetime point give the values of the four coordinates of that point in a certain coordinate system). One can define on $\kappa$-Minkowski a notion of Fourier transform, by expanding a generic scalar field $\phi(x)$ in \emph{ordered plane waves}~\cite{agostini}:
\begin{equation} 
\phi (x) = \int_{\mathbbm{R}^4} d^4 k \, \tilde \phi_r (k) \, e^{i k_i x^i} e^{i k_0 x^0} =  \int_{\mathbbm{R}^4} d^4 k \, \tilde \phi_l (k) \, e^{i k_0 x^0} e^{i k_i x^i}  =  \int_{\mathbbm{R}^4} d^4 k \, \tilde \phi_w (k) \, e^{i k_0 x^0 + i k_i x^i}  \,, \label{KappaFourierTransform}
\end{equation}
using the commutation relations~(\ref{kappaMinkowskiCommutationRelations}) we can prove the following relationships between waves written with different ordering conventions~\cite{flaaaaaaa}:
\begin{equation}\label{ChangeOfBasePlaneWaves}
e^{i k_0 x^0} e^{i k_i x^i}  = e^{i e^{\frac{k_0}{\kappa}} k_i x^i} e^{i k_0 x^0} \,,
\qquad
e^{i k_0 x^0 + i k_i x^i}  = e^{i \left(\frac{e^{k_0/\kappa} - 1}{k_0/\kappa}\right)  k_i x^i} e^{i k_0 x^0}  \,,
\end{equation}
and therefore the Fourier transform coefficients corresponding to differently-ordered waves are related to each other by coordinate transformations of momentum space:
\begin{equation}\label{DiffeoFourierCoefficients}
\tilde \phi_r (q_0 , {\vec q}) =  e^{-3 q_0/\kappa} \tilde \phi_l \left(q_0 , e^{-q_0/\kappa} {\vec q} \right) =
\frac{|q_0/\kappa|^3}{|e^{q_0/\kappa} -1|^3} \tilde \phi_w \left(q_0 ,\frac{q_0/\kappa}{(e^{q_0/\kappa} -1)} {\vec q} \right)  \,.
\end{equation}

We can find out how the $\kappa$-Poincaré  group acts on a scalar field by using the homomorphism property~\eqref{Coaction_kappa_Poincare_group} of the left coaction:
\begin{equation}
\Delta_L[\phi(x)] = \phi(\Delta_L[x]) =  \int_{\mathbbm{R}^4} d^4 k \, \tilde \phi_r (k) \, e^{i k_i \Delta_L[x^i]} e^{i k_0 \Delta_L[x^0]} 
\end{equation}
the right-ordered plane waves transform in the following way:
\begin{equation}\label{abbellliiii}
e^{i k_i (\Lambda^i{}_\nu \otimes x^\nu +  a^i \otimes \1)}
e^{i k_0 (\Lambda^0{}_\nu \otimes x^\nu +  a^0 \otimes \1)}
=
e^{i \lambda_i[k,\Lambda] \otimes x^i} e^{i \lambda_0[k,\Lambda] \otimes x^0} e^{i k_i a^i \otimes \1} e^{i k_0 a^0 \otimes \1} 
\end{equation}
where $\lambda_\mu[k,\Lambda]$ are four complicated, nonlinear function of $k_\mu$ and $\Lambda^\mu{}_\nu$, which we calculate explicitly (in the 1+1-dimensional case) in Appendix~\ref{AppendiceLorentzTransf}.

At first order in $\omega^\mu{}_\nu = \log \Lambda^\mu{}_\nu $ and $a^\mu$:
\begin{equation}\label{FirstOrderTransformOfPlaneWave}
 e^{i k_i \Delta_L[x^i]} e^{i k_0 \Delta_L[x^0]} 
 =
 \1 \otimes  e^{i k_i x^i} e^{i k_0 x^0} + \omega^\mu{}_\nu \otimes M^\nu{}_\mu \triangleright e^{i k_i x^i} e^{i k_0 x^0}
+ a^\mu \otimes P_\mu \triangleright e^{i k_i x^i} e^{i k_0 x^0}
+ \dots \,,
\end{equation}
the operators  $M^\mu{}_\nu$ and $P_\mu$ are the Lorentz and translation generators of the \emph{$\kappa$-Poincar\'e algebra,} which is dual to the quantum group. The symbol $\triangleright$ refers to an action of the $\kappa$-Poincar\'e algebra on $\mathcal A$.  From the last formula we deduce that, at first order in the transformation parameters, a scalar field transforms as
\begin{equation}\label{Exponential_formula_kP_group}
\Delta_L[\phi(x)] = \1 \otimes \phi +  \omega^\mu{}_\nu \otimes M^\nu{}_\mu \triangleright \phi
+ a^\mu \otimes P_\mu \triangleright  \phi  +  \dots 
\end{equation}
The operators $M_{\mu\nu}$ and $P_\mu$ close a Hopf algebra  $U_\kappa[\mathfrak{g}]$, which is dual to $\mathbbm{C}_\kappa[\mathcal G]$. Its commutators are\footnote{$N_i = M_{0i}$, $R_i = \frac 1 2 \epsilon_{ijk} M_{jk}$.}
\begin{equation}\label{kP-Algebra-commutators}
\begin{aligned}
[P_\mu, P_\nu] &= 0  \,, ~~~  [R_j , P_0] = 0 \,, ~~~  [R_j , P_k] = i \varepsilon_{jkl} P_l \,, ~~~  [R_j , N_k] = i \varepsilon_{jkl} N_l \,, ~~~  [R_j , R_k] = i \varepsilon_{jkl} R_l \,,
\\
[N_j , P_0] &= i P_j \,, \qquad [N_j , P_k] = \frac{i}{2} \delta_{jk} \left(   \kappa \left( 1 - e^{- 2 P_0/\kappa}\right)  + \frac{1}{\kappa}  |\vec{P}|^2 \right) - \frac{i}{\kappa} P_j P_k \,, \qquad [N_j,N_k] = -i \varepsilon_{jkl} R_l \,.
\end{aligned}
\end{equation}
Notice that $\left. \frac{\partial \lambda_\rho[P,\Lambda]}{\partial \Lambda^\mu{}_\nu} \right|_{\Lambda^\mu{}_\nu=\delta^\mu{}_\nu} =i [M^\mu{}_\nu , P_\rho ]$, and therefore the nonlinear commutators $[M^\mu{}_\nu , P_\rho ]$ encode the action of infinitesimal Lorentz transformations on momentum space. The coproducts, antipodes and counits of $U_\kappa[\mathfrak{g}]$ are\footnote{For a nice introduction to $\kappa$-Minkowski and $\kappa$-Poincar\'e see for instance \cite{review} and \cite{flaaaaaaa} for a more formal exposition. }
\begin{equation}
\begin{aligned}
\Delta [P_j] &=  P_j \otimes \1 +  e^{- P_0/\kappa} \otimes P_j \,, \qquad \Delta [P_0]  = P_0 \otimes \1 + \1 \otimes P_0 \,,
\\
\Delta [R_j] &= R_j \otimes \1 +  \1 \otimes R_j \,, \qquad  \Delta [N_k]  =  N_k \otimes \1 + e^{-P_0/\kappa} \otimes N_k  + \frac{i}{\kappa} \varepsilon_{klm}  P_l \otimes R_m \,,
\end{aligned}
\end{equation}
\begin{equation}
S[P_0] =  - P_0  \,, \qquad S[P_j] =  -e^{P_0/\kappa} P_j  \,, \qquad  S[R_j] =  - R_j    \,, \qquad  S[N_j] =  -e^{P_0/\kappa}N_j + \frac{i}{\kappa} \varepsilon_{jkl} e^{P_0/\kappa} P_k R_l  \,,
\end{equation}
\begin{equation}
\varepsilon[P_0] =  0 \,, \qquad  \varepsilon[P_j] =  0 \,, \qquad  \varepsilon[R_j] =  0  \,, \qquad \varepsilon[N_j]  = 0  \,.
\end{equation}
The coproducts, antipode and counit are deformations of the relations $\Delta[t] = t \otimes \1 + \1 \otimes  t$, $S[t] = -t$ and $ \varepsilon[t]=0$ which identify the Lie-algebra generator $t$ as a differential operator. In particular $\Delta[t]$ encodes the Leibniz rule for acting on products of functions, through the relation $t \triangleright (f \, g) = \mu \circ [\Delta[t] \triangleright (f \otimes g)] = (t \triangleright f) g + f (t \triangleright g)$.  In the case of  $U_\kappa[\mathfrak{g}]$ the coproduct above implies that only $P_0$ and $R_i$ will respect the Leibniz rule when acting on noncommutative products of functions. $P_j$ and $N_j$ will behave differently, \emph{e.g.} $P_j \triangleright (\phi \psi ) =  (P_j \triangleright \phi) \psi + (e^{-P_0/\kappa} \triangleright \phi) (P_j \triangleright \psi )$. 

The commutators~(\ref{kP-Algebra-commutators}) leave invariant the following function of the momentum generators:
\begin{equation}\label{kappa-Casimir}
\square_\kappa = - 4 \kappa^2  \sinh^2 \left( \frac{P_0}{2 \kappa} \right) + e^{\frac{P_0}\kappa} |\vec P |^2 \,,
\end{equation}
$\square_\kappa$ is a high-energy deformation of the quadratic Casimir of the Poincar\'e algebra. In fact, expanding in powers of $\kappa^{-1}$:
\begin{equation}\label{BicrossCasimir}
\square_\kappa = - P_0^2 + |\vec P |^2 + \frac{1}{\kappa} P_0 |\vec P |^2  + \mathcal{O}(\kappa^{-2}) \,,
\end{equation}
which is indistinguishable from $- P_0^2 + |\vec P |^2$ for $P_0 \ll \kappa$.

\subsection{The $\kappa$-momentum space}

\begin{figure}[h!]\center
\begin{tikzpicture}
\node[inner sep=0pt] (im1) at (0,0)
    {\includegraphics[width=.25\textwidth]{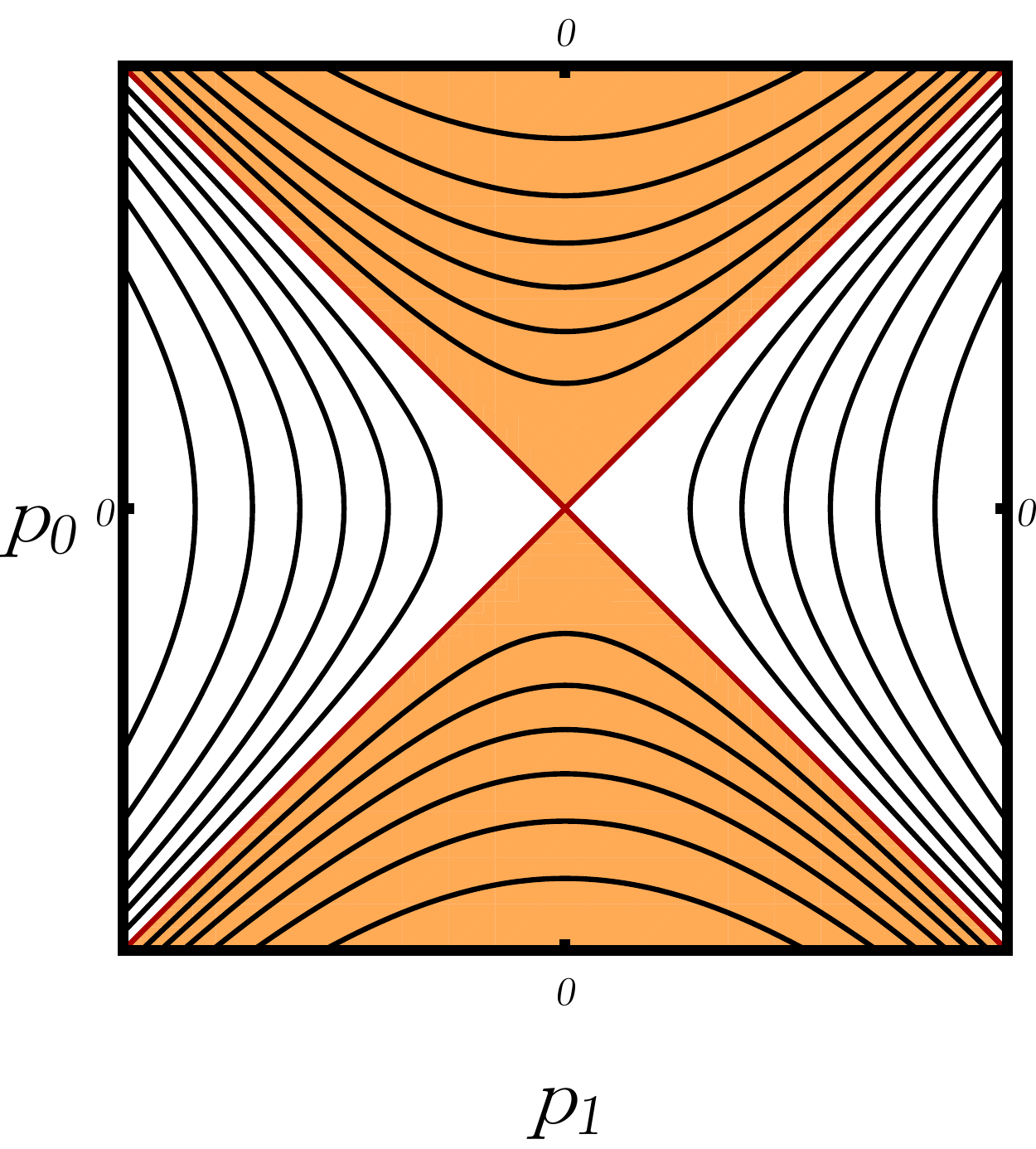}};
\node[inner sep=0pt] (im2) at (8,0)
    {\includegraphics[width=.6\textwidth]{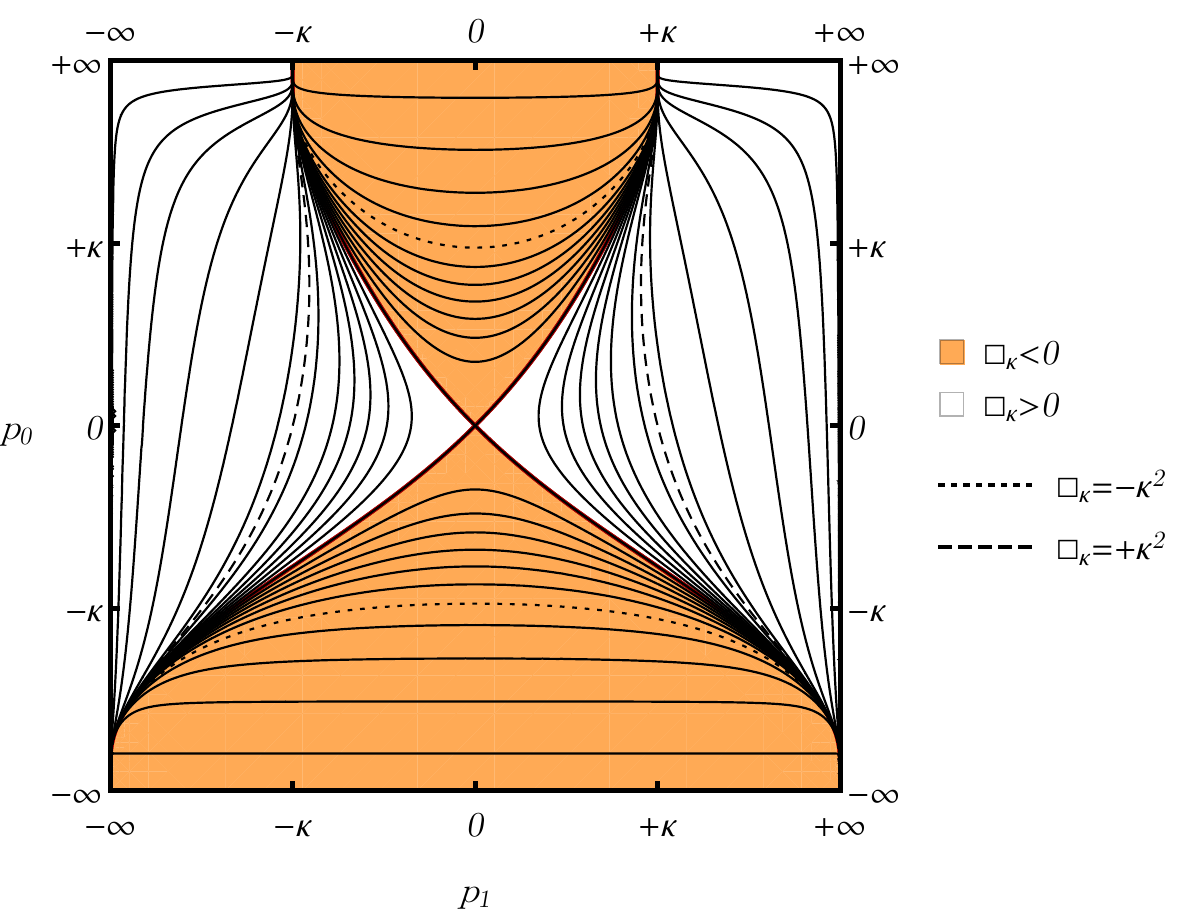}};
\draw[-,thick] (1.9,-1.5) -- (7.07,0.27);
\draw[-,thick] (1.9,1.95) -- (7.07,0.27);
\end{tikzpicture}
\caption{The $\square_\kappa = \text{\it const.}$ curves in momentum space in 1+1 dimensions. Both the vertical ($p_0$) and the horizontal ($p_1$) axes have been compactified by taking the $arctan$ of the variable. in a sufficiently small neighbourhood of the origin $p_\mu=0$ the diagram is indistinguishable from the mass-shell hyperboloids of the momentum space of waves on the commutative Minkowski spacetime.  This diagram is a compactified version of the diagram first appeared in Fig. 2 of \cite{majid22}. The orange region corresponds to negative $\square_\kappa$, which in the $\kappa \to \infty$ limit coincides with waves with positive squared mass. The white region, instead, tends to tachyonic (negative squared mass) waves in the  $\kappa \to \infty$ limit. 
Notice how the `future-directed' ($p_0>0$) and `past-directed' ($p_0<0$) mass shells are differently shaped. In particular, the future-directed one is spatially bounded $|p_1| < \kappa$ while the past-directed one is not. To give a sense of scale we plotted differently the on-shell curves with $\square_\kappa = \pm \kappa^2$.}
\label{Bicrossproduc_atlas_1}
\end{figure}

The time-to-the-right-ordered plane waves $e^{i p_i x^i} e^{i p_0 x^0}$ are eigenfunctions of the momentum operators $P_\mu$:
\begin{equation}
P_\mu \triangleright e^{i p_i x^i} e^{i p_0 x^0} = p_\mu \, e^{i p_i x^i} e^{i p_0 x^0} \,.
\end{equation}
Applying the $\kappa$-deformed Casimir operator~(\ref{kappa-Casimir}) to a plane wave:
\begin{equation}
\square_\kappa  \triangleright e^{i p_i x^i} e^{i p_0 x^0} = \left( - 4 \kappa^2  \sinh^2 \left( \frac{p_0}{2 \kappa} \right) + e^{\frac{p_0}\kappa} |\vec k |^2 \right) e^{i p_i x^i} e^{i p_0 x^0} \,,
\end{equation}
we can introduce a notion of `on-shell' wave, which satisfies a $\kappa$-deformed version of the Klein--Gordon equation:
\begin{equation}
- 4 \kappa^2  \sinh^2 \left( \frac{p_0}{2 \kappa} \right) + e^{\frac{p_0}\kappa} |\vec k |^2 = \text{const.} \,.
\end{equation}
Now consider the on-shell curves, in the space of momentum eigenvalues $p_\mu \in \mathbbm{R}^4$, associated to constant values of $\square_\kappa$. In Fig.~\ref{Bicrossproduc_atlas_1} we see that these curves are a high-energy deformation of the on-shell curves of Minkowski space.

\begin{figure}[h!]\center
\includegraphics[height=0.35\textheight]{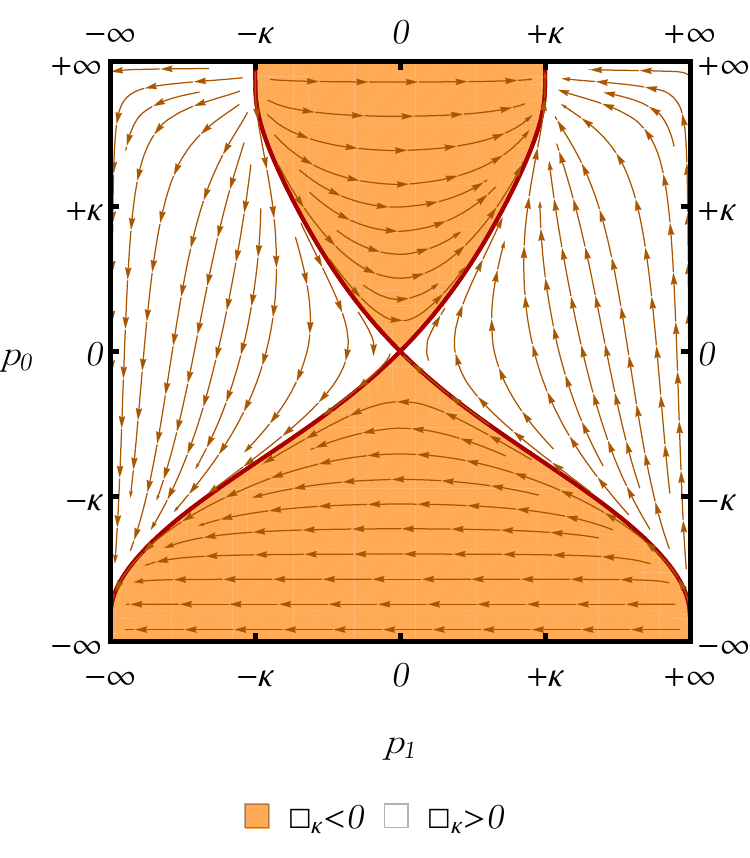}~~~~~
\includegraphics[height=0.35\textheight]{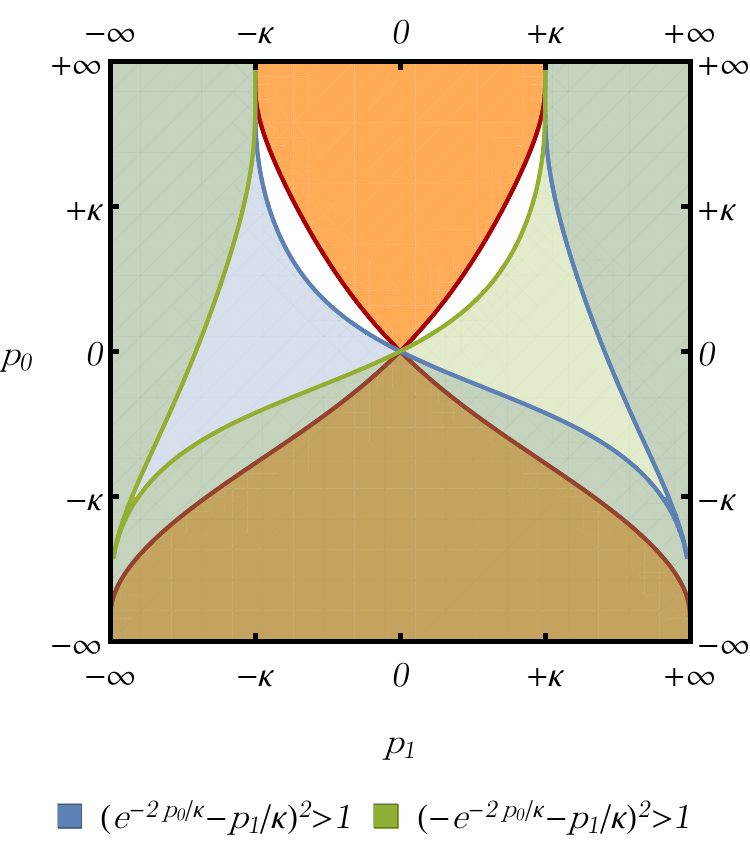}
\caption{Left: The Lorentz vector field on momentum space. Right: The regions where a critical rapidity exists. In the superposition between the blue and the green regions, there exist two values of the rapidity that make the boosted momenta diverge.}
\label{Bicrossproduc_atlas_vectorfield}
\end{figure}

We can study the action of Lorentz transformations on momentum space. In  1+1D, boost transformations of eq.\eqref{abbellliiii} can be written as (for the proof, see Appendix~\ref{AppendiceLorentzTransf}):
\begin{equation}\label{1+1DLorentzTransform}
\begin{gathered}
\lambda_0 [\xi,p] = p_0 + \kappa \log \left[ \left( \cosh {\frac \xi 2} + {\frac{p_1} \kappa} \sinh {\frac \xi 2} \right)^2 - e^{-2 p_0/\kappa} \sinh^2 {\frac \xi 2} \right] \,,
\\
\lambda_1 [\xi,p] = \kappa \frac{ \left( \cosh {\frac \xi 2} + {\frac{p_1} \kappa} \sinh {\frac \xi 2} \right) \left( \sinh {\frac \xi 2} + {\frac{p_1} \kappa} \cosh {\frac \xi 2} \right) -  e^{-2 p_0/\kappa} \cosh {\frac \xi 2}  \sinh {\frac \xi 2} }{\left( \cosh {\frac \xi 2} + {\frac{p_1} \kappa} \sinh {\frac \xi 2} \right)^2 - e^{-2 p_0/\kappa} \sinh^2 {\frac \xi 2} } \,.
\end{gathered}
\end{equation}
We plot the flux of the above transformation (\emph{i.e.} the vector field $\frac{\partial \lambda_\mu}{\partial \xi} \frac{\partial }{\partial p_\mu} $)  in Fig.~\ref{Bicrossproduc_atlas_vectorfield} (on the left).

Notice that both components of Eq.~(\ref{1+1DLorentzTransform}) have a divergence at
\begin{equation}
\left( \cosh {\frac \xi 2} + {\frac{p_1} \kappa} \sinh {\frac \xi 2} \right)^2 - e^{-2 p_0/\kappa} \sinh^2 {\frac \xi 2} = 0\,,
\end{equation}
which means
\begin{equation}
\coth {\frac \xi 2}  = \pm e^{2 p_0 / \kappa} - \frac{p_1}{\kappa}\,.
\end{equation}

Now, the image of $\coth$ is $(-\infty,-1) \cup (1 , +\infty) $, and therefore the above equation admits a solution in $\xi$ only wherever $|\pm e^{2 p_0 / \kappa} - \frac{p_1}{\kappa}| > 1$. This completely excludes the positive-frequency mass shell, and part of the $\square_\kappa >0$ white region of Fig.~\ref{Bicrossproduc_atlas_1}. We plot the two regions $| e^{2 p_0 / \kappa} - \frac{p_1}{\kappa}| > 1$ and $|- e^{2 p_0 / \kappa} - \frac{p_1}{\kappa}| > 1$ in Fig.~\ref{Bicrossproduc_atlas_vectorfield} (on the right). The existence of this `critical rapidity' was first noticed by Majid~\cite{majid22}. As can be seen in Fig.~\ref{Bicrossproduc_atlas_vectorfield}, the positive-frequency $\square_\kappa <0$  shell
does not have a critical rapidity: in there, any finite value of $\xi$ corresponds to a finite boosted momentum. All the other regions of momentum space, however, have this issue, and  the rapidities can only take either a finite interval or an interval bounded from above or below, because with a finite value of the rapidity one is boosted to the boundary of momentum space. This is in principle a serious issue which might spoil any theory based on this symmetry group of the equivalence between inertial observers. Tackling this issue will be an important part of our results.

\begin{figure}[b!]\center
\includegraphics[height=0.3\textheight]{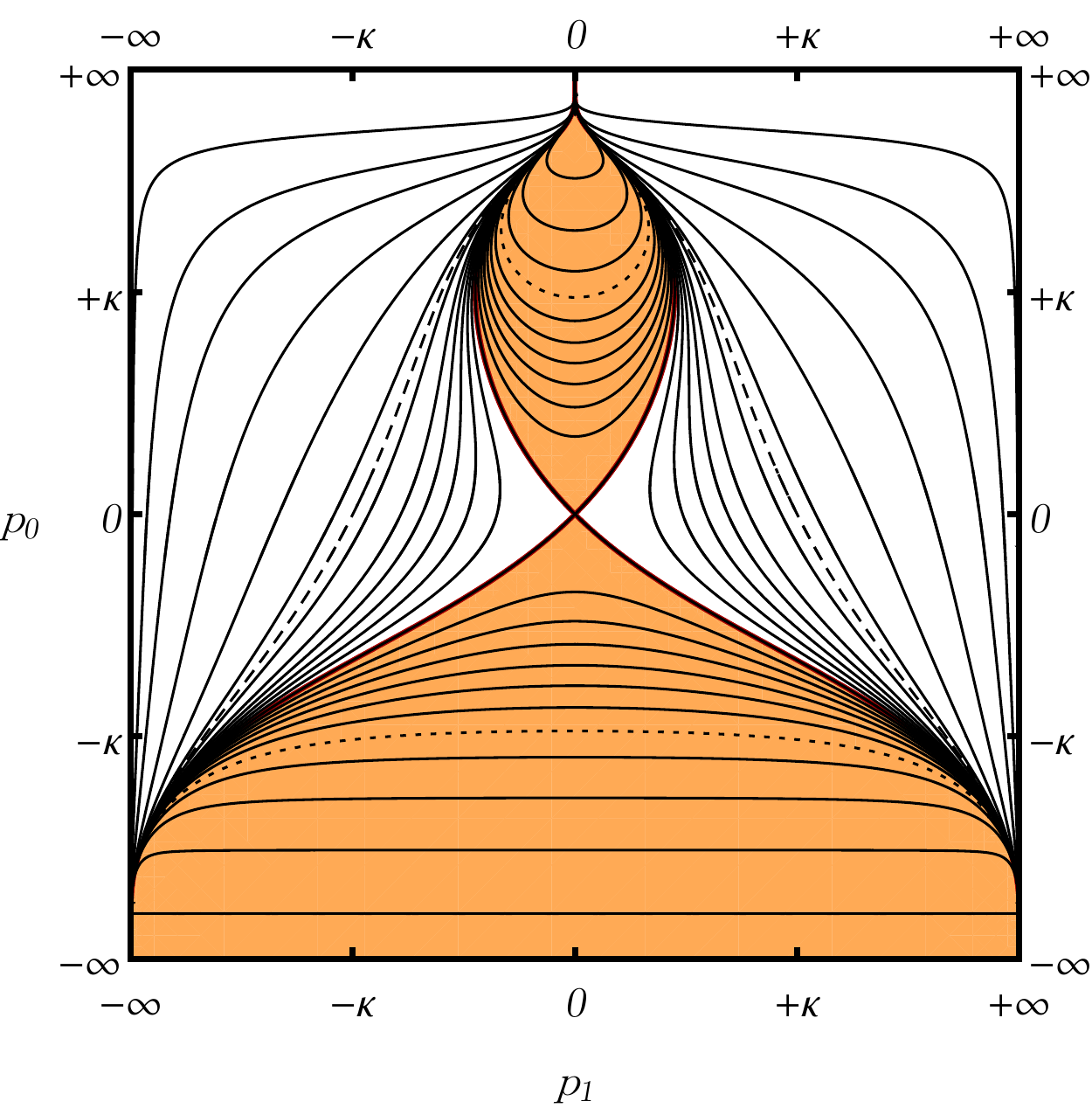}~\includegraphics[height=0.3\textheight]{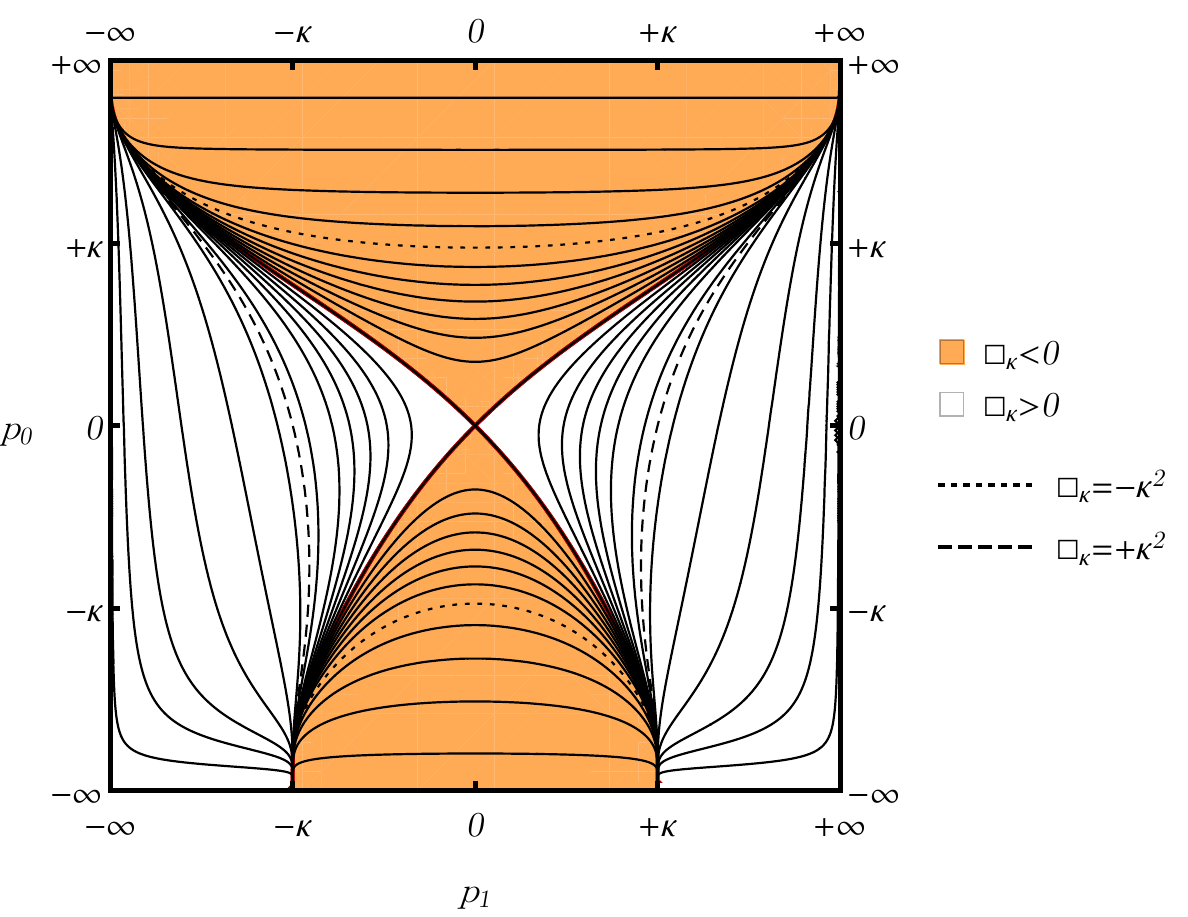}
\caption{The $\square_\kappa = \text{\it const.}$ curves in Weyl-ordered coordinates (left), and time-to-the-left-ordered coordinates (right).}
\label{Bicrossproduc_atlas_2}
\end{figure}

As we remarked in~(\ref{ChangeOfBasePlaneWaves}), changing the ordering coincides with a nonlinear redefinition of the eigenvalue $k_\mu$, for example $k_i \to e^{\frac{k_0}{\kappa}} k_i$, $k_0 \to k_0$ to go from the right-ordered to the left-ordered plane waves. This, in turn, means that the left-ordered plane waves are eigenfunctions of the  operators $P_i^l = e^{\frac{P_0}{\kappa}} P_i$, which are just a different basis for the  $U_\kappa[\mathfrak{g}]$ algebra. In Fig.~\ref{Bicrossproduc_atlas_2} we plot the on-shell curves in the Weyl-ordered and the time-to-the-left-ordered coordinates. We can see that most qualitative features of these diagrams depend on the ordering choice, and therefore on the Hopf algebra basis.

\subsection{The geometry of momentum space}

More than one author couldn't resist the temptation to interpret expressions like (\ref{BicrossCasimir}) literally, as relations between the physical (observable) energy and momentum carried by a wave in a noncommutative spacetime. Then one would be led to calculate things like the group velocity of the wave as  the slope of the on-shell curves. $\partial p_0/\partial p_1$, and deduce, for example, that the positive-frequency waves of Fig.~\ref{Bicrossproduc_atlas_1} have an infinite group velocity as the spatial momentum approaces $|p_1| \to \kappa$. The group velocity of negative-frequency waves instead would become infinite only as $|p_1| \to \infty$. However, if we adopted the convention of ordering our waves with the time to the left,  positive-and negative-frequency waves would swap their behaviours, as suggested by the diagram on the right-hand-side of Fig.~\ref{Bicrossproduc_atlas_2}. Even more queer would be the behaviour of the Weyl-ordered waves:   the left-hand-side of Fig.~\ref{Bicrossproduc_atlas_2} suggests that positive-frequency waves have a multi-valued dispersion relation, that associates two different energies to each spatial momentum (and the group velocity would diverge at a value of $|p_1|$ that is even smaller than $\kappa$).

It is of course unacceptable that our physical conclusions should depend on the convention we adopt to order our operators. Considering different ordering choices, however, suggests the way out of this problem: changing ordering corresponds to making a \emph{general coordinate transformation} on our momentum space. Then we should be looking for statements that are coordinate-independent: \emph{those that regard the pseudo-Riemannian geometry of momentum space.}

Kowalski--Glikman was the first to observe that the momentum space of $\kappa$-Minkowski has the geometry of a de Sitter manifold~\cite{firstds}.  Here we would like to repeat the argument presented in~\cite{fla}, which starts with the observation that there exists a change of basis:
\begin{equation}\label{DefEta0Eta1}
\eta_0 = \kappa \sinh \frac{P_0}\kappa + \frac 1 {2\kappa} e^{\frac{P_0}\kappa} |\vec P |^2 \,, 
\qquad
\eta_i = e^{\frac{P_0}\kappa} P_i \,, 
\end{equation}
such that  $\eta_\mu$ closes a Poincar\'e algebra with  the Lorentz generators $M_{\mu\nu}$:
\begin{equation}\label{PoincareAlgebraEtaCoordinates}
[M_{\mu\nu} , \eta_\rho] = i \, \eta_{\mu\rho} \eta_\nu - i \, \eta_{\nu\rho} \eta_\mu \,, 
\end{equation}
as can be explicitly deduced from the definition of $\eta_\mu$ and the commutation relations~(\ref{kP-Algebra-commutators}). The generators $\eta_\mu$ transform classically under Lorentz transformations. However they do not close a Hopf algebra: their coproducts are
\begin{equation}
\Delta[\eta_0] = \eta_0 \otimes e^{P_0/\kappa} + e^{-P_0/\kappa} \otimes \eta_0 + {\frac 1 \kappa} e^{-P_0/\kappa} \eta_i \otimes \eta_i \,,
\qquad
\Delta[\eta_i] = \eta_i \otimes e^{P_0/\kappa} + \1 \otimes \eta_i \,,
\end{equation}
\begin{equation}
S[\eta_0] = - \eta_0 + e^{- P_0 / \kappa} |\vec \eta |^2   \,,
\qquad
S[\eta_i] =  - e^{- P_0 / \kappa} \eta_i \,,
\end{equation}
but we cannot express $e^{P_0/\kappa}$ uniquely in terms of $\eta_0$, $\eta_1$. The reason is that Eq.(\ref{DefEta0Eta1}) as a transformation $(P_0,P_i) \to (\eta_0,\eta_i)$ is not injective, because it admits two inversions:
\begin{equation}
e^{\frac{P_0}\kappa} = \eta_0 \pm \sqrt{\eta_0^2-|\bm{\eta}|^2-\kappa^2} \,,
\qquad
P_i = \eta_i \frac{\kappa \left(\eta_0 \mp \sqrt{\eta_0^2 - |\bm{\eta}|^2-\kappa^2}\right)}{|\bm{\eta}|^2-\kappa^2} \,,
\end{equation}
so the coalgebra does not close. If we now introduce a new algebra element:
\begin{equation}
\eta_4 = \kappa \cosh \frac{P_0}\kappa - \frac 1 {2\kappa} e^{\frac{P_0}\kappa} |\vec P |^2\,,
\end{equation}
the coordinate transformation $(P_0,P_i) \to (\eta_0,\eta_i,\eta_4)$ becomes injective (although it is not surjective, because the target space is 5-dimensional)
\begin{equation}\label{Introduction_of_eta4}
e^{\frac{P_0}\kappa} = \frac{\eta_0 + \eta_4}{\kappa} \,,
\qquad
P_i = \eta_i \frac{ \kappa \left(\eta_0 - \eta_4\right)}{|\bm{\eta}|^2-\kappa^2} \,,
\end{equation}
and then the coproducts close on the $(\eta_0,\eta_i,\eta_4)$ basis:
\begin{equation}\label{eta_a_coproducts}
\begin{gathered}
\Delta[\eta_0] = \eta_0 \otimes \left(\eta_0 + \eta_4\right) + \frac{\kappa}{\eta_0 + \eta_4} \otimes \eta_0 + \frac{\eta_i}{\eta_0 + \eta_4} \otimes \eta_i \,,
~~~
\Delta[\eta_i] = {\frac 1 \kappa} \eta_i \otimes \left(\eta_0 + \eta_4\right)  + \1 \otimes \eta_i \,,
\\
\Delta[\eta_4] = \eta_4 \otimes \left(\eta_0 + \eta_4\right) - \frac{\kappa}{\eta_0 + \eta_4} \otimes \eta_0 - \frac{\eta_i}{\eta_0 + \eta_4} \otimes \eta_i \,,
\end{gathered}
\end{equation}
as well as the antipodes:
\begin{equation}\label{eta_a_antipodes}
S[\eta_0] = - \eta_0 + \frac{|\vec \eta |^2}{\eta_0 + \eta_4}   \,,
\qquad
S[\eta_i] =  - \frac{\kappa \, \eta_i}{\eta_0 + \eta_4} \,,
\qquad
S[\eta_4] = \eta_4 \,.
\end{equation}

\begin{figure}[t!]\center
\includegraphics[height=0.4\textheight]{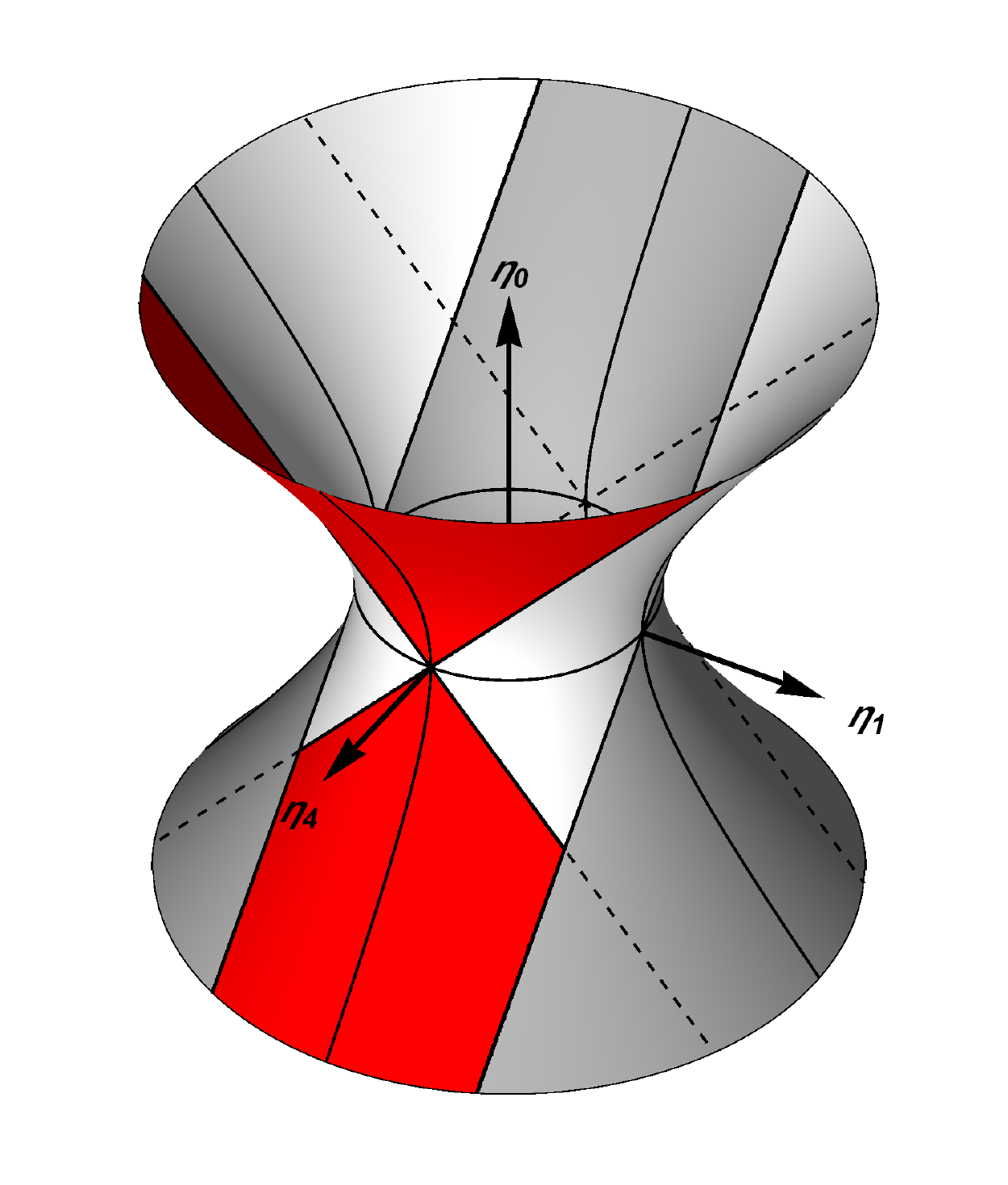}\!\!\!\!\!\!\!\!\!\!\!\!\!\!\!\!\!\raisebox{-0.5cm}{\includegraphics[height=0.47\textheight]{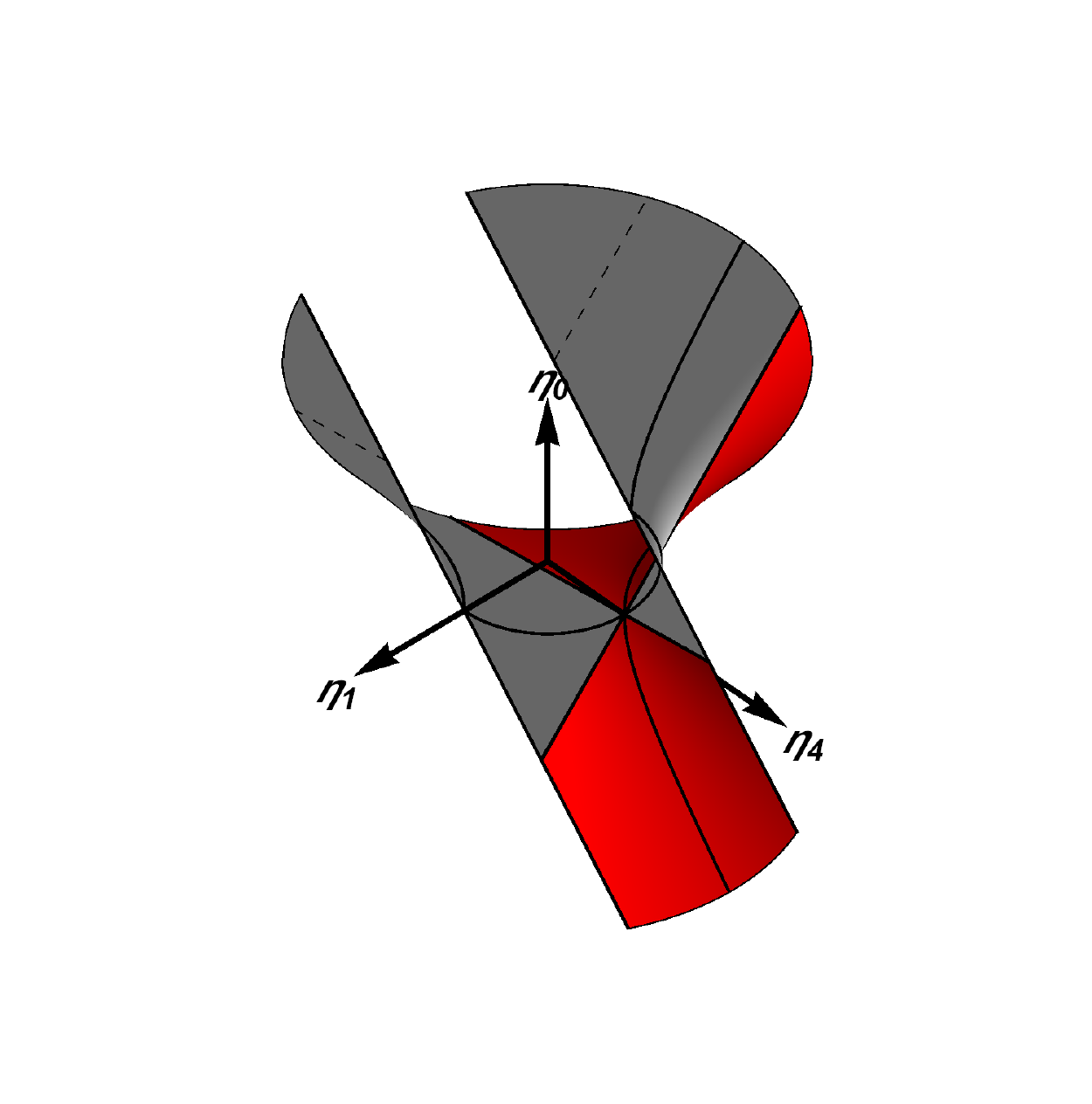}}
\vspace{-1cm}
\caption{Left: $\kappa$-Minkowski's de Sitter momentum space in embedding coordinates $\eta_a$. The dark grey area is not covered by the bicrossproduct coordinates. The red and white areas correspond, respectively, to $\square_\kappa <0$ and $\square_\kappa >0$.
Right: The part of de Sitter space that is covered by bicrossproduct coordinates, from a different perspective
}
\label{dS_momentum_space}
\end{figure}

Notice that the definitions of $\eta_a = \eta_a(P_0, \vec P)$, $a=0,\dots,4$, are such that the five $\eta_a$ are not independent: they satisfy an algebraic relation:
\begin{equation}\label{dSequation}
\eta_0^2 - |\bm{\eta}|^2 - \eta_4^2 = -\kappa^2 \,,
\end{equation}
which is the equation for the 4D de Sitter hyperboloid in 5-dimensional embedding coordinates. Together with the counits:
\begin{equation}
\varepsilon[\eta_0] = 0 \,, \qquad
\varepsilon[\eta_i] = 0 \,, \qquad
\varepsilon[\eta_4] = \kappa \,, 
\end{equation}
the coproducts~(\ref{eta_a_coproducts}) and antipodes~(\ref{eta_a_antipodes}) satisfy the Hopf algebra axiom~(\ref{HopfAxiom}) only \emph{on-shell}, that is, after imposing Eq.~(\ref{dSequation}).

We conclude that $\eta_a = \eta_a(P_0, \vec P) : \mathbbm{R}^4 \to dS_4$ are a coordinatization of de Sitter space. Differently-ordered bases just correspond to different ways of coordinatizing de Sitter space with $\mathbbm{R}^4$. In other words, different bases are related by diffeomorphisms.
The metric induced on $dS$ by the flat ambient metric:
\begin{equation}
\diff  s^2 = - \diff  \eta_0^2 + \diff  \eta_1^2 + \diff  \eta_2^2 + \diff  \eta_3^2+ \diff  \eta_4^2 = - \diff  P_0^2 + e^{2 P_0/\kappa} \left(\diff P_1^2 + \diff P_2^2 + \diff P_3^2 \right) \,,
\end{equation}
is, in right-ordered bicrossproduct coordinates, the de Sitter metric in \emph{comoving coordinates:}
\begin{equation}\label{dS_metric_comoving}
g_{\mu\nu} = \text{diag} \left(-1, e^{2 P_0/\kappa} , e^{2 P_0/\kappa} , e^{2 P_0/\kappa} \right)\,,
\end{equation}
so we may conclude that right-ordered coordinates are comoving coordinates on momentum space.

Notice now that the coordinates $\eta_a = \eta_a(P_0, \vec P)$ only cover half of de Sitter space. In fact from Eq.~(\ref{Introduction_of_eta4}) we see that, for $P_0$ real, $\eta_0 + \eta_4 = \kappa e^{\frac{P_0}\kappa} >0$. The region $\eta_0 + \eta_4$ cuts the de Sitter hyperboloid in the way shown in Fig.~\ref{dS_momentum_space}. 

\begin{figure}[t!]\center
\includegraphics[height=0.4\textheight]{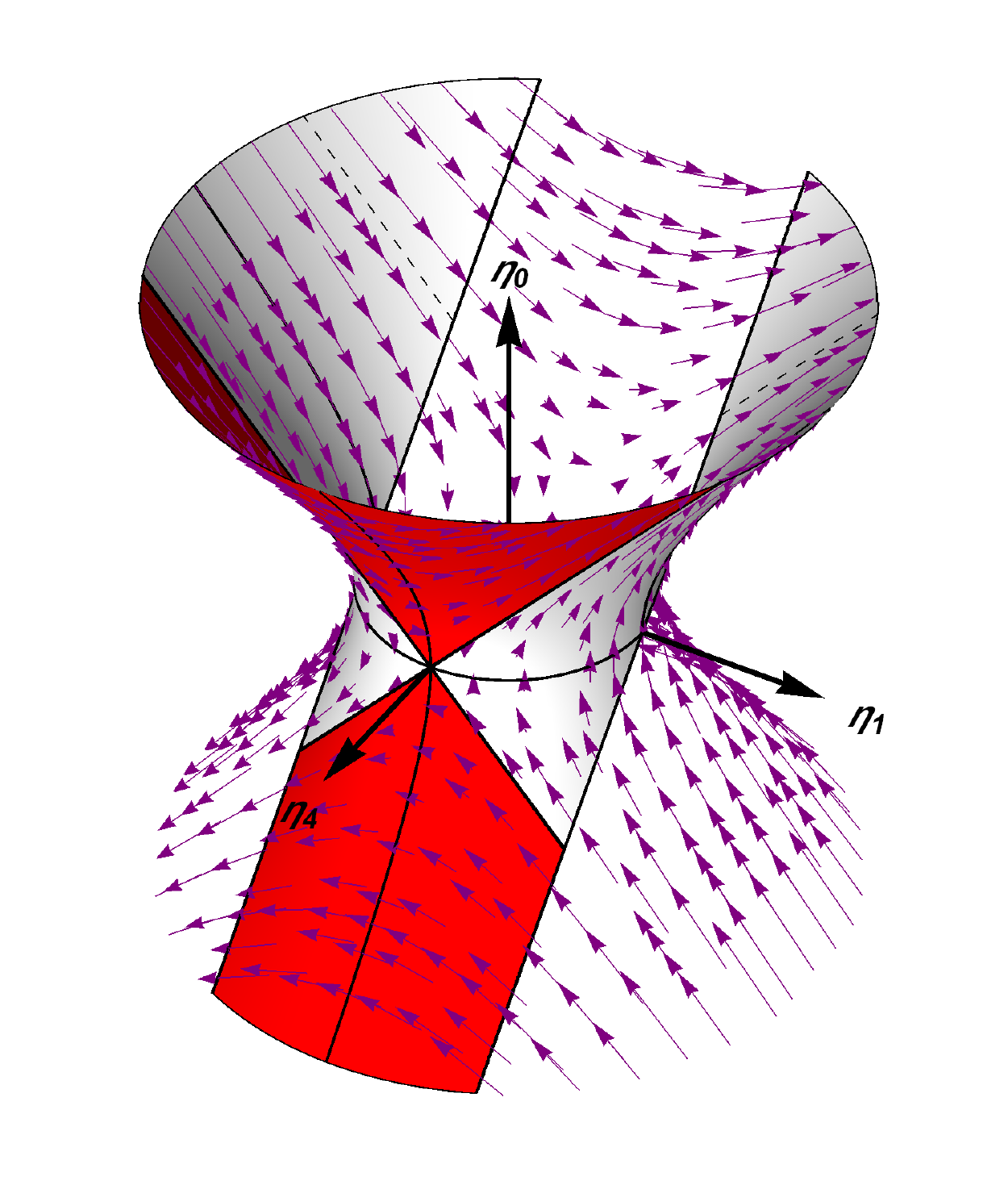}~~~
\includegraphics[height=0.4\textheight]{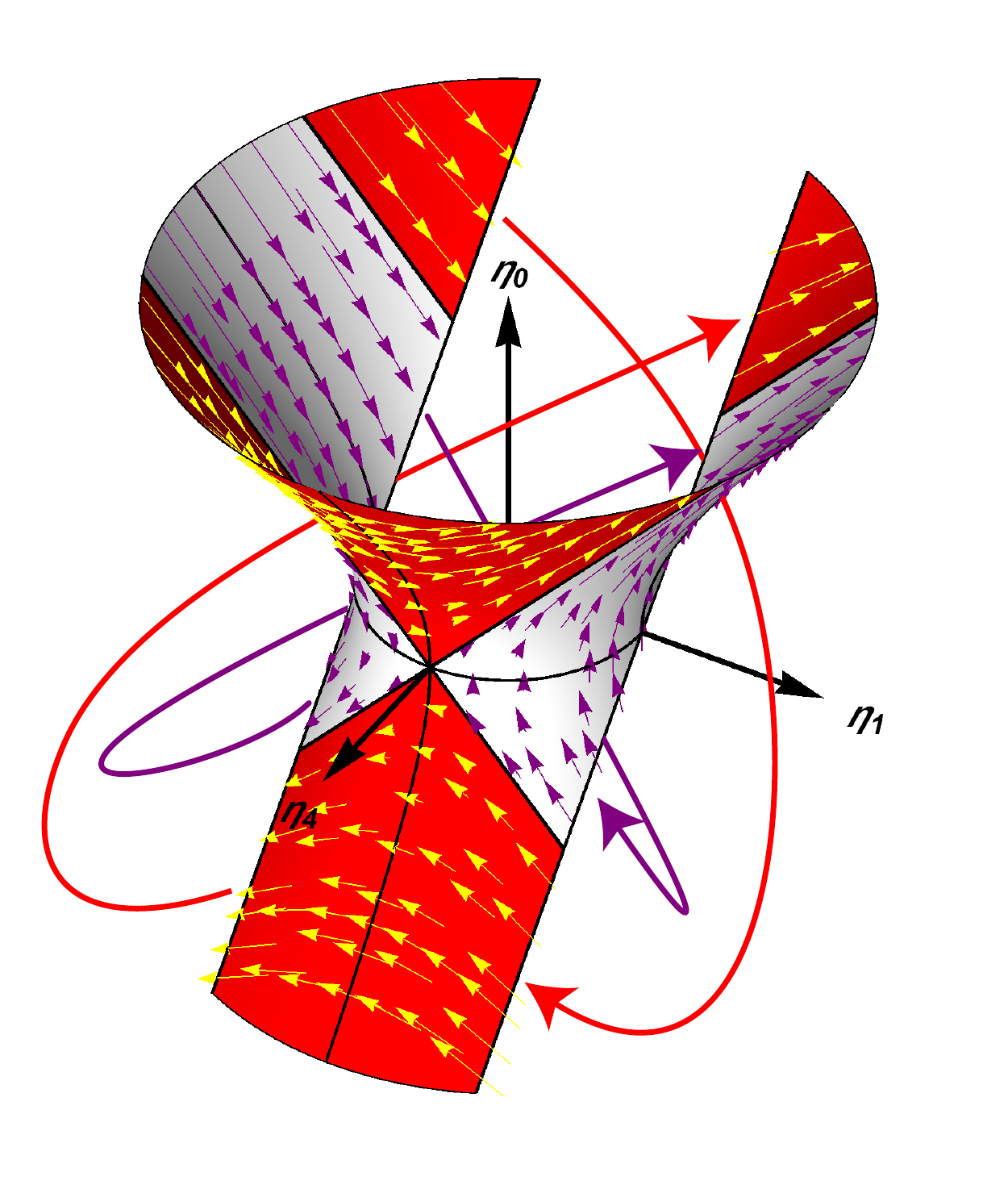}
\caption{Left: Lorentz flow on de Sitter momentum space. It is apparent that a finite boost can bring a point outside of the region covered by the bicrossproduct coordinates. Right: Elliptic boundary conditions on the patch of $dS$ momentum space covered by the bicrossproduct coordinates, and how to continue the Lorentz vector flow through the boundaries. Notice how the region that is connected by Lorentz transformations to the past light cone of the origin is now in red.
}
\label{Elliptic_dS_momentum_space_Lorentz_flow}
\end{figure}

Eq.~(\ref{PoincareAlgebraEtaCoordinates}) implies that the Lorentz transformations act in an undeformed way on the $\eta_\mu$ coordinates. The $\eta_4$ coordinate, on the other hand, is Lorentz-invariant: in fact it is a function of the Casimir~(\ref{kappa-Casimir}):
\begin{equation}
\eta_4 = \kappa - \frac{\square_\kappa}{2\kappa} \,.
\end{equation}
Then the Lorentz group acts on the de Sitter momentum space as the group of isometries that stabilize the `origin' point\footnote{Notice how the coordinates of the `origin' point are given by the counits of $\eta_a$: ~$\varepsilon(\eta_4) = \kappa$, $\varepsilon(\eta_\mu)=0$.} $\eta_4 = \kappa$, $\eta_\mu=0$ (see Fig.\ref{Elliptic_dS_momentum_space_Lorentz_flow}). It is obvious that these transformations do not leave the region covered by the bicrossproduct coordinates, $\eta_0+\eta_4 >0$, invariant. Moreover it takes a finite rapidity to bring a point in the region $\eta_0+\eta_4 >0$ outside of it, unless this point is in the future light cone of the `origin' -- the only region which is closed under Lorentz transformations.

The region $\eta_0 + \eta_4>0$ is sent to its complement by reflections $\eta_a \to - \eta_a$. Consequently, if we quotient de Sitter space by reflections, we obtain a manifold that is covered uniquely by the bicrossproduct coordinates. This manifold is a known and well-studied example of non-time-orientable solution of Einstein's equations: it is called `elliptic de Sitter spacetime'~\cite{ellds, FKKKN},  $dS/
\mathbbm{Z}_2$. This manifold is motivated by the fact that the whole $dS$ hyperboloid is `too much' as a cosmological model. In fact, unlike what happens in Minkowski spacetime, an observer located on a timelike curve can't be in causal contact with more than `half' of de Sitter space. The causal domain of a timelike curve is precisely a region of the same form as $\eta_0 + \eta_4>0$ (all these causal domains can be obtained by rotating $\eta_0 + \eta_4>0$ around the $\eta_0$ axis. They are parametrized only by the endpoint of their defining worldline on the boundary of $dS$ space.\footnote{One can convince themselves of this by observing that the past lightcone of a point on a timelike worldline is included in the past lightcone of any future point on the worldline. Therefore to find the causally-connected region to the whole worldline it is sufficient to consider the past lightcone of the `last' point on the worldline at the infinite boundary of $dS$ spacetime. This is given, in the ambient space, by a half-space delimited by a 45$^\circ$ plane which intersects the centre of the $dS$ hyperboloid. One such plane is $\eta_0 + \eta_4 = 0$.}

Both $dS$ and $dS/
\mathbbm{Z}_2$ locally solve Einstein's equations with a positive cosmological constant, under the assumption of spacetime homogeneity. What distinguishes them is the assumed topology. Similarly, in realizing that the $\kappa$-Poincar\'e momentum space has a $dS$ geometry, we used the local structure of its symmetries (the $\kappa$-Poincar\'e algebra) while tacitly making an assumption regarding its global topology, which in truth we are free to choose. The global behaviour of the Lorentz flow on momentum space reveals that the standard $dS$ topology $\mathbbm{R} \times S_3$ is not adequate, as it leads to a singular Lorentz flow. Assuming the $dS/\mathbbm{Z}_2$ topology solves this issue, because now the Lorentz flow lines close.

Now, under the elliptic identification the Casimir $\square_\kappa $ is not continuous: it is negative in the orange region of Fig.~\ref{Bicrossproduc_atlas_1}, but it changes sign when a Lorentz transformation brings us from the `past light cone' region through its boundary into the negative-$\eta_4$ region.
The way the elliptic identification is defined, a Lorentz transformation bringing us through the boundary of the bicrossproduct patch will preserve the absolute value of $\eta_4$ but flip its sign. 
$\square_\kappa $  is therefore not a good generalization of the D'Alembert operator, as it fails to be Lorentz-invariant at the global level. In particular, upon crossing the boundary of the bicrossproduct patch, $\square_\kappa \to 4 \kappa^2 - \square_\kappa$. Then it is obvious that any function of the quantity
\begin{equation}
\mathcal{C}_\kappa = \left( 1 - \frac{\square_\kappa}{4 \kappa^2} \right) \square_\kappa
\end{equation}
will be genuinely Lorentz-invariant all along any Lorentz orbit.
The above function on momentum space will be our choice for D'Alembert operator. It is negative-definite in the on-shell regions and unbounded. Moreover it reduces to $\square$ in the $\kappa \to \infty$ limit. This operator was already proposed as the generalization of the D'Alembert operator in~\cite{koso} and used also in other words, like~\cite{FKKKN}. However, the authors of~\cite{koso} had a different reason to introduce $\mathcal C_\kappa$: one obtains the same operator by defining the D'Alembert operator as $\eta_\mu \eta^\mu$, where the four $\eta_\mu$ are those defined in~\eqref{DefEta0Eta1}.

\usetikzlibrary{shapes.misc}

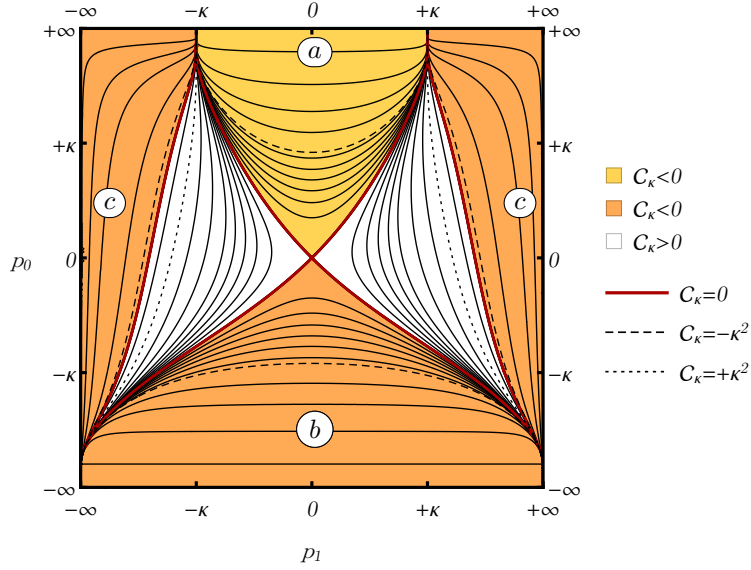
\begin{figure}[t!]\center
\begin{tikzpicture}
\node[inner sep=0pt] (im2) at (8,0)
    {\includegraphics[width=.6\textwidth]{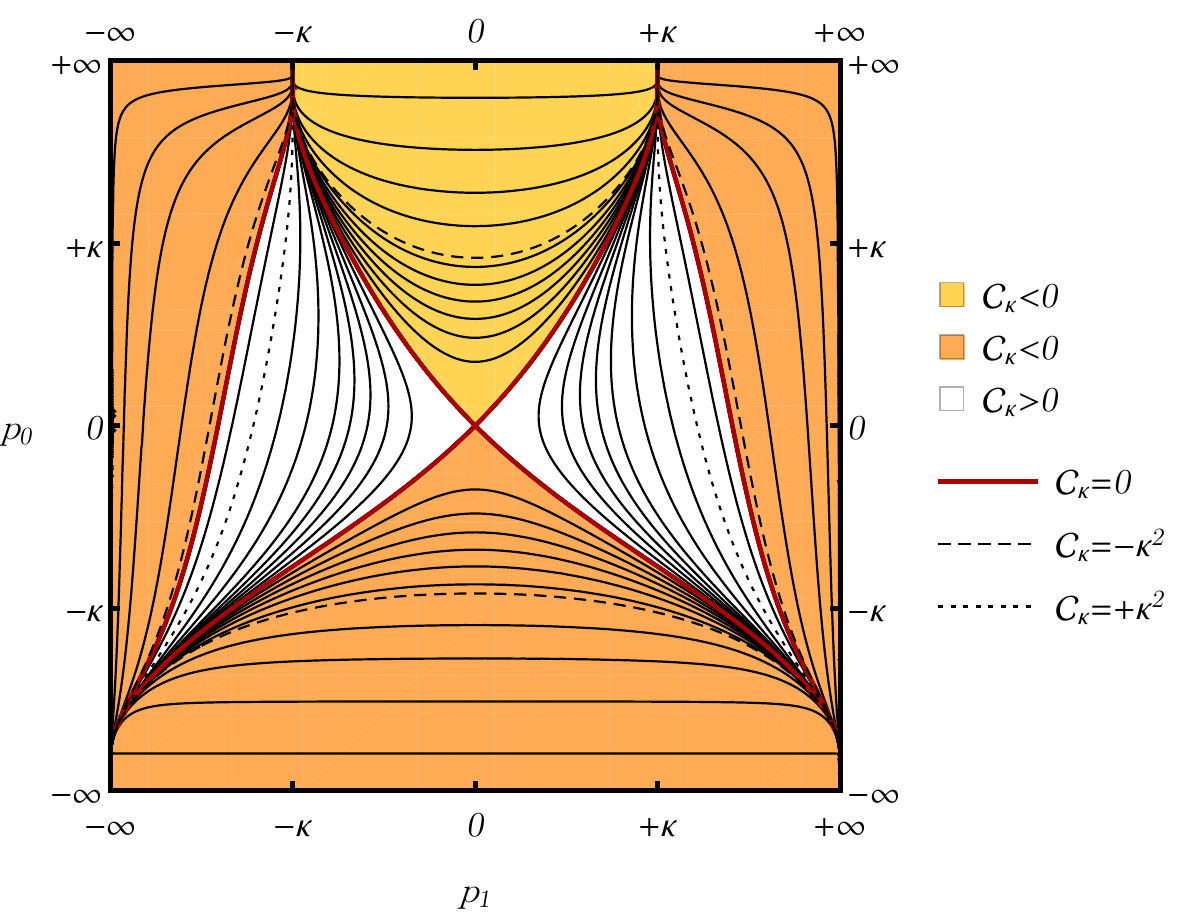}};
\node[draw, fill=white, rounded rectangle] at (7.05,3) {$a$};
\node[draw, fill=white, rounded rectangle] at (7.05,-2) {$b$};
\node[draw, fill=white, rounded rectangle] at (7.05-2.7,1) {$c$};
\node[draw, fill=white, rounded rectangle] at (7.05,-2) {$b$};
\node[draw, fill=white, rounded rectangle] at (7.05+2.7,1) {$c$};
\end{tikzpicture}
\caption{Momentum space in bicrossproduct coordinates. The negative-$\mathcal C_\kappa$ region (the one with real mass) has been divided into its two connected components under Lorentz transformations: the positive-frequency one `$a$' (in yellow) and the negative-frequency one, `$b$' and `$c$' (in orange). Moreover some special curves have been highlighted: the zero-mass one (in red), one with minimal value of $\mathcal C_\kappa=-\kappa^2$  and the one with $\mathcal C_\kappa = + \kappa^2$.}
\label{Bicrossproduc_atlas_Casimir2}
\end{figure}

\section{Free quantum $\kappa$-Klein--Gordon field}\label{FreeQuantumKleinGordonFieldSection}

Now that we know that momentum space is a pseudo-Riemannian manifold, we can observe that the noncommutative functions written in Fourier transform w.r.t. a basis of ordered plane waves, as in~(\ref{KappaFourierTransform}), have Fourier coefficients $\tilde \phi_r$ or  $\tilde \phi_w$ that transform, under a change of ordering (that is, a diffeomorphism of momentum space), as \emph{scalar densities}. Eq.~(\ref{DiffeoFourierCoefficients}) expresses this fact: the factors 
$e^{-3 q_0/\kappa}$ and 
$\frac{|q_0/\kappa|^3}{|e^{q_0/\kappa} -1|^3} $
are nothing else than the Jacobians $\det \left( 
\partial q_\mu / \partial q'_
\nu \right)$ of the coordinate transformations
$(q'_0 , {\vec q}') =  \left( q_0 , e^{-q_0/\kappa}  {\vec q} \right)$ and, respectively,
$(q'_0 , {\vec q}') =\left(q_0 ,q_0{\vec q} / (\kappa e^{q_0/\kappa} - \kappa) \right)$.
Then we can write the Fourier coefficients, \emph{e.g.} in the right-ordered coordinates, as a scalar field on momentum space times a volume density 
\begin{equation}
\sqrt{-g(k)} = \sqrt{- \det g_{\mu\nu}(k)} = e^{3 \frac{k_0}{\kappa}}\,,
\end{equation}
and a scalar field $\phi(x)$ can be written, in right-ordered momentum space coordinates:
\begin{equation}\label{RightFourierTransform}
\phi (x) = \int d^4 k \sqrt{-g(k)} \phi_r (k) \, e^{i k_i x^i} e^{i k_0 x^0} \,,
\end{equation}
where $g_{\mu\nu}$ is the $dS$ metric in comoving coordinates introduced in Eq.~(\ref{dS_metric_comoving}).

%

Our noncommutative generalization of the complex Klein--Gordon equation will be
\begin{equation}
\mathcal{C}_\kappa \triangleright \phi = - m^2 \, \phi \,, \qquad \mathcal{C}_\kappa \triangleright \phi^\dagger = - m^2 \, \phi^\dagger \,,
\end{equation}

A solution to the above equations of motion can be written as
\begin{equation}\label{4Dintegral_KG_solution}
\phi (x) = \frac 1 {2m} \int \diff^4 p \sqrt{-g(p)} \sqrt{ - g^{\mu\nu}(p) \frac{\partial \mathcal{C}_\kappa}{\partial p^\mu} \frac{\partial \mathcal{C}_\kappa}{\partial p^\nu} } \delta \left( \mathcal{C}_\kappa(p) + m^2\right) \, \phi_r(p) \, : e^{i p_\mu x^\mu} : \,, 
\end{equation}
where the expression $\sqrt{ - g^{\mu\nu}(p) \frac{\partial \mathcal{C}_\kappa}{\partial p^\mu}\frac{\partial \mathcal{C}_\kappa}{\partial p^\nu} }$ makes the integral invariant under reparametrizations of the on-shell curve, and the numerical factor $\frac 1 {2m}$ ensures that the expression has the right commutative limit. The previous expression is invariant under diffeomorphisms in momentum space.

We can solve the on-shell condition $\mathcal{C}_\kappa(p) = - m^2$ with respect to the $p_0$ coordinate, and rewrite the delta function term as
\begin{equation}
\begin{aligned}
\frac{1}{2m}\sqrt{ - g^{\mu\nu}(p) \frac{\partial \mathcal{C}_\kappa}{\partial p^\mu} \frac{\partial \mathcal{C}_\kappa}{\partial p^\nu} } \delta (\mathcal{C}_\kappa + m^2) =  \frac{\sqrt{\kappa ^2+m^2}}{\kappa} \delta (\mathcal{C}_\kappa + m^2 ) \\
= \frac{1}{2\sqrt{m^2 + |\vec p|^2}}  \left[ \delta (p_0 - \omega^+(|\vec p|) ) + \delta (p_0 - \omega^-(|\vec p|) ) \right],
\end{aligned}
\end{equation}
with 
\begin{equation}
\omega^\pm(|\vec p |)   = - \frac{1}{2} \kappa  \log \left(1 +\frac{2 m^2+| \vec p |^2  \mp 2 \sqrt{\left(\kappa ^2+m^2\right) \left(m^2+| \vec p |^2\right)}}{\kappa ^2}\right),
\end{equation}
which tends to the usual  $\pm\sqrt{\vec{p}^2+m^2}$ in the $\kappa\to \infty$ limit.
Notice that the $\omega^+(|\vec p |)$ solutions split into two, one set defined only in the region $|\vec p |<\kappa$  and the other in $|\vec p | > \kappa$. This is because at $|\vec p |=\kappa$,  $\omega^-(|\vec p |)$ has a (coordinate) singularity:
\begin{equation}
\lim_{|\vec p | \to \kappa} \omega^+(|\vec p |)   =  + \infty \,.
\end{equation}
The two parts of $\omega^+(|\vec p |)$ belong to the regions $a$ (when $|\vec p | < \kappa$) and $c$ (when $|\vec p | > \kappa$) of momentum space. For this reason, when integrating Eq.~(\ref{4Dintegral_KG_solution}) w.r.t $p_0$ we write
\begin{equation}\label{3Dintegral_KG_solution}
\begin{aligned}
\phi (x) =& \int_{|\vec p | < \kappa} \frac{\diff^3 p \,  e^{\frac{3\omega^+(|\vec p |)}{\kappa}}}{2 \sqrt{m^2 + |\vec p|^2}}  \, \phi_a({\vec p}) \,  e^{i {\vec p } \cdot  {\vec x}}  e^{i \omega^+(|\vec p |) x^0} 
+
\int_{\mathbbm{R}^3}  \frac{\diff^3 p \, e^{\frac{3\omega^-(|\vec p |)}{\kappa}}}{2 \sqrt{m^2 + |\vec p|^2}}  \, \phi_b({\vec p}) \,  e^{i {\vec p } \cdot  {\vec x}}  e^{i \omega^-(|\vec p |) x^0} +
\\
& \int_{|\vec p | > \kappa} \frac{\diff^3 p \, e^{\frac{3\omega^+(|\vec p |)}{\kappa}}}{2 \sqrt{m^2 + |\vec p|^2}}  \, \phi_c({\vec p}) \,  e^{i {\vec p } \cdot  {\vec x}}  e^{i \omega^+(|\vec p |) x^0} \,.
\end{aligned} 
\end{equation}
The coefficients $\phi_a(\vec p)$ are closed under Lorentz flow, while the coefficients $\phi_b(\vec p)$ and $\phi_c(\vec p)$ flow into each other.

The delta function~(\ref{4Dintegral_KG_solution}) should not be recasted in terms of an integral over a contour on the complex plane of $p_0$ (as was done in early works like \cite{koso,koso2}). In fact, in doing so, one necessarily breaks the invariance under diffeomorphisms of momentum space, which has been one of the guiding principles of our current analysis. A manifestation of this problem is the fact that the function $(\mathcal{C}_\kappa(p) + m^2)$ has infinitely many zeroes in the complex plane of $p_0$:
\begin{equation}
p_0 = \omega^\pm(|\vec p |) + i \, \pi \, \kappa \, n \,, \qquad n \in \mathbbm{Z} \,,
\end{equation} 
this is due to the fact that $\mathcal{C}_\kappa(p)$ depends on $p_0$ only through the function $e^{\frac{p_0}{2\kappa}}$ and therefore it is periodic in the imaginary direction, with period $\pi$. But a simple change of variable, \emph{e.g.} $q= e^{\frac{p_0}{2 \kappa}}$, removes all of the non-real zeroes from the complex plane of $q$. It is clear then that complexifying one coordinate is a non-diffeomorphism-invariant operation.

In \cite{koso,koso2} the complex zeroes of $\mathcal{C}_\kappa(p)$ were interpreted as poles of the Green function, and a summation over this infinite tower of poles was proposed. We believe that such a proposal violates one of the basic symmetries of the theory (namely, the independence on the coordinate system in momentum space), and therefore it is not tenable. Fortunately one is not obligated to use residues and contour integrals for what we are concerned with in this paper, and we will be able to use more geometric alternatives, which are explicitly invariant under general changes of coordinates in momentum space.

A classical scalar field is an element of the algebra $\mathcal A$. In the non-interacting case, scalar fields are quantized by simply replacing the coefficients $a(\vec p)$ and $b(\vec p)$ with operators on a Hilbert space $\mathcal H$ (Fock space). The natural generalization of this to the noncommutative case is to take quantum fields to be elements of $\mathcal A \otimes \mathcal H$, and replace $\phi_a(\vec p)$, $\phi_b(\vec p)$ and $\phi_c(\vec p)$ with operators on $\mathcal H$.
Then this will be our definition of noncommutative quantum scalar field:
\begin{equation}\label{3Dintegral_quantum_field}
\begin{aligned}
\hat \phi (x) =& \int_{|\vec p | < \kappa} \frac{\diff^3 p \,  e^{\frac{3\omega^+(|\vec p |)}{\kappa}}}{2 \sqrt{m^2 + |\vec p|^2}}  \, \ahat({\vec p}) \,  e^{i {\vec p } \cdot  {\vec x}}  e^{i \omega^+(|\vec p |) x^0} 
+
\int_{\mathbbm{R}^3}  \frac{\diff^3 p \, e^{\frac{3\omega^-(|\vec p |)}{\kappa}}}{2 \sqrt{m^2 + |\vec p|^2}}  \,  \bhat^\dagger ({\vec p}) \,  e^{i {\vec p } \cdot  {\vec x}}  e^{i \omega^-(|\vec p |) x^0} +
\\
& \int_{|\vec p | > \kappa} \frac{\diff^3 p \, e^{\frac{3\omega^+(|\vec p |)}{\kappa}}}{2 \sqrt{m^2 + |\vec p|^2}}  \, \chat^\dagger ({\vec p}) \,  e^{i {\vec p } \cdot  {\vec x}}  e^{i \omega^+(|\vec p |) x^0} \,.
\end{aligned} 
\end{equation}

\subsection{Hermitian conjugacy}

$\mathcal{C}_\kappa \triangleright \phi = - m^2 \, \phi$  implies $\mathcal{C}_\kappa \triangleright \phi^\dagger = - m^2\phi^\dagger$, because the Hermitian conjugate acts on~(\ref{4Dintegral_KG_solution}) by replacing $:e^{i p_\mu x^\mu}:$ with $:e^{i S(p_\mu) x^\mu}:$, and the Klein--Gordon operator is invariant under antipode: $\mathcal{C}_\kappa(S(p)) = \mathcal{C}_\kappa(p)$.
To find how Hermitian conjugacy acts on creation and annihilation operators, we first write
\begin{equation}\label{Phi_dagger_1}
\begin{aligned}
\hat \phi^\dagger (x) =& \int_{|\vec p | < \kappa} \frac{\diff^3 p \,  e^{\frac{3\omega^+(|\vec p |)}{\kappa}}}{2 \sqrt{m^2 + |\vec p|^2}}  \, \ahat^\dagger ({\vec p }) \,  e^{i S_+({\vec p }) \cdot  {\vec x}}  e^{-i \omega^+(|\vec p |) x^0} 
+
\int_{\mathbbm{R}^3}  \frac{\diff^3 p \, e^{\frac{3\omega^-(|\vec p |)}{\kappa}}}{2 \sqrt{m^2 + |\vec p|^2}}  \, \bhat ({\vec p }) \,  e^{i S_-({\vec p }) \cdot  {\vec x}}  e^{-i \omega^-(|\vec p |) x^0} +
\\
& \int_{|\vec p | > \kappa} \frac{\diff^3 p \, e^{\frac{3\omega^+(|\vec p |)}{\kappa}}}{2 \sqrt{m^2 + |\vec p|^2}}  \, \chat ({\vec p }) \,  e^{i S_+ ({\vec p }) \cdot  {\vec x}}  e^{-i \omega^+(|\vec p |) x^0} \,,
\end{aligned} 
\end{equation}
where
\begin{equation}
{\vec S}_\pm({\vec p }) = - e^{\frac{\omega^\pm({\vec p})}{\kappa}} {\vec p} \,.
\end{equation}
Now we would like, in all of the three integrals in~(\ref{Phi_dagger_1}), to make the following change of variables:
\begin{equation}
{\vec S}_\pm(\vec p) =  \vec q \,,
\end{equation}
the following relations:
\begin{equation}\label{SofSrelations}
\begin{aligned}
&{\vec S}_-\left({\vec S}_+({\vec p }) \right)= {\vec p }  ~~ \text{if} ~ |\vec{p}| < \kappa\,,&
&{\vec S}_+\left({\vec S}_-({\vec p }) \right)= {\vec p }  ~~  \forall ~ \vec{p} \in \mathbbm{R}^3 \,,&
&{\vec S}_+\left({\vec S}_+({\vec p }) \right)= {\vec p }  ~~ \text{if} ~ |\vec{p}| > \kappa\,,&
\end{aligned}
\end{equation}
which are useful, respectively, in region $a$, $b$ and $c$, 
allow us to invert  ${\vec S}_\pm(\vec p) =  \vec q $ for $\vec{p}$:
\begin{equation}
{\vec S}_+(\vec p) =  \vec q  ~~ \Rightarrow ~~ 
\left\{ \begin{aligned}
|\vec{p}| < \kappa \qquad {\vec p} = {\vec S}_-(\vec{q}) \,, ~~ \vec{q} \in \mathbbm{R}^3 \,,
\\
|\vec{p}| > \kappa \qquad {\vec p} = {\vec S}_+(\vec{q}) \,, ~~ |\vec{q}| > \kappa
\end{aligned}\right. \,,
\qquad 
{\vec S}_-(\vec p) =  \vec q  ~~ \Rightarrow ~~ {\vec p} = {\vec S}_+(\vec{q}) \,,  ~~ |\vec{q}| < \kappa
\end{equation}
So in region $a$ we have to make the substitution ${\vec p} = {\vec S}_-(\vec{q})$, where now the integration domain for $\vec{q}$ is all of $\mathbbm{R}^3$. In region $b$  we replace  ${\vec p} = {\vec S}_+(\vec{q})$, while integrating $\vec{q}$ only in the region $|\vec{q}| < \kappa$. Finally, in $c$ too the transformation is ${\vec p} = {\vec S}_+(\vec{q})$, but the integration region is $|\vec{q}| > \kappa$. These transformations act on the frequencies in the following way:
\begin{equation}\label{Omega_S_relations}
\begin{aligned}
 &\omega^- ({\vec S}_+(\vec{q}))=-\omega^+(\vec{q}) ~~ \text{if} ~  |\vec{q}| < \kappa  \,,&
 \!\!
 &\omega^+ ({\vec S}_-(\vec{q}))= -\omega^-(\vec{q}) ~~ \forall ~ \vec{q} \in \mathbb{R}^3 \,,&
 \!\!
 & \omega^+ ({\vec S}_+(\vec{q}))=-\omega^+(\vec{q}) ~~ \text{if} ~   |\vec{q}|>\kappa \,,&
\end{aligned}
\end{equation}
again, the three relations above are useful, respectively, in region $a$, $b$ and $c$.
Finally, the integration measure transforms in the following way:
\begin{equation}
\frac{\diff^3 p}{2 \sqrt{m^2 + |\vec p|^2}}
\xrightarrow[\vec p \to {\vec S}_\pm(\vec q)]{} \frac{\diff^3 q \,  e^{\frac{- 3\omega^\pm(|\vec q |)}{\kappa}}}{2 \sqrt{m^2 + |\vec q |^2}} \,,
\end{equation}
so, applying the coordinate transformations to  the three integrals in~(\ref{Phi_dagger_1}), the expression for the Hermitian conjugate scalar field becomes
\begin{equation}\label{Phi_dagger_2}
\begin{aligned}
\hat \phi^\dagger (x) = & \int_{\mathbbm{R}^3} \frac{\diff^3 q}{2 \sqrt{m^2 + |\vec q|^2}}  \ahat^\dagger ({\vec S}_-({\vec q })) \,  e^{i {\vec q} \cdot  {\vec x}}  e^{i \omega^-(|\vec q |) x^0} +
\int_{|\vec q | < \kappa}  \frac{\diff^3 q}{2 \sqrt{m^2 + |\vec q|^2}} \bhat ({\vec S}_+({\vec q })) \,  e^{i {\vec q } \cdot  {\vec x}}  e^{i \omega^+(|\vec q |) x^0} +
\\
&\int_{|\vec q | > \kappa} \frac{\diff^3 q}{2 \sqrt{m^2 + |\vec q|^2}}   \chat ({\vec S}_+({\vec q })) \,  e^{i {\vec q } \cdot  {\vec x}}  e^{i \omega^+(|\vec q |) x^0} \,.
\end{aligned} 
\end{equation}

\section{Pauli--Jordan Function}\label{PauliJordanSec}

Scalar field theory quantization on $\kappa$-Minkowski was studied by several authors in the past. Among others, we believe the most remarkable approaches are the ones pursued in~\cite{koso,FKKKN,marcianoo,arzanooooo}. Each of those results contributed to our present understanding of field theory on $\kappa$-Minkowski. For example~\cite{koso} was the first to consider the Green functions and the Pauli--Jordan function. \cite{FKKKN} exploited the de Sitter geometry of momentum space to define field theory in a Lorentz invariant way, and was the first to observe that the half-cover of de Sitter space offered by the bicrossproduct coordinates is sent to its complement by the elliptic identification map.

Early studies like \cite{koso,koso2} lacked the present understanding of the geometry of momentum space. This led to choices that we do not consider physical, for example in~\cite{koso} the positive- and negative-frequency modes in the Pauli--Jordan function are distinguished by the sign of $p_0$, a choice that is not Lorentz-covariant in our setting. In~\cite{koso2} on the other hand, the tower of complex solutions of the $\kappa$-deformed on-shell relation  we discussed in Sec.~\ref{FreeQuantumKleinGordonFieldSection} was considered a physical feature. This, as we argued in the preceding Sections, is incompatible with momentum-space diffeomorphism invariance.

More recent studies, \emph{e.g.}~\cite{FKKKN,arzanooooo} take into account the de Sitter geometry of momentum space,  however they made use of Hamiltonian/canonical methods which are not explicitly spacetime-covariant. Expressions involving only-space integrals are hard to make sense in $\kappa$-Minkowski, turning out to be basis-dependent. In general one can note that $\kappa$-Minkowski commutation rules imply uncertainty relations of the kind:
\begin{equation}
\delta t \,\delta x \sim \frac{1}{2 \kappa}\langle x\rangle,
\end{equation}
thus choosing a state on the algebra~\eqref{kappaMinkowskiCommutationRelations} such that $\delta t=0$  necessarily sets $\langle x\rangle=0$. As a consequence, defining equal-time canonical commutation rules such as
\begin{equation}
[\dot{\phi}(0, \vec{x}), \phi(0, \vec{y})]=-i \delta^3(\vec{x}-\vec{y}),
\end{equation}
is an ill-defined procedure.

An approach that is devoid of any of the problems listed above is~\cite{MicheleAntonino1,marcianoo} and following works~\cite{Arzano2007,MicheleDarioStatistics,Arzano2011}. These papers use a covariant symplectic form to define the canonical structure of the theory. This, being defined on the space of solutions, does not require a hypersurface of simultaneity or space-only integrals. The problem of the non-closure of the region  `$b$' under Lorentz transformations is solved, in~\cite{marcianoo}, by restricting the Hilbert space to the `$a$' region, and demanding reality conditions on the field. This was extended to a complex field in~\cite{arzanooooo}, where the  non-closure of  region  `$b$' under Lorentz transformations was solved by defining the Lorentz transformations in region  `$b$' as the image, under antipode, of the Lorentz transformations on region `$a$' (which, we recall, leaves  region `$a$' closed).

In the present paper, we will follow a different approach, motivated by the request of invariance under coordinate transformations in momentum space. This immediately exclude the possibility of using Hamiltonian/canonical quantization methods, which rely on splitting time and 3-dimensional hypersurfaces of simultaneity. We have to rely on a covariant quantization scheme. In order to implement our momentum-space general coordinate invariance, we find the best-suited approach is to define the Pauli--Jordan function, an approach already explored in~\cite{koso} (without however attempting to implement coordinate invariance).

In the commutative case the covariant scalar field commutators $[\hat \phi(x),  \hat  \phi(y)]= i\Delta_\text{PJ}(x, y)$ are given by
\begin{equation}\label{commpj}
\begin{aligned}
 i\Delta_\text{PJ}(x, y)=i\Delta_\text{PJ}(x-y)&= \int \frac{\diff^3 p}{2\sqrt{\vec{p}^2+m^2}}\left(e^{ip\cdot(x-y)}- e^{-ip\cdot(x-y)}\right)\\&=\int \diff^4 p ~ \text{sign}(p_0)~ \delta(p^2-m^2)e^{ip\cdot(x-y)},
\end{aligned}
\end{equation}
the $\Delta_\text{PJ}$ function satisfies
\begin{equation}\label{propppp}
(\square_x+m^2)\Delta_\text{PJ}(x,y) = (\square_y+m^2)\Delta_\text{PJ}(x,y) = 0, 
\,.
\end{equation}
Moreover $\Delta_\text{PJ}$ is antisymmetric in its two variables, $\Delta_\text{PJ}(y,x)=-\Delta_\text{PJ}(x,y)$, and it is zero outside of the light cone:
\begin{equation}
\Delta_\text{PJ}(x,y) = 0  \qquad \text{if} \qquad (x-y)^2 > 0 \,,
\end{equation}
this last property is the most interesting one from the physical point of view, because it allows to introduce a notion of light cone in a purely field-theoretical fashion.

We want to generalize $\Delta_\text{PJ}$ to $\kappa$-Minkowski. First of all we need a generalization of the notion of \emph{bilocal function}. The only such generalization we are aware of was introduced in~\cite{2point}: a function of two points is assumed to be an element of $\mathcal A \otimes \mathcal A$. The basis elements of the algebra $\mathcal A \otimes \mathcal A$ generalize the notion of two independent points $z$ and $y$:
\begin{equation}
z^\mu = x^\mu \otimes \1 \,, \qquad y^\mu = \1 \otimes x^\mu \,.
\end{equation}
With this construction we can generalize the properties (\ref{propppp}) when $\Delta_\text{PJ}(z,y) \in \mathcal{A} \otimes \mathcal{A}$:
\begin{equation}
\left( (\mathcal{C}_\kappa + m^2)  \otimes \text{id} \right) \triangleright \Delta_\text{PJ}(z,y) = \left( \text{id} \otimes  (\mathcal{C}_\kappa + m^2)  \right) \triangleright \Delta_\text{PJ}(z,y) = 0 \,,
\end{equation}
The following expression satisfies the condition above and generalizes the Pauli--Jordan function in a way that is diffeomorphism-invariant on momentum space:
\begin{equation}\label{Noncommutative_Pauli-Jordan_4D}
i \, \Delta_\text{PJ} (z,y) =  \frac 1 {2m} \int \diff^4 p \sqrt{-g(p)} \varepsilon(p) \sqrt{ - g^{\mu\nu}(p) \frac{\partial \mathcal{C}_\kappa}{\partial p^\mu} \frac{\partial \mathcal{C}_\kappa}{\partial p^\nu} } \delta (\mathcal{C}_\kappa + m^2 ) : e^{i p_\mu z^\mu} e^{i S(p)_\mu y^\mu} : \,,
\end{equation}
where $:~:$ refers to time-to-the-right ordering, and:
\begin{equation}
\varepsilon(p) = \left\{ \begin{array}{ll}
+1  \qquad &\text{in region} ~ a
\\
-1 &\text{in region} ~ b
\\
-1 &\text{in region} ~ c
\end{array}\right. \,.
\end{equation}
The form of the $\varepsilon$ function has been chosen in order to be invariant under the Lorentz flow, which connects regions $b$ and $c$.
The function $\Delta_\text{PJ}$ introduced here is also invariant under $\kappa$-Poincar\'e transformations, understood in the following sense: $z^\mu \to \Lambda^\mu{}_\nu \otimes z^\nu + a^\mu \otimes \1 \otimes \1$,  $y^\mu \to \Lambda^\mu{}_\nu \otimes y^\nu + a^\mu \otimes \1 \otimes \1$, that is:
\begin{equation}
z^\mu = x^\mu \otimes \1  \to \Lambda^\mu{}_\nu \otimes x^\nu \otimes \1 + a^\mu \otimes \1 \otimes \1 \,, \qquad y^\mu = \1 \otimes x^\mu \to \Lambda^\mu{}_\nu \otimes \1 \otimes x^\nu \otimes \1 + a^\mu \otimes \1 \otimes \1 \,.
\end{equation}
We can show that the exponentials $ : e^{i p_\mu z^\mu} e^{i S(p)_\mu y^\mu} : $ transform in the following way:
\begin{equation}
e^{i p_i z^i} e^{i p_0 z^0} e^{i S(p)_i y^i} e^{i S(p)_0 y^0}  \to
 e^{i \lambda_i[\xi,p] \otimes z^i} e^{i \lambda_0[\xi,p] \otimes z^0}  e^{i S[\lambda_i[\xi,p]]\otimes  y^i} e^{i S[\lambda_0[\xi,p]] \otimes y^0} \,,
\end{equation}
where $\lambda_\mu[\xi,p]$ is the 3+1-D generalization of the Lorentz transformation of momenta~(\ref{1+1DLorentzTransform}). The expression~(\ref{Noncommutative_Pauli-Jordan_4D}) is invariant under Lorentz transformations, so changing variables $p_\mu \to \lambda_\mu[\xi,p]$ allows to prove that
\begin{equation}\label{Invariance_PJ_under_kPoincare}
\Delta_\text{PJ}(\Lambda \otimes z + a \otimes \1 \otimes \1 , \Lambda \otimes y + a \otimes \1 \otimes \1) = \1 \otimes \Delta_\text{PJ}(z,y) \,, 
\end{equation}
\emph{i.e.} our noncommutative Pauli--Jordan function is $\kappa$-Poincar\'e invariant.


Solving the delta function in~(\ref{Noncommutative_Pauli-Jordan_4D}) we get 
\begin{equation}\label{Noncommutative_Pauli-Jordan_3D}
\begin{aligned}
i \Delta_\text{PJ}(z,y) 
= & \int_{|\vec p|<\kappa} \!\!\!\!\!\!\!   \diff^3 p  \frac{ e^{3 \frac{\omega^+ }\kappa}  e^{i   {\vec p} \cdot {\vec z} - i e^{\frac{\omega^+ }\kappa} {\vec p} \cdot {\vec y}  }  e^{i \omega^+ (z^0 - y^0) }  }{ 2 \sqrt{ \left(m^2+| \vec p |^2\right)}} - \int_{\mathbbm{R}^3} \diff^3 p  \frac{ e^{3 \frac{\omega^- }\kappa}  e^{i {\vec p} \cdot {\vec z}} e^{- i  e^{\frac{\omega^- }\kappa} {\vec p} \cdot {\vec y}}  e^{i \omega^- (z^0 - y^0) }  }{ 2 \sqrt{\left(m^2+| \vec p |^2\right)}} 
\\
&- \int_{|\vec p|>\kappa}  \!\!\!\!\!\!\! \diff^3 p  \frac{ e^{3 \frac{\omega^+ }\kappa}  e^{i {\vec p} \cdot {\vec z}} e^{- i  e^{\frac{\omega^+ }\kappa} {\vec p} \cdot {\vec y}}  e^{i \omega^+ (z^0 - y^0) }  }{ 2 \sqrt{ \left(m^2+| \vec p |^2\right)}}
\,.
\end{aligned}
\end{equation}
The Pauli--Jordan-like function we defined is Poincar\'e invariant but it is not Hermitian nor antisymmetric under exchange of $z$ and $y$. However it satisfies the following conjugacy relation:
\begin{equation}
\Delta_\text{PJ}^\dagger (z,y) = \Delta_\text{PJ}(y,z) \,,
\end{equation}
because 
\begin{equation}
\left(  : e^{i p_\mu z^\mu} e^{i S(p)_\mu y^\mu} : \right)^\dagger =
 : e^{i S(p)_\mu z^\mu} e^{i p_\mu y^\mu} : =  : e^{i p_\mu y^\mu} e^{i S(p)_\mu z^\mu}  : \,.
\end{equation}
This property is exactly what we expect from a field commutator:
\begin{equation}
\left( [ \hat \phi(z),  \hat \phi^\dagger(y) ] \right)^\dagger = [  \hat \phi(y),  \hat \phi^\dagger(z)] \,.
\end{equation}

\subsection{Covariant commutation relations}

Consider now the following commutation relations:
\begin{equation}\label{CovariantCommutators}
[ \hat \phi(z),  \hat \phi^\dagger(y) ]
= i \, \Delta_\text{PJ} (z,y)\,,
\qquad 
[ \hat \phi(z),  \hat \phi(y) ] =[ \hat \phi^\dagger(z),  \hat \phi^\dagger(y) ]= 0 \,,
\end{equation}
where of course we interpret $\hat \phi(z)$ as $ \hat \phi(x) \otimes \1$ and  $\hat \phi(y)$ as $\1 \otimes \hat \phi(x)$.
We can deduce the corresponding commutation relations for the creation and annihilation operators, $\ahat(\vec{p})$, $\bhat(\vec{p})$, $\chat(\vec{p})$,  $\ahat^\dagger(\vec{p})$, $\bhat^\dagger(\vec{p})$, $\chat^\dagger(\vec{p})$, by replacing the expression~(\ref{3Dintegral_quantum_field}) for $\hat \phi(z)$ and~(\ref{Phi_dagger_2}) for $\hat \phi^\dagger(y)$ into~(\ref{CovariantCommutators}).
A detailed calculation is presented in Appendix~\ref{AppendiceCommRel}. The final result is:
\begin{equation}\label{Creation_annihilation_operators_algebra}
\begin{aligned}
&[\ahat(\vec{p}) ,  \ahat^\dagger (\vec{k}) ] = 2 e^{-\frac{3\omega^+({\vec k})}{\kappa}} \sqrt{m^2 + |{\vec S}_+({\vec k})|^2} \, \delta^{(3)}[\vec{p} - \vec{k}  ] \,,
\\
&[\bhat(\vec{p}) ,  \bhat^\dagger (\vec{k}) ] =   2 e^{-\frac{3\omega^-({\vec k})}{\kappa}} \sqrt{m^2 + |{\vec S}_-({\vec k})|^2}   \, \delta^{(3)}[\vec{p} - \vec{k} ] \,,
\\
&[\chat(\vec{p}) ,  \chat^\dagger  (\vec{k}) ] =  2 e^{-\frac{3\omega^+({\vec k})}{\kappa}} \sqrt{m^2 + |{\vec S}_+({\vec k})|^2}  \, \delta^{(3)}[\vec{p} - \vec{k} ] \,,\end{aligned}
\end{equation}

with all the other commutators vanishing.

Except for a momentum-dependent weight that is necessary to make it Lorentz-invariant, the algebra above is a standard creation and annihilation operator algebra.
One can check that the algebra~(\ref{Creation_annihilation_operators_algebra}) is covariant under $\kappa$-deformed Lorentz transformations~(\ref{1+1DLorentzTransform}). In fact the delta function transforms like the inverse of $\diff^3 p$, and $\diff^3 p / \sqrt{m^2 + |S_\pm(\vec{p})|^2}$ transforms like the inverse of the volume element $\sqrt{- g(p)}=e^{-\frac{3\omega^\pm({\vec p})}{\kappa}} $. Therefore, if $\ahat (\vec{p})$, $\bhat (\vec{p})$, $\chat (\vec{p})$, $\ahat^\dagger(\vec{k})$, $\bhat^\dagger (\vec{k})$ and $\chat^\dagger(\vec{k})$ transform like scalar fields on momentum space, the algebra ~(\ref{Creation_annihilation_operators_algebra}) is left invariant.

The algebra~(\ref{Creation_annihilation_operators_algebra}) suggests a natural form for the number operator:
\begin{equation}
\begin{aligned}\label{Number_operator}
\hat N  =& \int_{|\vec p | < \kappa} \frac{\diff^3 p \, e^{\frac{3\omega^+({\vec p})}{\kappa}}}{2 \sqrt{m^2 + |{\vec S}_+(\vec p)|^2}}  \, \ahat^\dagger ({\vec p}) \, \ahat (\vec{p})
+
\int_{\mathbbm{R}^3}  \frac{\diff^3 p \, e^{\frac{3\omega^-({\vec p})}{\kappa}}}{2 \sqrt{m^2 + |\vec {\vec S}_-(\vec p)|^2}}  \, \bhat^\dagger (\vec p) \, \bhat ({\vec p})
\\
+ &\int_{|\vec p | > \kappa} \frac{\diff^3 p \, e^{\frac{3\omega^+(|\vec q |)}{\kappa}}}{2 \sqrt{m^2 + |{\vec S}_+(\vec p)|^2}}  \,  \chat^\dagger (\vec p) \, \chat  ({\vec p})  \,,
\end{aligned} 
\end{equation}
its commutation relations with the creation and annihilation operators are perfectly standard:
\begin{equation}
\begin{aligned}\label{Number_operator_commutations}
&[\hat N , \ahat^\dagger (\vec{p}) ] = \ahat^\dagger (\vec{p})  \,,& &[\hat N , \ahat (\vec{p}) ] = - \ahat (\vec{p})  \,, 
\\
&[\hat N , \bhat^\dagger (\vec{p}) ] = \bhat^\dagger (\vec{p})  \,,& &[\hat N , \bhat (\vec{p}) ] = - \bhat(\vec{p})  \,, 
\\
&[\hat N , \chat^\dagger(\vec{p}) ] = \chat^\dagger (\vec{p})  \,,& &[\hat N , \chat (\vec{p}) ] = - \chat (\vec{p})  \,,
\end{aligned}
\end{equation}
and moreover $\hat N$ is explicitly Hermitian:  $\hat N^\dagger = \hat N$. This proves that we can define the Fock space exactly like in the commutative case.

In the rest of the present paper we are interested in the consequence for causality of the Pauli--Jordan function, so we won't dwell on the consequences of the algebra~(\ref{Creation_annihilation_operators_algebra}).  It is worth mentioning, however, that the construction of the Fock space implementing the right statistics, with (anti-)symmetrized multiparticle states, is a particularly complex challenge in $\kappa$-Minkowski, and several papers discussed the issue, \emph{e.g.}~\cite{kappaStatistics4,kappaStatistics3,kappaStatistics2,kappaStatistics1}.  The issue was initially noticed in~\cite{Kosinski2001}, and the first paper to propose a solution is the already mentioned~\cite{marcianoo}, where a deformed creation and annihilation operator algebra was introduced, in which commuting two operators changes the momentum of the particles they create/annihilate. Such commutation rules are capable of reproducing the $\kappa$-Poincar\'e coproduct rules. However, if a standard/undeformed (anti-)symmetrization rule is assumed when defining multiparticle states, these will not have a well-defined total momentum. A deformed statistics is necessary to define multiparticle states which are eigenstates of the total momentum. The (anti-)symmetrization proposed in~\cite{marcianoo} however is not covariant, and cannot be applied to massive fields. These further issues were addressed in~\cite{MicheleDarioStatistics}, where the difficulty of satisfying Lorentz covariance in such a $\kappa$-deformed Fock space is shown to reduce to the problem of finding closed form, all-order expression for the R matrix of $\kappa$-Poincar\'e, which is still lacking. In the case of deformations of the Poincar\'e algebra which do admit a closed-form R matrix, however, the problem is solvable: for example in the case of the Lorentz double in 2+1 dimensions~\cite{Michele3DgravityFock}, where the deformed Fock space is related to the phenomenon of `flux metamorphosis' of non-abelian anyons.

\section{Minimal-uncertainty states on the $\kappa$-Minkowski algebra}\label{MinimalUncertaintyStatesSec}

We now turn to developing the machinery that will allow us to extract physical predictions from the noncommutative Pauli--Jordan function we introduced in Sec.~\ref{PauliJordanSec}. $\Delta_\text{PJ}(z,y)$ in fact is not a traditional function of two variables. It is rather an element of the nonabelian algebra $\mathcal A \otimes \mathcal A$, and as such does not admit a single numerical value. By introducing a representation of $\mathcal A \otimes \mathcal A$, $\Delta_\text{PJ}(z,y)$  can be made into an operator on a Hilbert space $\mathcal H$, and then one could invoke Born's rule to argue that a measurement involving $\Delta_\text{PJ}(z,y)$ will deliver one of its eigenvalues with probability given by the squared norm of the value of the wavefunction on that eigenvalue. Of course we are talking about the state of the background quantum geometry, and invoking Born's rule in this case is likely unwarranted, as we lack the whole interpretational edifice of standard quantum mechanics, which is based on a vast empirical basis. This situation is, however, the daily bread of the quantum gravity researcher, and we have to make do with the mathematical structures at our disposal, without the power of experimental guidance, for the moment.
Therefore we resort to the structure that is more likely to provide a translation from a quantum operator,  $\Delta_\text{PJ}(z,y)$, into definite outcomes: the expectation value on a state of the  Hilbert space $\mathcal H$. The logic is the following: we suppose that the underlying quantum geometry is in some unknown state that cannot be determined within our theory, and the observables that depend on $\Delta_\text{PJ}(z,y)$\footnote{Which could be, for example, the arrival time of a particle in a detector (see below).} will on average give outcomes that are compatible with a commutative Pauli--Jordan function of value $\langle\Delta_\text{PJ}(z,y)\rangle$. The first goal is then to introduce a representation of our noncommutative algebra~\eqref{kappaMinkowskiCommutationRelations}.

In~\cite{dabrow} the irreducible representations of the $\kappa$-Minkowski algebra $\mathcal{A}$  are constructed,\footnote{Larger realizations have been considered in the literature, for example  involving deformed phase spaces
\cite{Amelino-Camelia2011,Meljanac2017}.} and they involve (for simplicity let us consider the 1+1-dimensional case) representing $\hat x^0$ as derivative operator\footnote{From now on noncommuting coordinates will be represented with a hat.}, $\hat x^0 = - \frac{i}{\kappa} \frac{d}{d x}$ and $\hat x^1$ as (plus or minus) the exponential of a multiplicative operator, $\hat x^1 = \pm \frac{e^{x}}{\kappa}$ (notice that the placement of the constant $\kappa$ is unambiguously fixed by the request that $\hat x^\mu$ have dimensions of length while the `pregeometric' variable $x$ is dimensionless). These two are the nondegenerate representations, and one must complete them with a degenerate representation in which  $\hat x^1$ is in an eigenstate with eigenvalue zero, and $\hat x^0$ has a real spectrum. These three representations can be put together by representing instead $\hat x^1$ as a multiplicative operator (in this case we use the variable $q$) and $\hat x^0$ as a dilatation operator:
\begin{equation}
\hat{x}^0= \frac{i}{\kappa} q \frac{\diff}{\diff q} \,, \qquad \hat{x}^1=\frac{q}{\kappa} \,, ~~~ q \in \mathbb{R}.
\end{equation}
In this way we are simultaneously considering the positive, negative and zero parts of the spectrum of $\hat x^1$. Indeed on any test function $f$ the commutation relations~(\ref{kappaMinkowskiCommutationRelations}) are respected:
\begin{equation}
[\hat{x}^0, \hat{x}^1]\triangleright f=\left[\frac{i}{\kappa} q \frac{\diff}{\diff q}, \frac{q}{\kappa}\right]f =\frac{i}{\kappa} q \frac{\diff}{\diff q}\left(\frac{q}{\kappa}f\right)-q\frac{i}{\kappa} \frac{q}{\kappa} \frac{\diff}{\diff q} f = i \frac{q}{\kappa^2}f= \frac{i}{\kappa}\hat{x}^1 \triangleright f \,.
\end{equation}
The representation we have introduced however has a problem:
if the Hilbert space of functions over which this representation acts is $\mathscr{L}^2(\mathbb{R})$ with inner product 
\begin{equation}
(\psi_1, \psi_2)=\int_\mathbb{R} \diff q\, \psi^*_1 (q) \psi_2(q) \,, \qquad \psi \in \mathscr{L}^2 \,, 
\end{equation}
then $\hat{x}^0$ is not hermitian. This can be fixed by shifting the representation as 
\begin{equation}
\frac{i}{\kappa} q \frac{\diff}{\diff q} \to \frac{i}{\kappa} q \frac{\diff}{\diff q}+\frac{i}{2\kappa},
\end{equation}
which can always be done since our representation is defined up to an additive constant. Now 
\begin{equation}
\begin{aligned}
(\psi_1, \hat{x}^0\triangleright \psi_2)&= \int_\mathbb{R} \diff q\,\psi_1^*(q) \left(\frac{i}{\kappa} q \frac{\diff}{\diff q} +\frac{i}{2\kappa} \right)\psi_2(q)\\&= \int_\mathbb{R} \diff q\, \left(\left(\frac{i}{\kappa} q \frac{\diff}{\diff q} +\frac{i}{2\kappa} \right)\psi_1(q)\right)^* \psi_2(q) =(\hat{x}^0\triangleright \psi_1, \psi_2),
\end{aligned}
\end{equation}
as can be easily seen.
The representation we introduced can be easily generalized to 3+1 dimensions as
\begin{equation}
\hat{x}^0= \frac{i}{\kappa} \sum_{i=1}^3 q_i \frac{\diff}{\diff q_i} + \frac{3 i }{2 \kappa} \,, \qquad \hat{x}^i=\frac{q^i}{\kappa} \, \qquad q_i \in \mathbb{R}^3 \,,
\end{equation}
but for the rest of the paper we will only consider, for simplicity, the 1+1 dimensional case without losing anything essential.

\subsection{Semiclassical states}\label{SemiclassicalStatesSubsec}

As we said before, we think of the Hilbert space $\mathcal H$ as the set of states of the underlying quantum geometry. $\kappa$-Minkowski is supposed to be an \emph{effective} description of matter fields propagating on a quantum gravity background. By its nature an effective theory cannot calculate everything it depends on within its own framework: some input is needed. In our case this input is represented by the element of $\mathcal H$ which represents the quantum state of our noncommutative coordinates. Of the infinitely many such states present in $\mathcal H$, many represent very non-classical situations: one could have a superposition of macroscopically distant points (\emph{i.e.} the wavefunction of $\hat x^1$ is peaked around points that are far away). Or the uncertainty on one of the coordinates could be enormous. If the underlying quantum theory of gravity that is supposed to admit $\kappa$-Minkowski as `ground state' produces such queer states, then we could probably rule it out with the observation that we experience, within a very small error margin, classical Minkowski spacetime as the geometric background on which known physics unfolds. Then we can, on physical ground, rule out most states in   $\mathcal H$ as unphysical and focus on those that resemble more closely a classical geometry.
We then formulate the following \emph{semiclassicality conditions} for states in $\mathcal H$: 1. the wavefunction needs to be localized, in the sense that the amplitudes need to fall off fast at a few variances away from the expectation values. 2. None of the variances of the time and spatial coordinates should be too large. This can be achieved by requiring that the squared sum of the variances $(\delta x^0)^2+(\delta x^1)^2$ is near its theoretical minimum.
These variances are  constrained by the uncertainty relations:
\begin{equation}
\delta x^0 \delta x^1 \geq \frac{\langle x^1 \rangle}{2 \kappa} \,,
\end{equation}
it is clear that the values of $\delta x^0$ and $\delta x^1$ that minimize the squared sum $(\delta x^0)^2+(\delta x^1)^2$ are those such that\footnote{Notice that measurability limits of the form~\eqref{SqrtUncertainty} have been already conjectured in~\cite{youngamelino}, based on independent physical arguments that did not involve at all commutation relations of the form of $\kappa$-Minkowski.}
\begin{equation}\label{SqrtUncertainty}
\delta x^0 \sim \delta x^1 \sim \sqrt{\frac{\langle x^1 \rangle}{2\kappa}} \,.
\end{equation}
Within the above constraint, we still have at our disposal a vast class of wavefunctions that have good semiclassical properties. Only a more fundamental theory can single out a particular function, so, if we are working within an effective field theory framework, we should stay agnostic with regards to the particular choice of function, and study the behaviour of our observables on all semiclassical wavefunctions. Universal features, \emph{i.e.} features that are independent of the choice of function, can be deemed `emergent' properties of the theory, and can be legitimately claimed to be physical.
In the following, we will consider Gaussian wavefunctions, which allow to perform almost all calculations analytically, and impose the semiclassicality conditions within this restricted class of functions.

Consider a Gaussian state $|\psi \rangle$ with wavefunction
\begin{equation}\label{gaussss}
\psi(q; \langle x^0\rangle, \langle x^1\rangle)\equiv\psi(q)=\left(\frac{2}{\pi \sigma^2}\right)^{\frac{1}{4}}
e^{-\frac{(q-\kappa \langle x^1\rangle)^2}{\sigma^2}-i\frac{\langle x^0\rangle}{\langle x^1\rangle} q},
\end{equation}
 $\langle x^0\rangle$ and $\langle x^1\rangle$ here are numerical parameters.
Its $\mathscr{L}^2$ norm is $1$ 
and the expectation values $\langle \psi| \hat{x}^0|\psi
\rangle$ and $\langle \psi |\hat{x}^1|\psi \rangle$ are
\begin{equation}\label{exp}
\langle \hat{x}^0 \rangle =\int_\mathbb{R} \diff q\, \psi^*(q)  \left(\frac{i}{\kappa} q \frac{\diff}{\diff q} +\frac{i}{2\kappa} \right)\psi (q) =\langle x^0\rangle \,, \qquad  \langle \hat{x}^1 \rangle =\int_\mathbb{R} \diff q \,\psi^*(q) \frac{q}{\kappa} \psi (q) =\langle x^1\rangle \,,
\end{equation}
The variances are 
\begin{equation}\label{var}
\begin{aligned}
(\delta \hat{x}^0)^2 &=\langle(\hat{x}^0)^2\rangle-\langle \hat{x}^0\rangle^2 = \left(\frac{\langle x^1\rangle}{\sigma}\right)^2+\frac{1}{4\kappa^2}\left(2+\left(\frac{\langle x^0\rangle\sigma}{\langle x^1\rangle}\right)^2  \right) \,,
\\
(\delta \hat{x}^1)^2 &=\langle(\hat{x}^1)^2\rangle-\langle \hat{x}^1\rangle^2= \frac{\sigma^2}{4\kappa^2}\,.
\end{aligned}
\end{equation}
We can find an optimal balance between the uncertainty in $\hat{x}^0$ and that in $\hat{x}^1$ by minimizing the sum $(\delta \hat{x}^0)^2+(\delta \hat{x}^1)^2$
\begin{equation}
\frac{\diff}{\diff \sigma}((\delta \hat{x}^0)^2+(\delta \hat{x}^1)^2)=
\frac{\diff}{\diff \sigma}\left(
\left(\frac{\langle x^1\rangle}{\sigma}\right)^2+\frac{1}{4\kappa^2}\left(2+\left(\frac{\langle x^0\rangle\sigma}{\langle x^1\rangle}\right)^2  \right)+ 
\frac{\sigma^2}{4\kappa^2} \right)=0, 
\end{equation}
this equation selects a specific value of $\sigma$:
\begin{equation}
\sigma^4=\frac{4\kappa^2{\langle x^1\rangle}^4}{{\langle x^0\rangle}^2+{\langle x^1\rangle}^2}.
\end{equation}
Close to the classical light-cone, \emph{i.e.} when $\langle \hat{x}^0\rangle\sim \langle \hat{x}^1\rangle$, we have  $\sigma^2 \sim 2 \kappa \langle x^1\rangle$, and the uncertainties are $ (\delta \hat{x}^0)^2 \sim  \frac{3 \langle x^1 \rangle}{2 \sqrt{2} \kappa}$, $(\delta \hat{x}^1)^2 \sim \frac{\langle x^1 \rangle}{2 \sqrt{2} \kappa}$.

Let us look into some real-world numbers, and imagine we are considering an astrophysical event, \emph{e.g.} a Gamma Ray Burst which is highly localized in space and time, compared to the space and time distances at which the event is located (Gamma Ray Bursts can be located as far back in time as several billion years, and their temporal resolution can be of the order of the second).
Assuming that the deformation parameter $\kappa$ is equal to the Planck energy, $\kappa \sim E_p \sim 10^{28} eV$, then its inverse is (with the appropriate powers of $\hbar$ and $c$, which we omit) the Planck length, $\frac{1}{\kappa}\sim L_p \sim 10^{-35} m$. Now consider a point near the classical light-cone, at 2 billion light years from the origin, \emph{i.e.} the expectation value of $\hat x^1$ is 
$\langle \hat{x}^1\rangle \sim 2 \times 10^9 ly$, or, in terms of the Planck length, $\langle \hat{x}^1\rangle  \sim 10^{60}L_p$. Our semiclassical Gaussian state in this case would have
\begin{equation}
\delta \hat{x}^1  \sim \sqrt{\frac{\langle x^1 \rangle L_p}{2 \sqrt{2}}} \sim 6 \times 10^{-6} m \,, \qquad \delta \hat{x}^0 \sim  \sqrt{3\frac{\langle x^1 \rangle L_p}{2 \sqrt{2}}} \sim  3 \times 10^{-14} s  \,.
\end{equation}
So we can prepare a semiclassical state of one point centred on $\langle \hat{x}^1 \rangle \sim 2$ billion light years, $\langle \hat{x}^0 \rangle \sim 2$ billion years, and $\delta \hat{x}^0 \sim 30$ femtoseconds, $\delta \hat{x}^1 \sim 6$ microns. This state has a microscopic
space and time uncertainty while the expectation values are of the order of billion (light) years.

Now, if the expectation value of the Pauli--Jordan function on such a geometric state has a tail
outside of the classical light cone $\langle \hat{x}^1 \rangle = \langle \hat{x}^0 \rangle$ that is exponentially suppressed over a range of the order
of the 10 femtoseconds (or tens of microns), we can say that all the fuzziness of the light cone is
due to the intrinsic uncertainty of the geometric state, and there is no effect due to propagation
over cosmological distances. Instead, if it is suppressed over a much greater range (\emph{e.g.} over time intervals of the order of one second, as is necessary in order to reveal such effects with astrophysical sources like Gamma Ray Bursts~\cite{21}),
we can say that we expect photons or neutrinos from distant localized sources to arrive with a
measurable uncertainty in the time of arrival, whose origin is to be traced back to the quantum fluctuations of the underlying spacetime. We now turn to evaluating the Pauli--Jordan function on a semiclassical Gaussian state of two coordinates $\hat y^\mu$ and $\hat z^\mu$, which will allow us to study its dependence on the four free parameters of the state: the expectation values $\langle y^0 \rangle$, $\langle z^0 \rangle$, $\langle y^1 \rangle$ and $\langle z^1 \rangle$.

\subsection{Expectation value of the Pauli--Jordan function around the light cone}

We now consider a representation of the (1+1 dimensional version of the) tensor product algebra~(\ref{kappaMinkowskiCommutationRelations}):
\begin{equation}
\hat{z}^0= \frac{i}{\kappa}  q_z \frac{\diff}{\diff q_z} + \frac{i }{2 \kappa} \,, \qquad \hat{z}^1=\frac{q_z}{\kappa} \,, \qquad q_z \in \mathbb{R} \,,
\qquad
\hat{y}^0= \frac{i}{\kappa}  q_y \frac{\diff}{\diff q_y} + \frac{i }{2 \kappa} \,, \qquad \hat{y}^1=\frac{q_y}{\kappa} \,, \qquad q_y \in \mathbb{R} \,.
\end{equation}
We are interested in calculating matrix elements 
such as $
\langle \psi_z,\psi_y | e^{ik\hat{z}}e^{i \omega^+(k)\hat{z}^0}|\psi_z,\psi_y \rangle$, where:
\begin{equation}\label{pj}
\begin{aligned}
\Delta_\text{PJ}(\hat{z}, \hat{y}) 
= & \int_{|p|<\kappa} \!\!\!\!\!\!\!   \diff p  \frac{ e^{ \frac{\omega^+ }\kappa}  e^{i   p \hat{z} - i e^{\frac{\omega^+ }\kappa} p \hat{y}  }  e^{i \omega^+ (\hat{z}^0 - \hat{y}^0) }  }{ 2 \sqrt{  m^2+ p^2}} - \int \diff p  \frac{ e^{ \frac{\omega^- }\kappa}  e^{i  p \hat{z}} e^{- i  e^{\frac{\omega^- }\kappa} p \hat{y}}  e^{i \omega^- (\hat{z}^0 - \hat{y}^0) }  }{ 2 \sqrt{  m^2+ p^2}} 
\\
&- \int_{|p|>\kappa}  \!\!\!\!\!\!\! \diff p  \frac{ e^{ \frac{\omega^+ }\kappa}  e^{i  p \hat{z}} e^{- i  e^{\frac{\omega^+ }\kappa} p\hat{y} }  e^{i \omega^+ (\hat{z}^0 - \hat{y}^0) }  }{ 2 \sqrt{  m^2+ p^2}}=:\Delta_a (\hat{z}, \hat{y}) +\Delta_b(\hat{z}, \hat{y})+\Delta_c (\hat{z}, \hat{y})  
\,,
\end{aligned}
\end{equation}
and the states $| \psi_z,\psi_y \rangle $ are products of Gaussian states:
\begin{equation}
| \psi_z,\psi_y \rangle  = \psi(q_z; \langle z^0 \rangle ,\langle z^1 \rangle)  \psi(q_y; \langle y^0 \rangle ,\langle y^1 \rangle) \,. 
\end{equation}
Notice that the $\hat{z}^0$ operator is just the dilatation operator plus a phase shift:
\begin{equation}
e^{i \omega^+(p)\hat{z}^0} \triangleright \psi(q_z)= e^{i \omega^+(p)\left(\frac{i}{\kappa} q_z \frac{\diff}{\diff q_z}+\frac{i}{2\kappa}\right)}\psi(q_z)=
e^{-\frac{\omega^+(p)}{2\kappa}} \psi\left(e^{-\frac{\omega^+}{\kappa}}q_z\right),
\end{equation}
while $
e^{ip\hat{z}^1}\triangleright \psi(q_z)=e^{ikq_z}\psi(q_z).
$
In order to calculate, for instance, $\langle \psi_z,  \psi_y| \Delta_a(\hat{z}, \hat{y})|\psi_z, \psi_y\rangle$ we need 
\begin{equation}
f_z (p):=
\langle \psi_z | e^{ip \hat{z}^1}e^{i \omega^+(p)\hat{z}^0}|\psi_z \rangle = \int_\mathbb{R} \diff q_z\, \psi^*(q_z) e^{ipq_z}e^{-\frac{\omega^+(p)}{2\kappa}} \psi\left(e^{-\frac{\omega^+}{\kappa}}q_z\right) \,,
\end{equation} 
and
\begin{equation}
f_y(p):=
\langle \psi_y | e^{-i p \hat{y}^1 e^{\frac{\omega^+(p)}{\kappa}}}e^{-i \omega^+(p)\hat{y}^0}|\psi_y \rangle = \int_\mathbb{R} \diff q_y\, \psi^*(q_y) e^{-ipq_ye^{\frac{\omega^+(p)}{\kappa}}}e^{\frac{\omega^+(p)}{2\kappa}} \psi\left(e^{\frac{\omega^+}{\kappa}}q_y\right),
\end{equation} 
which are
Gaussian
integrals and can
be readily
calculated
analytically, with parameters $\langle z^0\rangle$, $ \langle z^1\rangle$, $ \langle y^0\rangle$, $ \langle y^1\rangle$, $ m$, $ \kappa$ (and, if we don't want to specialize immediately to semiclassical Gaussian states, also the variances  $\sigma_z$ and  $\sigma_y$). After that one  can evaluate numerically
\begin{equation}
\langle \psi_z,  \psi_y| \Delta_\text{PJ}(\hat{z}, \hat{y})|\psi_z , \psi_y\rangle= \int_{-\kappa}^{\kappa} \diff p\,\frac{ e^{\frac{\omega^+}{\kappa}}}{2\sqrt{p^2+m^2}} f_z(p) f_y(p) \,,
\end{equation}
where the $f$ are some functions of momenta and  parameters.
An analogue procedure allows to calculate the expectation values  of $\Delta_b(\hat{z}, \hat{y})$ and $\Delta_c (\hat{z}, \hat{y})$ on $|\psi_z, \psi_y\rangle$.

\begin{figure}[h!]
\begin{center}
\includegraphics[height=0.35\textheight]{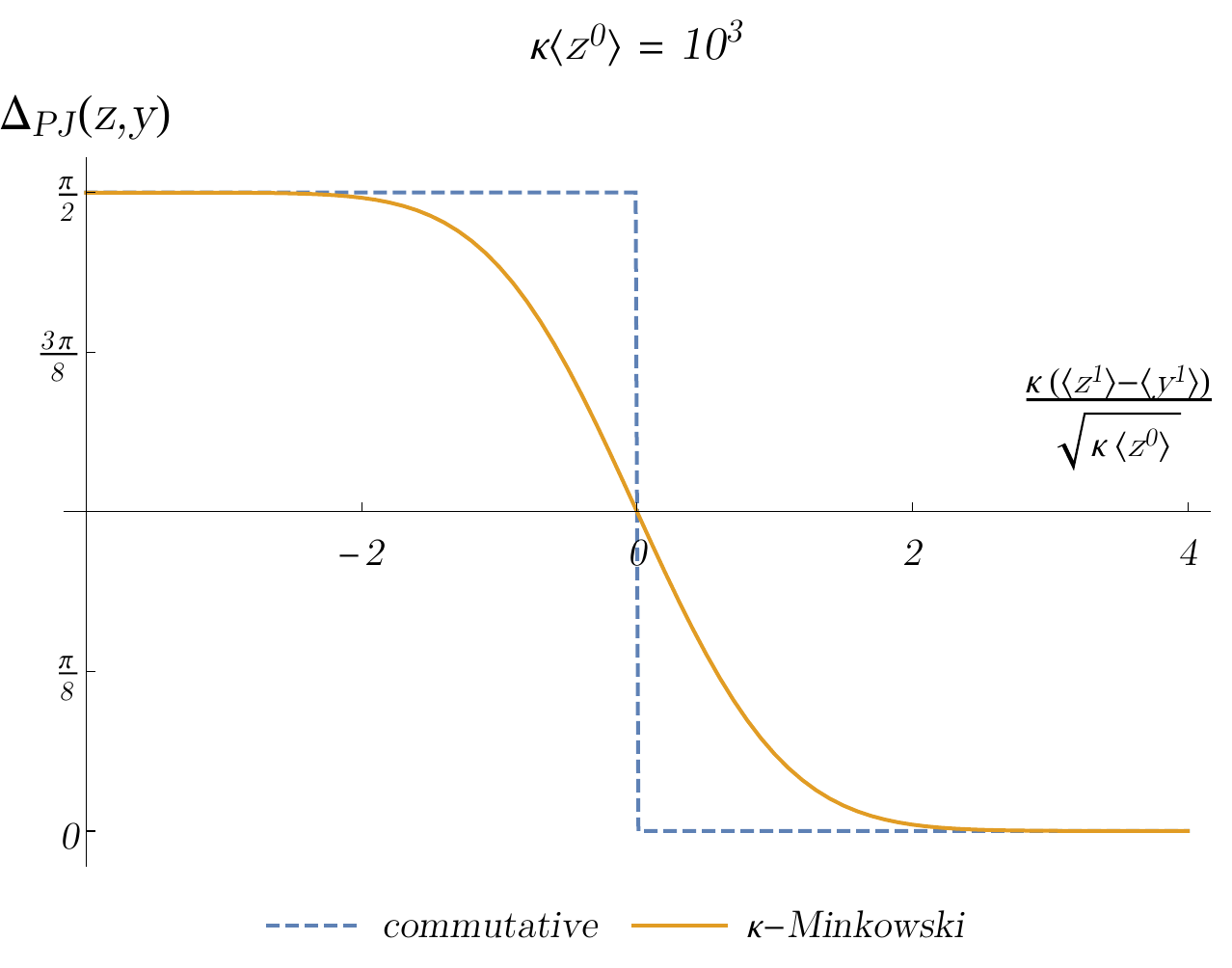}
\end{center}\caption{Pauli--Jordan function for Minkowski (dashed line) and $\kappa$-Minkowski (solid line) spacetimes. The $\langle y\rangle$ coordinates are fixed to a smaller value $\langle y^0 \rangle \sim  \langle y^1 \rangle \sim \frac{1}{100} \langle z^0 \rangle $, while the $\langle z^0 \rangle$ coordinate is fixed to a large value (in this case $10^3$ in units of $\kappa^{-1}$). The $\langle z^1 \rangle$ coordinate varies on the horizontal axis around the light-cone value  $\langle z^1 \rangle  = \langle z^0 \rangle$, over an interval of a few units of $\sqrt{ \langle z_0 \rangle / \kappa}$. 
In this simulation we are looking at unrealistically small distances (1000 Planck lengths), but realistic values (\emph{i.e.} $10^9~\text{light years}$ = $10^{60}$  Planck lengths) are impossible to work with on the calculator. Our strategy is therefore to start with small distances, and then see how the result evolves as we increase the distances by several orders of magnitude at the time (see Figure~\ref{TwoFigures} below).
}\label{OneFigure}
\end{figure}

\begin{figure}[h!]
\begin{center}
\includegraphics[height=0.24\textheight]{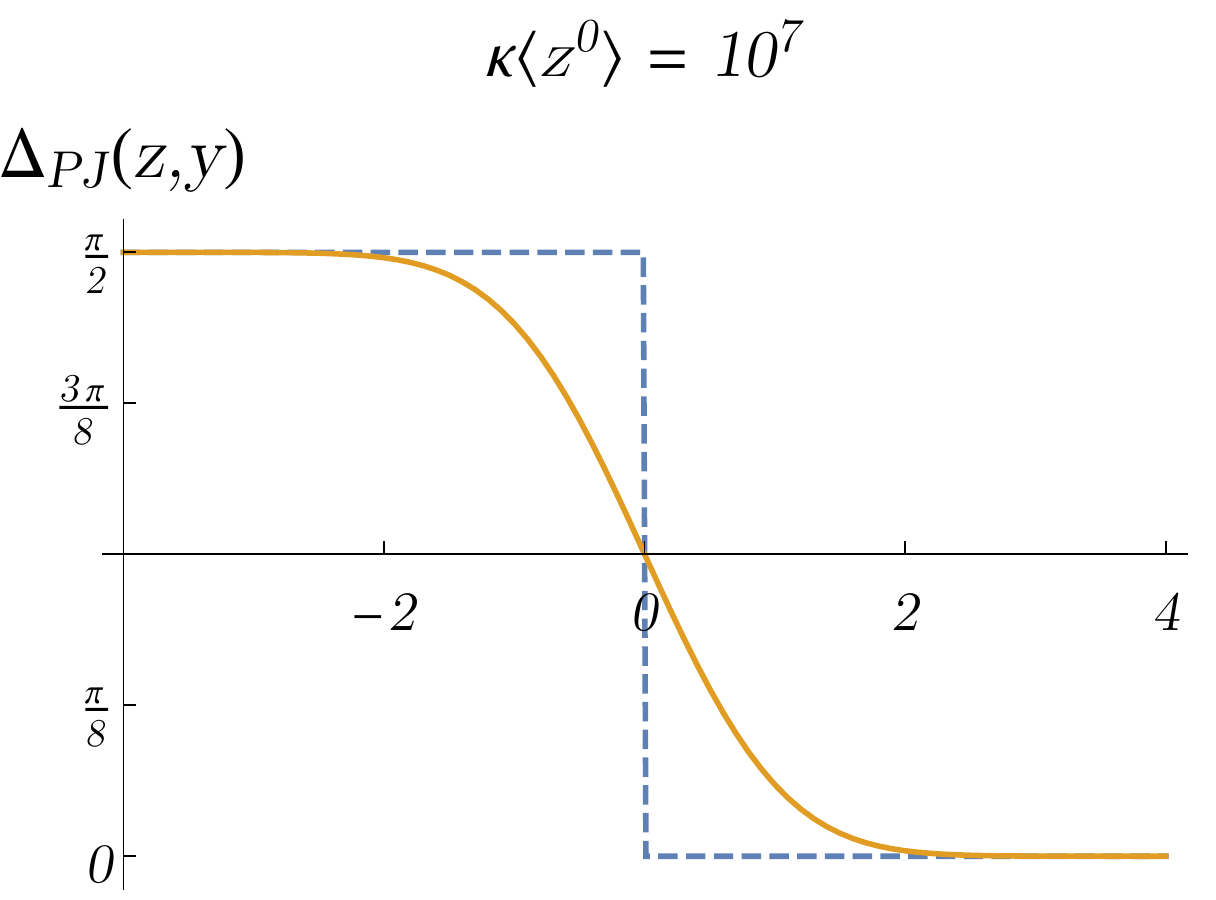}
~~~
\includegraphics[height=0.24\textheight]{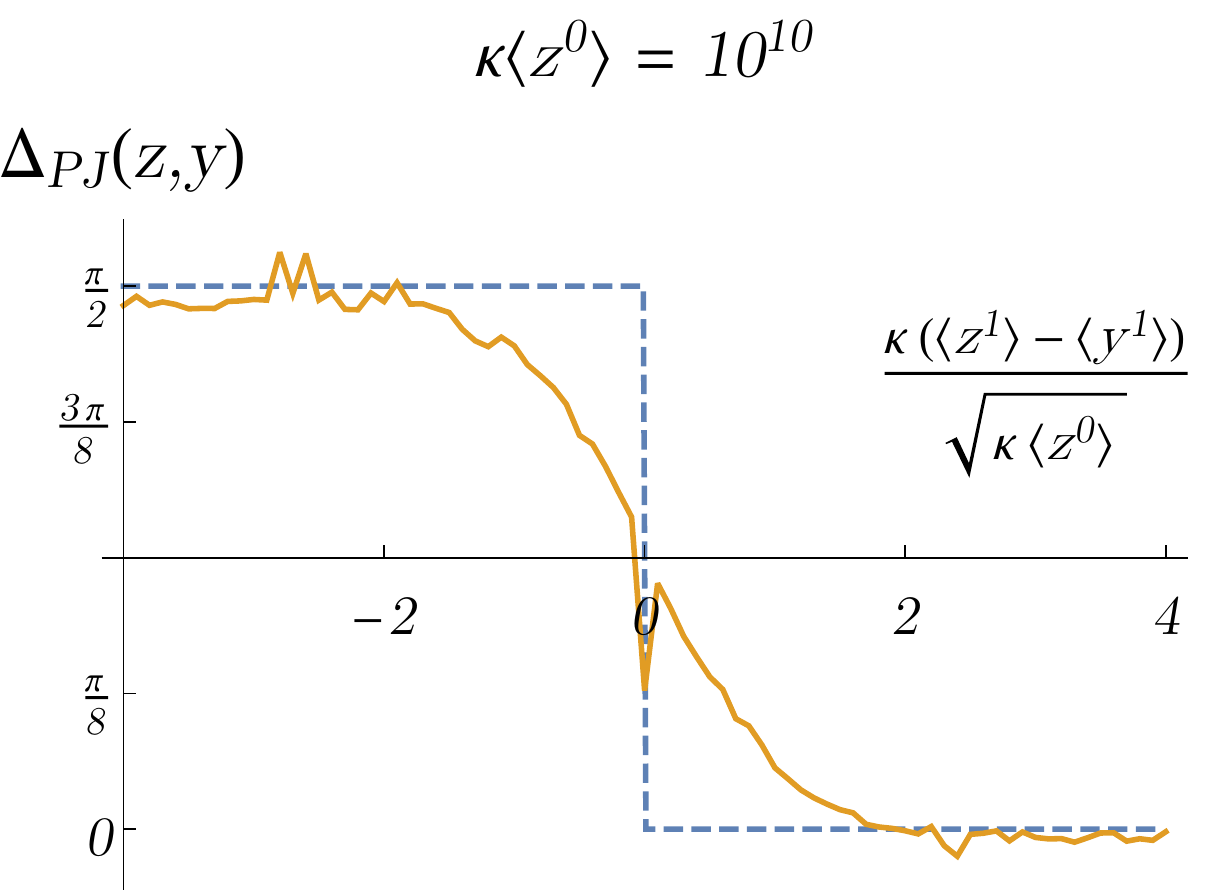}
\\
\includegraphics[height=0.032\textheight]{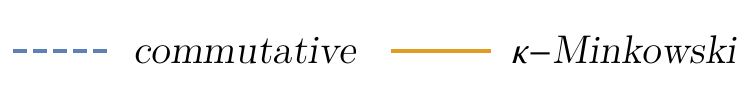}
\end{center}\caption{The Pauli--Jordan function calculated at $10^7$ and $10^{10}$ Planck units of distance from $\langle y^\mu\rangle$. The horizontal scale is units of  $\sqrt{\langle z^0 \rangle / \kappa}$. We see that in this scale the function has exactly the same shape, independently of the value of $\kappa \langle z^0 \rangle$ (the right-hand-side plot suffers from significant numerical errors, but it shows, near to the limits of our numerical precision, that the shape of the function is still the same). It is therefore safe to assume that the shape of the Pauli--Jordan function will be the same also for $\kappa \langle z^0 \rangle \sim 10^{60} L_p \sim 1 $ billion light years.}\label{TwoFigures}
\end{figure}

Our calculation reveals that $\Delta_\text{PJ}$ is real (or at least its imaginary part is smaller than the numerical error), just like in the commutative case. Moreover, when $\langle z^\mu \rangle - \langle y^\nu \rangle$ is spacelike the noncommutative Pauli--Jordan function is not zero; however it falls off exponentially away from the light cone, with a characteristic length of the order of $\sqrt{L_p \, \langle z^1 \rangle}$. We conclude that the light cone in $\kappa$-Minkowski spacetime is `blurry', and `spills out' of the classical light cone over a distance of the same order of magnitude of the minimal intrinsic uncertainty of semiclassical states on the underlying noncommutative geometry.
This is bad news for $\kappa$-Minkowski phenomenology based on Gamma Ray Bursts. For a point-like source placed at a distance $\langle z^1 \rangle$ of the order of one billion light years, one can expect photons to arrive with an uncertainty of the order of $\sqrt{L_p \, \langle z^1 \rangle}\sim 10^{-14} ~ s = 10 ~ fs$, but not more. This is 14 orders of magnitude less than what predicted by a linear relation of the kind~(\ref{LinearDispRel}) for $E = 1~TeV$ particles.

\section{Outlook and Conclusions}\label{ConclusionSec}

Our approach allowed us to define a quantum noninteracting scalar field theory on the $\kappa$-Minkowski noncommutative spacetime, in a way that is \textbf{1.} completely covariant,  \textbf{2.} invariant under diffeomorphism of momentum space and \textbf{3.} $\kappa$-Poincar\'e invariant.

\textbf{Point 1.} refers to the fact that our quantization prescription  does not depend on 3-dimensional structures like 3+1 decompositions of spacetime. Hamiltonian formulations or symplectic structures \cite{FKKKN,marcianoo,arzanooooo}  are inconsistent in $\kappa$-Minkowski due to its peculiar commutators $[x^0 ,x^i] = \frac i \kappa x^i$. These imply Heisenberg-like relations $\delta x^0 \delta x^i \geq \frac 1 {2\kappa} \langle x^i \rangle$ that do not allow the uncertainty on the time coordinate, $\delta x^0$, to be infinitely small: one therefore loses the notion of a spacelike hypersurface. Covariant commutators~\eqref{Creation_annihilation_operators_algebra} and the Pauli--Jordan function allowed us to quantize the theory in a 4-dimensional way, without being forced to make reference to any hypersurface of simultaneity (and therefore without referring to a particular inertial frame). There is already in the literature (see~\cite{koso} and~\cite{koso2}) an attempt at introducing Pauli--Jordan function (and the related Green functions). However, at the time, the understanding of $\kappa$-Minkowski's momentum space was somewhat limited, and this led the authors of~\cite{koso, koso2} to define their Green functions as contour integrals on the complexification of a particular timelike `energy' coordinate on momentum space. Complexifying this coordinate breaks diffeomorphism invariance, and introduces an infinite tower of poles of the Green functions. The authors interpreted this as a physical feature of the theory. With hindsight, we can exclude that these features have physical meaning, because they depend on the choice of a  particular set of coordinates on momentum space. This brings us to \textbf{Point 2.}, invariance under diffeomorphisms of momentum space. One of our basic principles is that only statements that are coordinate-independent on momentum space should be physical. Without such an assumption, the nonlinear structures introduced by spacetime noncommutativity would make the theory not predictable. One could draw all sorts of mutually-incompatible conclusions only by choosing different coordinate systems on momentum space. As observed in Sec.~\ref{Sec_k-Poincare_group_and_k-Minkowski} [\emph{e.g.} Eq.~(\ref{DiffeoFourierCoefficients})] each ordering prescription for $\kappa$-Minkowski plane waves corresponds to a different coordinate system on momentum space. Moreover each of these coordinate system corresponds to a choice of basis for momentum generators in the $\kappa$-Poincar\'e algebra. Just like in a Lie algebra one is free to make linear redefinitions of the generators (without changing the algebra itself), in a Hopf algebra this freedom extends to any nonlinear redefinition of the generators. If we are to use Hopf algebras to describe the symmetries of a noncommutative spacetime, we should not break their basic invariance properties when building physical models on such a framework.
\textbf{Point 3.} is of course the reasonable request that all of the equations of the theory, in particular the commutation relations, be left invariant by $\kappa$-Poincar\'e transformations. There is one subtlety in this point, something that in the commutative case one does not need to worry about. In the commutative case one can check independently the invariance under translations \textit{and} Lorentz transformations, this ensures invariance under the whole Poincar\'e group. In the noncommutative case one should look for invariance under a \textit{complete} $\kappa$-Poincar\'e transformation of the form~(\ref{Coaction_kappa_Poincare_group}). Invariance under such a transformation does not imply invariance under a `pure' Lorentz transformation of the form $x^\mu \to \Lambda^\mu{}_\mu \otimes x^\nu$. For example, to prove the $\kappa$-Poincar\'e invariance of the Pauli--Jordan function~(\ref{Invariance_PJ_under_kPoincare}) we have to use the commutation relations  $[a^\mu, \Lambda^\rho{}_\sigma]$ [see Eq.~(\ref{kappaPoincareGroup})]. Setting $a^\mu$ to zero would not deliver the desired result: the Pauli--Jordan function is not invariant under ``pure-Lorentz" transformations. This is an aspect of the so-called `no pure boost' property of the $\kappa$-Poincar\'e group (first noticed in~\cite{nopure}). The commutators~(\ref{kappaPoincareGroup}) imply uncertainties between $\Lambda^\mu{}_\nu$ and $a^\mu$ such that we cannot set $ \delta a^\mu=0$, not without making either  $\delta \Lambda^\mu{}_\nu$ divergent or by setting the expectation $\langle \Lambda^\mu{}_\nu \rangle = {\delta^\mu}_{\nu}$ (at least for some components of $\Lambda^\mu{}_\nu$). This means that we cannot fix the translation parameters to be zero with total certainty, and still be free to choose the Lorentz parameters arbitrarily.

The most advanced theoretical result that was enabled by our approach is the \textbf{definition of the creation and annihilation operators algebra}~(\ref{Creation_annihilation_operators_algebra}), and the \textbf{introduction of a number operator $\hat N$} with standard commutation rules~(\ref{Number_operator_commutations}). The number operator is Hermitian, and therefore we can construct a Fock space in the standard way, with a vacuum that is annihilated by all of the annihilation operators. The $n$-particle states are eigenstates of $\hat N$ with eigenvalue $n$, and are obtained by the successive application of creation operators.

The next step of our analysis would be to introduce an electric charge operator (recall that our scalar field is complex) and C, P and T conjugation operators. Then to study how they are represented on the creation and annihilation operator algebra. In the commutative case the charge operator is just a number operator with a minus sign in front of the integral on negative-frequency states. By analogy it seems natural to introduce a noncommutative charge operator as the same expression as~(\ref{Number_operator}), with a minus in front of the second and third integral. Such an expression is Hermitian and $\kappa$-Poincar\'e invariant. However in this case we anticipate a difficulty with the definition of charge conjugation: if the antipode map is used to define the charge conjugation operator, then there is an issue with Lorentz covariance. In fact the antipode maps region $a$ to region $b$, and  maps region $c$ onto itself. But Lorentz transformations connect regions $b$ and $c$ and this could lead to a non-Lorentz-invariant charge conjugation operator.
These issues deserve a focused analysis, which we postpone to future studies.

On the level of physical interpretation and, possibly, phenomenology, our study focused on a \textbf{prescription to extract observable features from our noncommutative Pauli--Jordan function.} In particular we were interested in understanding how to generalize to our noncommutative setting the well-known result that the \textbf{Pauli--Jordan function defines a light cone}. In ordinary QFT this function is zero on spacelike intervals.
To achieve this result we had to introduce a representation of the noncommutative algebra $\mathcal A$ of $\kappa$-Minkowski coordinates~(\ref{kappaMinkowskiCommutationRelations}), and a Hilbert space $\mathcal H$ on which this representation acts. The Pauli--Jordan function is a function of two variables (\emph{i.e.} an element of  $\mathcal A \otimes \mathcal A$), and therefore we need to consider the tensor product Hilbert space $\mathcal H \otimes \mathcal H$. There are many states on this Hilbert space, and our theory is not able to make any prediction regarding which of those will be physically realized. We then make an ansatz: our field theory on $\kappa$-Minkowski is supposed to be an effective description of matter on a quantum gravity background, once the gravitational degrees of freedom are integrated away (in analogy to what happens in 2+1 dimensions~\cite{matschull,freidel}). Then the particular state of the  quantum geometry can only be determined by the underlying quantum gravity theory, which we do not have access to. The best we can do is to make an educated guess regarding the properties of the states that are physically realized. The one we made is based on the insight that such states will have to be `as classical as possible', resembling commutative, classical Minkowski geometry as much as possible. This excludes those states in $\mathcal H$ in which the uncertainty of one of the coordinates is macroscopically large, for instance. We therefore introduce a notion of `semiclassical' states which minimize the squared sum of the uncertainties of the coordinates, \emph{i.e.} $(\delta \hat x^0)^2 + (\delta \hat x^1)^2$ is minimal. Among those states (which are still uncountably many) we may pick one and calculate the expectation value of the Pauli--Jordan function $\langle \Delta_\text{PJ}(\hat z , \hat y) \rangle$. The result is a commutative function that depends on the state, and we are interested in how it varies when the expectation value of the coordinates of the two points, $\langle \hat z^\mu \rangle$ and $\langle \hat y^\mu \rangle$, are varied. These expectation values now are commutative functions and define a classical lightcone (\emph{i.e.} when $\langle \hat z^\mu \rangle - \langle \hat y^\mu \rangle$ is lightlike). One can check whether  $\langle \Delta_\text{PJ}(\hat z , \hat y) \rangle$ vanishes outside of this classical light cone, or it possesses a `tail'. In the massless case it is easy to integrate  the commutative Pauli--Jordan function for a massless scalar field [Eq.~(\ref{commpj})] - in the 1+1 dimensional case - and get:
\begin{figure}[t!]
\begin{center}
\includegraphics[height=0.3\textheight]{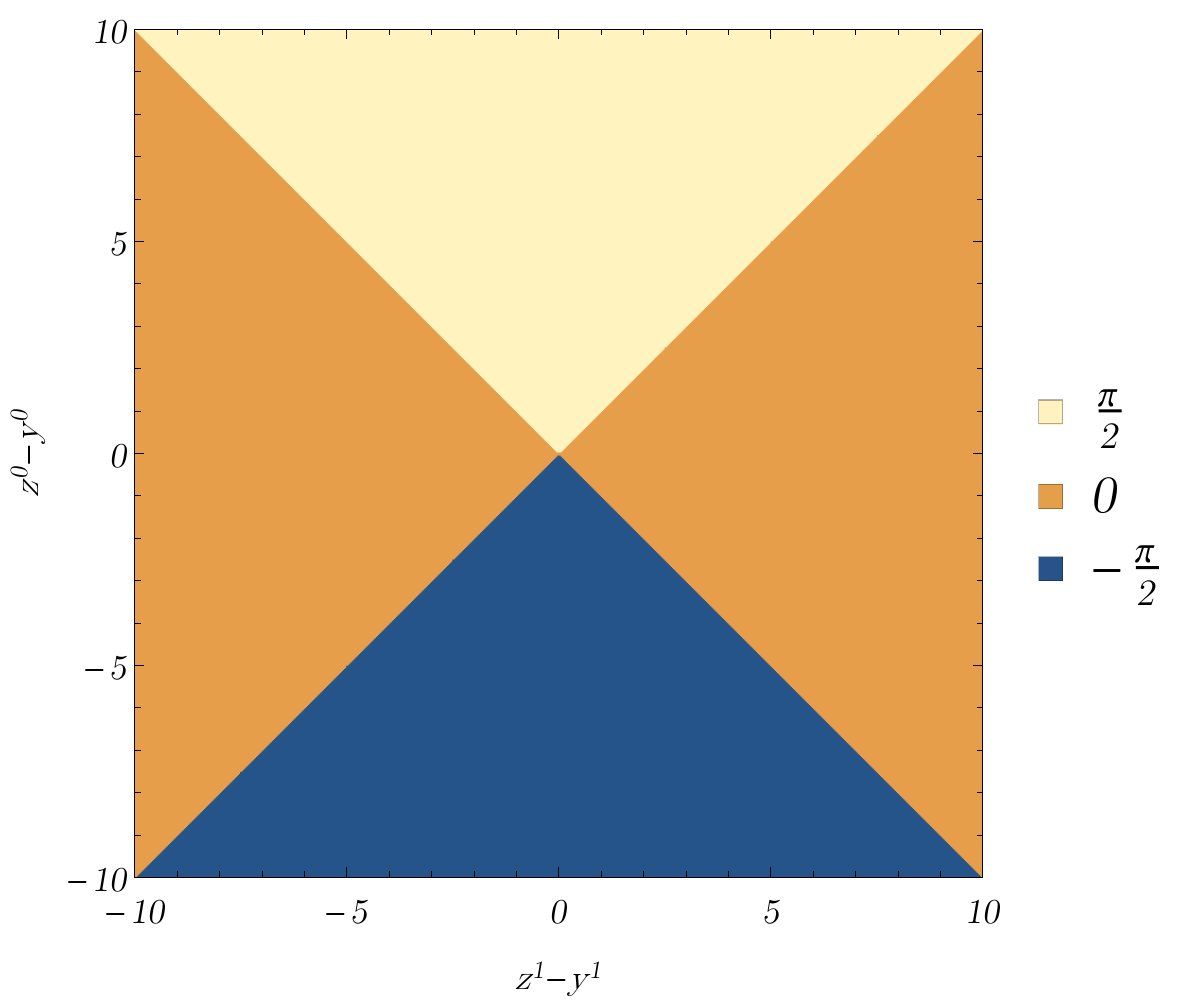}
\end{center}\caption{Contour plot of the Pauli--Jordan function  in the plane of space intervals $\langle z^1 \rangle - \langle y^1 \rangle$ vs. time intervals $\langle z^0 \rangle - \langle y^0 \rangle$, for a commutative massless scalar field (units are arbitrary, as the plot is scale-invariant).}\label{LightCone2D_0}
\end{figure}
\begin{figure}[b!]
\begin{center}
\includegraphics[height=0.19\textheight]{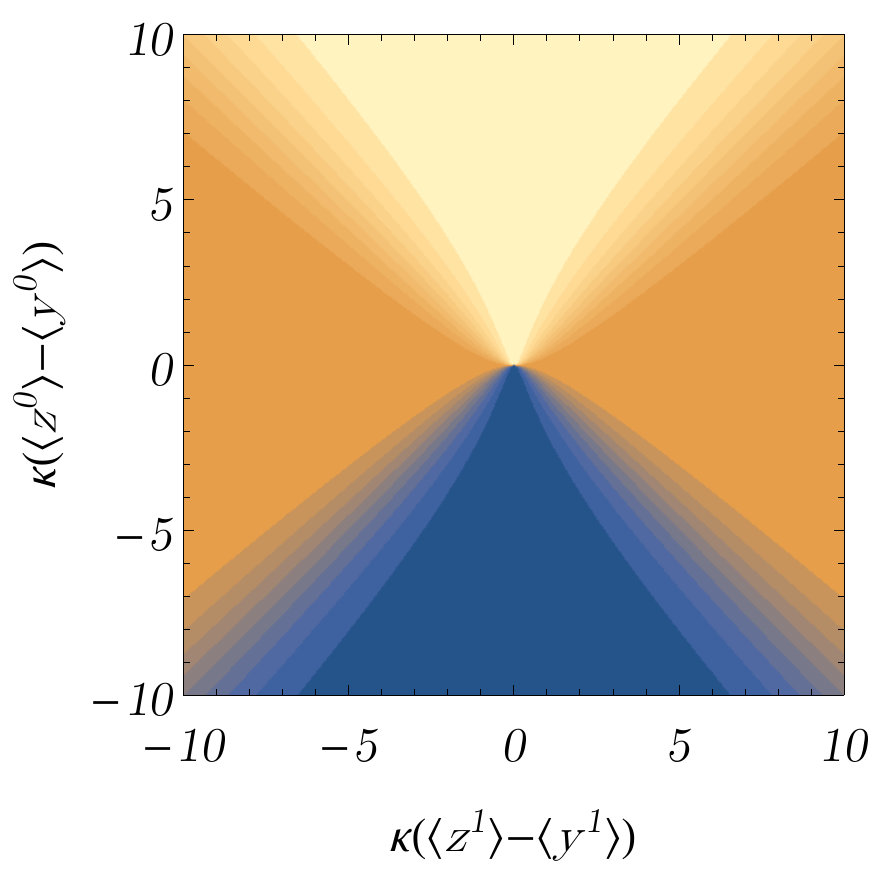}
\includegraphics[height=0.19\textheight]{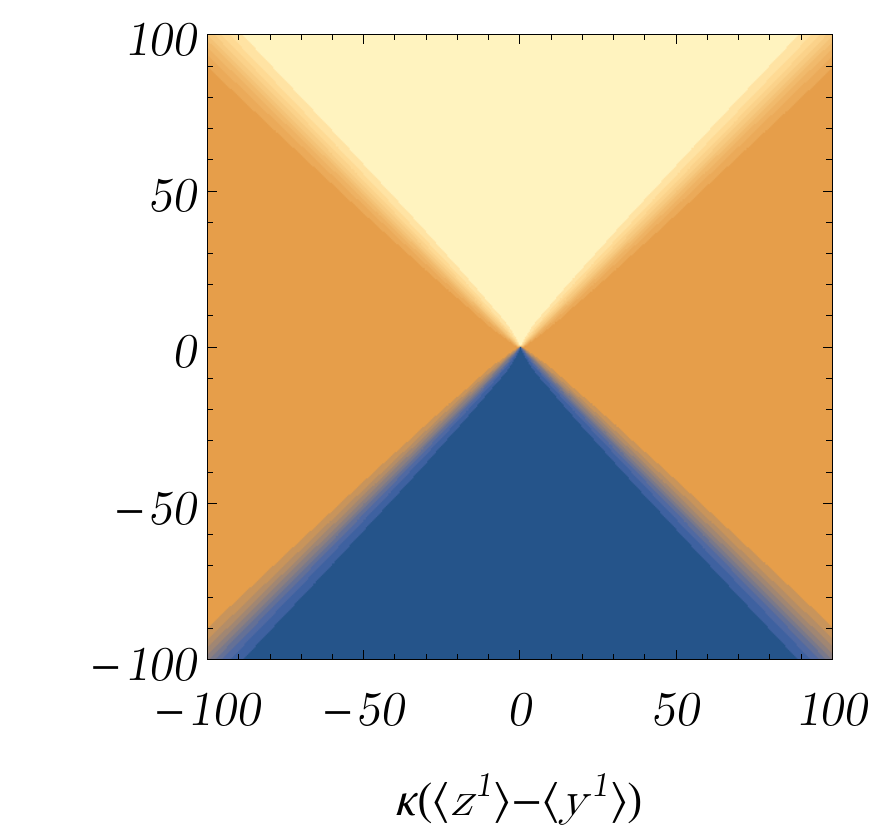}
\includegraphics[height=0.19\textheight]{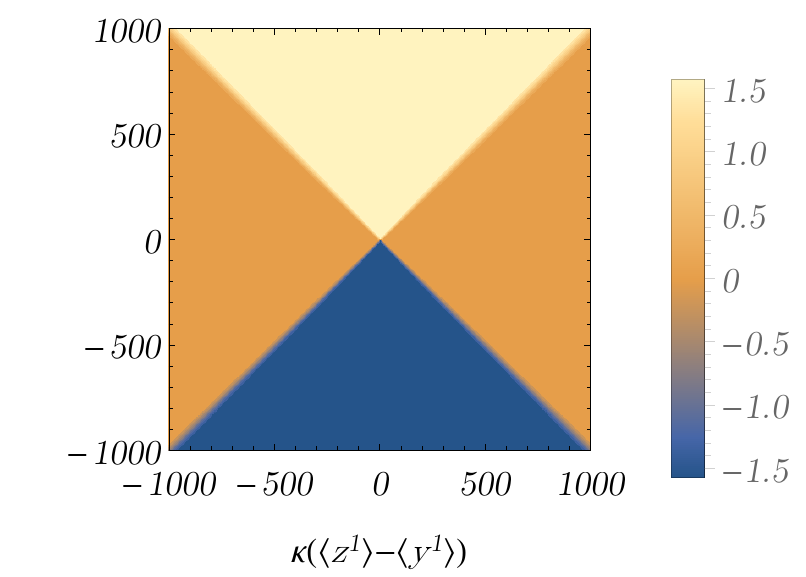}
\end{center}\caption{Same diagram as Fig.~\ref{LightCone2D_0} in the noncommutative case. The diagram on the left is zoomed in at the origin, and shows a region of $20\times20$ Planck units $\kappa^{-1}$. The central diagram shows $200\times200$ Planck units, while the one on the left shows $2000\times2000$.}\label{LightCone2D_1}
\end{figure}
\begin{equation}
\Delta_\text{PJ} (z,y) = \frac{\pi}{2} \frac{\text{sign}[\left(z^0-y^0\right) -\left(z^1-y^1\right) ] - \text{sign}[\left(z^0-y^0\right) + \left(z^1-y^1\right)]}{2} \,,
\end{equation}
we plot this function in the plane $(z^1-y^1)$ vs. $(z^0-y^0)$ in Fig.~\ref{LightCone2D_0}. Its main feature is that it is zero on lightlike intervals, while it is $\pm \frac \pi 2$ inside the light cone ($+\frac \pi 2$ in the future light cone and $-\frac \pi 2$ in the past one). The noncommutative generalization of this function is the expectation value $\langle \Delta_\text{PJ}(\hat z , \hat y) \rangle $, which we plot in~\ref{LightCone2D_1} (our numerical calculation revealed that $\langle \Delta_\text{PJ}(\hat z , \hat y) \rangle$ is real within the numerical error). The plots show how the border between the region where $\langle \Delta_\text{PJ}(\hat z , \hat y) \rangle$ is a nonzero constant and the region where it is zero gets `blurred'. The size of the region over which this blurring occurs depends on the distance from the origin: in the leftmost plot of  Fig.~\ref{LightCone2D_0} we are focusing on a region of size $20 \times 20$ Planck units $\kappa^{-1}$ (20 Planck lengths $\times$ 20 Planck times), and the blurring zone extends over several Planck units. The other two diagrams are zoomed out over, respectively, a  $200 \times 200$ and a $2000 \times 2000$ Planck units region. The blurring zone in those two cases extends over order $\sim 10$ and order $\sim 30 \sim \sqrt{1000}$ Planck units. We see that the blurring range goes like the square root of the distance from the origin in Planck units, or, in other words, as the geometric mean $\sqrt{L\kappa^{-1}}$ between the distance from the origin $L$ and the Planck scale $\kappa^{-1}$. This can be seen clearly in Fig.~\ref{OneFigure} and~\ref{TwoFigures}, where we plot a constant-$(\langle z^0 \rangle - \langle y^0 \rangle)$ slice of $\langle \Delta_\text{PJ}(\hat z , \hat y) \rangle$, centred upon the classical light cone, $(\langle z^1 \rangle - \langle y^1 \rangle) = (\langle z^0 \rangle - \langle y^0 \rangle)$, and rescaled by $\sqrt{\langle z^0 \rangle \kappa^{-1}}$. The three diagrams represent three choices of $\sqrt{\langle z^0 \rangle \kappa^{-1}}$ or $\sqrt{L\kappa^{-1}}$: $10^3$, $10^7$ and $10^{10}$, and it becomes apparent how, in these scale, the three slices of the  Pauli--Jordan function appear exactly identical. Therefore we can conclude that the blurring radius is always of order $\sqrt{L\kappa^{-1}}$.

To obtain our result we chose a particular example of `semiclassical' wavefunction: a Gaussian~(\ref{gaussss}). This allows to integrate analytically all of the functions that appear in $\langle \Delta_\text{PJ}(\hat z , \hat y) \rangle$, except the very last one, and permits us to produce particularly clean numerical results. The Gaussian wavefunction is characterized only by two parameters: the expectation values of  the time and space coordinates (the normalization being fixed to one and the variance being fixed by the semiclassicality condition). It is with respect to these two variables that we study the variation of $\langle \Delta_\text{PJ}(\hat z , \hat y) \rangle$. In general, a wavefunction satisfying the semiclassicality condition will be much more complicated than~(\ref{gaussss}), and will depend on a potentially infinite set of parameters. We conjecture that the dependence of $\langle \Delta_\text{PJ}(\hat z , \hat y) \rangle$ on all of these additional parameters will be weak, and the essential features of the function will only depend on the expectation values $\langle \hat  z^\mu \rangle$ and $\langle \hat  y^\mu \rangle$. We postpone the verification of our conjecture to further numerical analyses.

\textbf{For what concerns phenomenology,} the shape of our Pauli--Jordan function leads us to the  conclusion that there is an in-principle-observable violation of the classical light cone. The Pauli--Jordan function `spills out' of the classical light cone by an amount that depends on the geometric mean of the distance from the source and the Planck scale. This suggests that, on a $\kappa$-Minkowski background, a perfectly-localized signal emitted at a distance $L$ will be detected with an uncertainty in time and space of the order of $\sqrt{L \kappa^{-1}}$. Real-life number are quite small: for $L \sim$ 1 billion light-years we get an uncertainty in the time of arrival of the order of $10$ femtoseconds. If we had a sufficiently bright ultrashort pulse source (\emph{i.e.} a femtosecond laser) placed at one billion years from Earth, we would be able to detect the effect as a broadening of the pulse duration upon detection. Similarly, a distant source that is sufficiently localized in space (\emph{i.e.} a micrometer-scale source one billion light years away) would allow detection of our effect with present-day technology. Of course we are not aware of such precisely-localized sources in our universe, and the proposed effect remains well beyond the reach of current experiments. Our analysis however is useful to \textbf{settle a question that was debated since the introduction of the $\kappa$-Poincar\'e group: does $\kappa$-Poincar\'e predict in-vacuo dispersion of particles that can be detected with Gamma Ray Burst observations?} Our analysis is arguably  the most advanced of a long series of studies~(see for instance \cite{amelinowaves}) attempting to answer this question. This allows us to answer the question with more confidence than any previous study, and our answer is negative. The commutator between quantum fields at spacelike separated points becomes zero already at a distance of a few units of $\sqrt{L \kappa^{-1}}$ from the classical light cone, which for $L \sim$ 1 billion years and $\kappa^{-1} \sim L_p$ is $\sim 10$ $fs$. This is a time interval that is 14 orders of magnitude smaller than one second. So at a distance of one second from the light cone there can be no signal transmission. We indeed have an effect of in-vacuo dispersion  of the kind that is considered by the literature on  quantum-gravity phenomenology with Gamma Ray Burst~\cite{21,Fermi2009Nature}, but the effect is enormously smaller than what would be required to be detected using Gamma Ray Bursts. \textbf{Our conclusion is that Gamma Ray Bursts are not capable of constraining models of quantum fields propagating on a $\kappa$-Poincar\'e background.}

The present work opens several lines of future investigation.
QFT on a $\kappa$-Minkowski background should be further developed. We already remarked that the next natural step is to discuss discrete symmetries (C, P and T). Then an important issue is to understand conserved quantities, which provide the building blocks for asymptotic observables. The theory at that point would be developed enough to introduce interactions and develop dynamically nontrivial models.
For what regards phenomenology, the two most urgent extensions of our work are on one hand the exploration of the dependence of the light cone on the choice of state of the quantum geometry. As we remarked in Sec.~\ref{SemiclassicalStatesSubsec} a feature of our effective theories can be deemed observable only if it is robust under change of the choice of state. On the other hand, the Pauli--Jordan function is suited to discuss causality issues (\emph{i.e.} to tell whether two events can be causally connected, and, in our case, to exclude an apparently superluminal propagation of a certain magnitude). To make actual predictions regarding the waveform of signals propagating in vacuum we need the two-point function. This can be calculated using the techniques developed in our paper, and will be the subject of future works.

\section*{Acknowledgements}
This project has received funding from the European Union's research and innovation programme under a Marie Sk\l{}odowska-Curie grant through the  INdAM-COFUND-2012 programme of the Italian Institue of High Mathematics (INdAM).

\providecommand{\href}[2]{#2}\begingroup\raggedright\endgroup

\newpage

\appendix

\section{Finite Lorentz transformations in momentum space (in 1+1 dimensions)}
\label{AppendiceLorentzTransf}

In this Appendix, we want to derive Eq.~(\ref{1+1DLorentzTransform}), which we used first inEq.~Eq.~(\ref{abbellliiii}). This is the expression for the finite Lorentz transformations on momenta that is implied by the (non-commutative) way in which $\kappa$-Poincar\'e coordinates transform under the left coaction of the $\kappa$-Poincar\'e group:
\begin{equation}
\Delta_L \left[ x^\mu \right] = \Lambda^\mu{}_\nu \otimes x^\nu + a^\mu \otimes \1
~~~ \Rightarrow ~~~
\Delta_L \left[
e^{i p_i x^i} e^{i p_0 x^0} \right] =
e^{i \lambda_i[p,\Lambda] \otimes x^i} e^{i \lambda_0[p,\Lambda] \otimes x^0} e^{i p_i a^i \otimes \1} e^{i p_0 a^0 \otimes \1}  \,.
\end{equation}
In 1+1-dimensions, we can write the $\kappa$-Poincar\'e group  commutation relations~(\ref{kappaPoincareGroup}) [in the timelike case~(\ref{timelike_v})] as
\begin{equation}\label{1+1kPgroup}
[a^0 , a^1] = \frac i \kappa a^1  \,, \qquad 
[a^0 , \xi ] = \frac i \kappa \, \sinh \xi \,, \qquad 
[a^1 , \xi ] = \frac i \kappa \left( \cosh \xi -1 \right) \,,
\end{equation}
where $\xi$ is the rapdity, $\cosh \xi \Lambda^0{}_0 = \Lambda^1{}_1$, $\sinh \xi \Lambda^0{}_1 = \Lambda^1{}_0$. The commutation relations above can also be obtained from the request that the coaction $\Delta_L $ leaves the  $\kappa$-Minkowski commutation relations invariant, \emph{i.e.}:
\begin{equation}
[ \cosh \xi \, x^0 + \sinh \xi \, x^1 + a^0  ,
\cosh \xi \, x^1 + \sinh \xi \, x^0 + a^1 \, \1  ] = \frac i \kappa \left( \cosh \xi \, x^1 + \sinh \xi \, x^0 + a^1  \right) \,,
\end{equation}
where, for simplicity, we omitted the tensor product $\otimes$ between group elements and coordinates, and we understand $x^\mu$, $\xi$ and $a^\mu$ as belonging to the same algebra, in which $x^\mu$ commute with $\xi$ and $a^\mu$ (this is just the standard construction of a tensor product algebra).

To find the explicit expression of $\lambda_\mu[p,\Lambda]$, we need to first calculate the adjoint action of $e^{i p_0 a^0}$ and $e^{i p_1 a^1}$ on $\xi$. To do so, we notice that the following nonlinear changes of variable
\begin{equation}\label{DefEtaRho}
\eta =  \log \left( \tanh \frac \xi 2 \right) \,, \qquad 
\rho = - \coth \frac \xi 2 \,,
\end{equation}
make the commutator with $a^0$, resp. $a^1$, canonical: 
\begin{equation}
[ a^0 , \eta] = \frac i \kappa  \,, \qquad
[ a^1 , \rho] = \frac i \kappa  \,.
\end{equation}
Then the adjoint action  of $e^{i p_0 a^0}$ and $e^{i p_1 a^1}$ on these algebra elements is a translation
\begin{equation}
e^{i p_0 a^0} \eta \, e^{-i p_0 a^0} = \eta - \frac{p_0}{\kappa} \,,
\qquad
e^{i p_1 a^1} \rho \, e^{-i p_1 a^1} = \rho - \frac{p_1}{\kappa}
\end{equation}
using the homomorphism property of the adjoint action and the inverses of the relations~(\ref{DefEtaRho}):
\begin{equation}
 e^{i p_0 a^0} \xi \,  e^{-i p_0 a^0} =  2  \, \text{arctanh}\left(  e^{- \frac{p_0}{\kappa}} \tanh \frac \xi 2 \right) \,, \qquad
e^{i p_1 a^1}\xi \, e^{-i p_1 a^1} =  2 \, \text{arccoth} \left( \frac{p_1}{\kappa} + \coth \frac \xi 2  \right) \,.
\end{equation}
The homomorphism property implies that the above rules can be applied to an arbitrary function of $\xi$; in particular the following holds:
\begin{equation}\label{AdAction_a_on_xi}
\begin{aligned}
 e^{i p_0 a^0} f(\xi) = f(\tilde \xi)  e^{i p_0 a^0} \,, \qquad \tilde \xi = 2  \, \text{arctanh}\left(  e^{- \frac{p_0}{\kappa}} \tanh \frac \xi 2 \right) \,,
\\
e^{i p_1 a^1} f(\xi) =  f(\hat \xi)  e^{i p_1 a^1} \,, \qquad \hat \xi = 2 \, \text{arccoth} \left( \frac{p_1}{\kappa} + \coth \frac \xi 2  \right) \,.
\end{aligned}
\end{equation}

Now we are interested in the adjoint action of (an arbitrary function of) $\xi$ on an ordered exponential of $a^\mu$. Consider the following induction chain:
\begin{equation}
\begin{aligned}
e^{f(\xi)} a^0 &= \left( a^0 - \frac i \kappa \, \sinh \xi \, f'(\xi) \right) e^{f(\xi)} \,, \\
e^{f(\xi)} (a^0)^2 &= \left( a^0 - \frac i \kappa \, \sinh \xi \, f'(\xi) \right)^2 e^{f(\xi)} \,, \\
\vdots &
\\
e^{f(\xi)} (a^0)^n &= \left( a^0 - \frac i \kappa \, \sinh \xi \, f'(\xi) \right)^n e^{f(\xi)} \,.
\end{aligned}
\end{equation}
therefore
\begin{equation}
e^{f(\xi)} e^{i p_0 a^0} = e^{i p_0\left( a^0 - \frac i \kappa \, \sinh \xi \, f'(\xi) \right)} e^{f(\xi)} \,, 
\end{equation}
which can be written as an ad-action of $f(\xi)$ on $e^{i p_0 a^0}$:
\begin{equation}
e^{i p_0 a^0 + \frac {p_0} \kappa \, \sinh \xi \, f'(\xi) } = e^{f(\xi)} e^{i p_0 a^0} e^{-f(\xi)}   \,, 
\end{equation}
and, similarly, in the case of $e^{i p_1 a^1}$:
\begin{equation}
e^{i p_1  a^1 + \frac {p_1} \kappa \, ( \cosh \xi -1 ) \, f'(\xi) } = e^{f(\xi)} e^{i p_1 a^1} e^{-f(\xi)}   \,.
\end{equation}
The last two relations are true if $f(\xi)$ is a pure function  of $\xi$. We are interested in the case in which it is a linear combination of $x^0$ and $x^1$ which, despite commuting with $\xi$, do not commute with each other. We can apply the rule unchanged when the exponent contains $x^0$ or $x^1$ alone (because they commute with $\xi$ and $a^\mu$, so it is convenient to consider the following ordered plane wave:
\begin{equation}
e^{i k_1(\xi) x^1} e^{i k_0(\xi) x^0} \,,
\end{equation}
where the coefficients $k_\mu(\xi)$ are arbitrary functions of $\xi$. Computing the commutator with $a^0$
\begin{equation}
\begin{aligned}
e^{i k_1(\xi) x^1} e^{i k_0(\xi) x^0} a^0 &= e^{i k_1(\xi) x^1} \left( a^0 + \frac 1 \kappa \, \sinh \xi \, k_0'(\xi) x^0 \right) e^{i k_0(\xi) x^0} 
\\
&= \left[ a^0 + \frac 1 \kappa \, \sinh \xi \, \left( k_0'(\xi) x^0  + k_1'(\xi) x^1 +  \frac 1 \kappa k_0'(\xi) k_1(\xi) x^1 \right) \right] e^{i k_1(\xi) x^1} e^{i k_0(\xi) x^0} \,,
\end{aligned}
\end{equation}
where the last addend in the square brackets is a consequence of
\begin{equation}
e^{i k_1(\xi) x^1} x^0 = \left(x_0 + \frac 1 \kappa k_1(\xi) x^1 \right) e^{i k_1(\xi) x^1} \,,
\end{equation}
which is easy proved using the commutation relations $[x^0,x^1] = \frac i \kappa x^1$. A similar relation is satisfied by $a^1$:
\begin{equation}
\begin{aligned}
e^{i k_1(\xi) x^1} e^{i k_0(\xi) x^0} a^1 &= e^{i k_1(\xi) x^1} \left( a^1 + \frac 1 \kappa \, ( \cosh \xi -1 )\, k_0'(\xi) x^0 \right) e^{i k_0(\xi) x^0} 
\\
&= \left[ a^1 + \frac 1 \kappa \, ( \cosh \xi -1 ) \, \left( k_0'(\xi) x^0  + k_1'(\xi) x^1 +  \frac 1 \kappa k_0'(\xi) k_1(\xi) x^1 \right) \right] e^{i k_1(\xi) x^1} e^{i k_0(\xi) x^0} \,.
\end{aligned}
\end{equation}
These relation hold too on powers of $a^0$ or $a^1$, and consequently also on exponentials:
\begin{equation}
\begin{aligned}
&e^{i k_1(\xi) x^1} e^{i k_0(\xi) x^0} e^{ i p_1 a^1}  = e^{i p_1 \left[ a^1 + \frac 1 \kappa \, ( \cosh \xi -1 ) \, \left( k_0'(\xi) x^0  + k_1'(\xi) x^1 +  \frac 1 \kappa k_0'(\xi) k_1(\xi) x^1 \right) \right]} e^{i k_1(\xi) x^1} e^{i k_0(\xi) x^0} \,,
\\
&e^{i q_1(\xi) x^1} e^{i q_0(\xi) x^0} e^{ i p_0 a^0}  = e^{i p_0 \left[ a^0 + \frac 1 \kappa \, \sinh \xi \, \left( q_0'(\xi) x^0  + q_1'(\xi) x^1 +  \frac 1 \kappa q_0'(\xi) q_1(\xi) x^1 \right) \right]} e^{i q_1(\xi) x^1} e^{i q_0(\xi) x^0} \,.
\end{aligned}
\end{equation}
if we multiply both sides from the right by $e^{-i k_0(\xi) x^0}  e^{-i k_1(\xi) x^1}$  and re-order the exponentials using~(\ref{AdAction_a_on_xi}) we get two BCH formulas for our algebra:
\begin{equation}
\begin{aligned}
 e^{i p_1 \left[ a^1 + \frac 1 \kappa \, ( \cosh \xi -1 ) \, \left( k_0'  x^0  + k_1'  x^1 +  \frac 1 \kappa k_0'  k_1  x^1 \right) \right]}&=   e^{i k_1  x^1} e^{i k_0  x^0} e^{ i p_1 a^1} e^{-i k_0  x^0} e^{-i k_1  x^1} 
 \\
 &=   e^{i k_1  x^1} e^{i \left( k_0  - \hat k_0\right) x^0} e^{-i \hat k_1 x^1} e^{ i p_1 a^1}
\\
&=   e^{i \left( k_1 - e^{ -\left( k_0  - \hat k_0\right) / \kappa} \hat k_1\right) x^1} e^{i \left( k_0  - \hat k_0\right) x^0} e^{ i p_1 a^1} \,,
\\
e^{i p_0 \left[ a^0 + \frac 1 \kappa \, \sinh \xi \, \left( q_0'  x^0  + q_1'  x^1 +  \frac 1 \kappa q_0'  q_1  x^1 \right) \right]}  &=   e^{i q_1  x^1} e^{i q_0  x^0} e^{ i p_0 a^0} e^{-i q_0  x^0} e^{-i q_1  x^1} 
\\
&=   e^{i q_1  x^1} e^{i \left( q_0  - \tilde q_0 \right) x^0} e^{-i \tilde q_1 x^1}  e^{ i p_0 a^0}
\\
&=   e^{i \left( q_1  - e^{- \left( q_0  - \tilde q_0 \right) /\kappa} \tilde q_1 \right) x^1} e^{i \left( q_0  - \tilde q_0 \right) x^0}  e^{ i p_0 a^0}\,,
\end{aligned}
\end{equation}
where we used the relation
\begin{equation}
e^{ a \, x^0} e^{b \, x^1} = e^{e^{i a /\kappa} b \, x^1} e^{a \, x^0} \,,
\end{equation}
and the notation $\hat k_\mu = k_\mu(\hat \xi)$, $\tilde k_\mu = k_\mu(\tilde \xi)$ is pretty self-explanatory. Multiplying the first equation to the second from the left:
\begin{equation}\label{BKL_for_kappa_Poincare}
\begin{aligned}
 e^{i p_1 \left[ a^1 + \frac 1 \kappa \, ( \cosh \xi -1 ) \, \left( k_0'  x^0  + k_1'  x^1 +  \frac 1 \kappa k_0'  k_1  x^1 \right) \right]}
 e^{i p_0 \left[ a^0 + \frac 1 \kappa \, \sinh \xi \, \left( q_0'  x^0  + q_1'  x^1 +  \frac 1 \kappa q_0'  q_1  x^1 \right) \right]} 
 =
 \\
e^{i \left( k_1 - e^{ -\left( k_0  - \hat k_0\right) / \kappa} \hat k_1\right) x^1} e^{i \left( k_0  - \hat k_0\right) x^0} e^{ i p_1 a^1} 
e^{i \left( q_1  - e^{- \left( q_0  - \tilde q_0 \right) /\kappa} \tilde q_1 \right) x^1} e^{i \left( q_0  - \tilde q_0 \right) x^0}  e^{ i p_0 a^0} =
\\
e^{i \left( k_1 - e^{ -\left( k_0  - \hat k_0\right) / \kappa} \hat k_1\right) x^1} e^{i \left( k_0  - \hat k_0\right) x^0} 
e^{i \left( \hat q_1  - e^{- \left( \hat q_0  - \hat {\tilde q}_0 \right) /\kappa} \hat{\tilde q}_1 \right) x^1} e^{i \left( \hat q_0  - \hat{\tilde q}_0 \right) x^0} e^{ i p_1 a^1}  e^{ i p_0 a^0} =
\\
e^{i \left[ k_1 - e^{- \left( k_0  - \hat k_0\right) / \kappa} \hat k_1 + e^{- \left( k_0  - \hat k_0\right) /\kappa}\left( \hat q_1  - e^{- \left( \hat q_0  - \hat {\tilde q}_0 \right) /\kappa} \hat{\tilde q}_1 \right) \right] x^1}
e^{i \left( k_0  - \hat k_0 + \hat q_0  - \hat{\tilde q}_0 \right) x^0} e^{ i p_1 a^1}  e^{ i p_0 a^0}\,,
\end{aligned}
\end{equation}

The final step is to equate the left-hand-sides of~(\ref{BKL_for_kappa_Poincare}) with 
\begin{equation}
\begin{aligned}
\Delta_L \left[ e^{i p_1 x^1}  e^{i p_0 x^1} \right]
=
 e^{i p_1 \left( x^1 \cosh \xi + x^0 \sinh \xi \right) + i p_1 a^1}  e^{i p_0  \left( x^0 \cosh \xi + x^1 \sinh \xi \right) + i p_0 a^0}  \,.
\end{aligned}
\end{equation}
This can be achieved by fixing the functions $k_\mu(\xi)$, $q_\mu(\xi)$ through the following differential equations:
\begin{equation}
\left\{\begin{aligned}
&\frac 1 \kappa \, ( \cosh \xi -1 ) \,  k_0' = \sinh \xi \,,
\\
&\frac 1 \kappa \, ( \cosh \xi -1 ) \, \left( k_1'  +  \frac 1 \kappa k_0'  k_1 \right) = \cosh \xi \,,
\end{aligned}
\right.
\qquad
\left\{\begin{aligned}
&\frac 1 \kappa \, \sinh \xi \,  q_0' = \cosh \xi \,,
\\
&  \frac 1 \kappa \, \sinh \xi \, \left(  q_1' +  \frac 1 \kappa q_0'  q_1 \right)
= \sinh \xi \,,
\end{aligned}
\right.
\end{equation}
these are easily solved:
\begin{equation}
\left\{\begin{aligned}
&k_0 =  \kappa \left[ 2 \log \left( \sinh \frac \xi 2 \right) + c_1\right] \,,
\\
&k_1 =  \kappa \left[ \frac{\sinh \xi + c_2}{\cosh \xi -1} \right] \,,
\end{aligned}
\right.
\qquad
\left\{\begin{aligned}
&q_0 =  \kappa \left[ \log \left( \sinh \xi \right) + c_3 \right] \,,
\\
&q_1 =  \kappa \left[\frac{ \cosh \xi + c_4}{\sinh \xi}   \right] \,.
\end{aligned}
\right.
\end{equation}
so the conclusion is the formula
\begin{equation}
\begin{aligned}
\Delta_L \left[ e^{i p_1 x^1}  e^{i p_0 x^1} \right]
=e^{i \lambda_1 [\xi,p] \, x^1} e^{i \lambda_0 [\xi,p] \, x^0}  e^{ i p_1 a^1} e^{ i p_0 a^0}\,,
\end{aligned}
\end{equation}
where
\begin{equation}
\left\{
\begin{aligned}
\lambda_1 [\xi,p] &= k_1 - e^{ -\left( k_0  - \hat k_0\right) / \kappa} \hat k_1 + e^{- \left( k_0  - \hat k_0\right) /\kappa}\left( \hat q_1  - e^{- \left( \hat q_0  - \hat {\tilde q}_0 \right) /\kappa} \hat{\tilde q}_1 \right) \,,
\\
\lambda_0 [\xi,p] &=  k_0  - \hat k_0 + \hat q_0  - \hat{\tilde q}_0 \,,
\end{aligned}\right.
\end{equation}
in which
\begin{equation}
\left\{
\begin{aligned}
&\hat k_\mu = k_\mu [\hat \xi] = k_\mu \left[2 \, \text{arccoth} \left( \frac{p_1}{\kappa} + \coth \frac \xi 2  \right) \right] \,,
\\
&\tilde k_\mu = k_\mu [\tilde \xi] = k_\mu\left[2 \, \text{arctanh}\left(  e^{- \frac{p_0}{\kappa}} \tanh \frac \xi 2 \right)  \right]\,,
\\
&\hat{\tilde k}_\mu = k_\mu\left[2 \, \text{arctanh}\left(  e^{- \frac{p_0}{\kappa}} \tanh \frac {\hat \xi} 2 \right)  \right] = k_\mu\left[2 \, \text{arctanh}\left(  \frac{e^{- \frac{p_0}{\kappa}}}{\frac{p_1}{\kappa} + \coth \frac \xi 2 } \right)  \right]\,,
\end{aligned}\right.
\end{equation}
and analogously for $q_\mu$. An explicit calculation reveals that
\begin{equation}
\begin{gathered}
\lambda_0 [\xi,p] = p_0 + \kappa \log \left[ \left( \cosh {\frac \xi 2} + {\frac{p_1} \kappa} \sinh {\frac \xi 2} \right)^2 - e^{-2 p_0/\kappa} \sinh^2 {\frac \xi 2} \right] \,,
\\
\lambda_1 [\xi,p] = \kappa \frac{ \left( \cosh {\frac \xi 2} + {\frac{p_1} \kappa} \sinh {\frac \xi 2} \right) \left( \sinh {\frac \xi 2} + {\frac{p_1} \kappa} \cosh {\frac \xi 2} \right) -  e^{-2 p_0/\kappa} \cosh {\frac \xi 2}  \sinh {\frac \xi 2} }{\left( \cosh {\frac \xi 2} + {\frac{p_1} \kappa} \sinh {\frac \xi 2} \right)^2 - e^{-2 p_0/\kappa} \sinh^2 {\frac \xi 2} } \,,
\end{gathered}
\end{equation}
which is Eq.~(\ref{1+1DLorentzTransform}) which we used, initially, in Eq.~(\ref{abbellliiii}).

The transformation rule above can be found by integrating the vector flow on momentum space that is generated by the action of infinitesimal Lorentz transformations [Eq.~\ref{kP-Algebra-commutators}] on the momenta. This was first done in~\cite{amel}. Alternatively, one can use a matrix representation of the algebra~(\ref{1+1kPgroup}) and essentially do the same we did algebraically in this appendix, as was done in~\cite{majid22}.

\newpage

\section{Calculation of the creation and annihilation operator algebra}\label{AppendiceCommRel}

It is convenient to introduce the following three new operators:
\begin{equation}\label{StarredOperatorDef}
\dhat (\vec q) = e^{-\frac{3\omega^-(|\vec q |)}{\kappa}} \ahat^\dagger ({\vec S}_-({\vec q})) \,,
\qquad
\ehat (\vec q) = e^{-\frac{3\omega^+(|\vec q |)}{\kappa}} \bhat ({\vec S}_+({\vec q})) \,,
\qquad
\fhat (\vec q) = e^{-\frac{3\omega^+(|\vec q |)}{\kappa}} \chat ({\vec S}_+({\vec q})) \,,
\end{equation}
which makes $\hat \phi^\dagger (x)$ in~\eqref{Phi_dagger_2} take the same form as $\hat \phi(x)$ in Eq.~(\ref{3Dintegral_quantum_field}). Expliciting the mode expansion of the fileds in the commutation relations~\eqref{CovariantCommutators}:
\begin{equation}
\begin{aligned}
[ \hat \phi(z),  \hat \phi^\dagger(y) ]
&&= \int_{|\vec p | < \kappa} \frac{\diff^3 p \,  e^{\frac{3\omega^+(|\vec p |)}{\kappa}}}{2 \sqrt{m^2 + |\vec p|^2}}  \bigg{(} & \int_{\mathbbm{R}^3} \frac{\diff^3 q \,  e^{\frac{3\omega^-(|\vec q |)}{\kappa}}}{2 \sqrt{m^2 + |\vec q|^2}} [ \ahat ({\vec p}),\dhat ({\vec q })] \,  e^{i {\vec p } \cdot  {\vec z} + i {\vec q} \cdot  {\vec y}}  e^{i \omega^+(|\vec p |) z^0 + i \omega^-(|\vec q |) y^0} + 
\\
&& & \int_{|\vec q | < \kappa} \frac{\diff^3 q \,  e^{\frac{3\omega^+(|\vec q |)}{\kappa}}}{2 \sqrt{m^2 + |\vec q|^2}} [ \ahat ({\vec p}),\ehat ({\vec q })] \,  e^{i {\vec p } \cdot  {\vec z} + i {\vec q} \cdot  {\vec y}}  e^{i \omega^+(|\vec p |) z^0 + i \omega^+(|\vec q |) y^0} +
\\
&& & \int_{|\vec q | > \kappa} \frac{\diff^3 q \,  e^{\frac{3\omega^+(|\vec q |)}{\kappa}}}{2 \sqrt{m^2 + |\vec q|^2}} [ \ahat ({\vec p}),\fhat ({\vec q })] \,  e^{i {\vec p } \cdot  {\vec z} + i {\vec q} \cdot  {\vec y}}  e^{i \omega^+(|\vec p |) z^0 + i \omega^+(|\vec q |) y^0}  \bigg{)}
\\
&&+
\int_{\mathbbm{R}^3} \frac{\diff^3 p \,  e^{\frac{3\omega^-(|\vec p |)}{\kappa}}}{2 \sqrt{m^2 + |\vec p|^2}}  \bigg{(} & \int_{\mathbbm{R}^3} \frac{\diff^3 q \,  e^{\frac{3\omega^-(|\vec q |)}{\kappa}}}{2 \sqrt{m^2 + |\vec q|^2}} [ \bhat^\dagger({\vec p}),\dhat ({\vec q })] \,  e^{i {\vec p } \cdot  {\vec z} + i {\vec q} \cdot  {\vec y}}  e^{i \omega^-(|\vec p |) z^0 + i \omega^-(|\vec q |) y^0} + 
\\
&& & \int_{|\vec q | < \kappa} \frac{\diff^3 q \,  e^{\frac{3\omega^+(|\vec q |)}{\kappa}}}{2 \sqrt{m^2 + |\vec q|^2}} [ \bhat^\dagger ({\vec p}),\ehat ({\vec q })] \,  e^{i {\vec p } \cdot  {\vec z} + i {\vec q} \cdot  {\vec y}}  e^{i \omega^-(|\vec p |) z^0 + i \omega^+(|\vec q |) y^0} +
\\
&& & \int_{|\vec q | > \kappa} \frac{\diff^3 q \,  e^{\frac{3\omega^+(|\vec q |)}{\kappa}}}{2 \sqrt{m^2 + |\vec q|^2}} [ \bhat^\dagger ({\vec p}),\fhat ({\vec q })] \,  e^{i {\vec p } \cdot  {\vec z} + i {\vec q} \cdot  {\vec y}}  e^{i \omega^-(|\vec p |) z^0 + i \omega^+(|\vec q |) y^0}  \bigg{)}
\\
&&+
 \int_{|\vec p | > \kappa} \frac{\diff^3 p \,  e^{\frac{3\omega^+(|\vec p |)}{\kappa}}}{2 \sqrt{m^2 + |\vec p|^2}}  \bigg{(} & \int_{\mathbbm{R}^3} \frac{\diff^3 q \,  e^{\frac{3\omega^-(|\vec q |)}{\kappa}}}{2 \sqrt{m^2 + |\vec q|^2}} [ \chat^\dagger ({\vec p}),\dhat ({\vec q })] \,  e^{i {\vec p } \cdot  {\vec z} + i {\vec q} \cdot  {\vec y}}  e^{i \omega^+(|\vec p |) z^0 + i \omega^-(|\vec q |) y^0} + 
\\
&& & \int_{|\vec q | < \kappa} \frac{\diff^3 q \,  e^{\frac{3\omega^+(|\vec q |)}{\kappa}}}{2 \sqrt{m^2 + |\vec q|^2}} [ \chat^\dagger ({\vec p}),\ehat ({\vec q })] \,  e^{i {\vec p } \cdot  {\vec z} + i {\vec q} \cdot  {\vec y}}  e^{i \omega^+(|\vec p |) z^0 + i \omega^+(|\vec q |) y^0} +
\\
&& & \int_{|\vec q | > \kappa} \frac{\diff^3 q \,  e^{\frac{3\omega^+(|\vec q |)}{\kappa}}}{2 \sqrt{m^2 + |\vec q|^2}} [ \chat^\dagger ({\vec p}),\fhat ({\vec q })] \,  e^{i {\vec p } \cdot  {\vec z} + i {\vec q} \cdot  {\vec y}}  e^{i \omega^+(|\vec p |) z^0 + i \omega^+(|\vec q |) y^0}  \bigg{)} \,.
\end{aligned} 
\end{equation}
Comparing with~(\ref{Noncommutative_Pauli-Jordan_3D}) reveals that the three terms above should be identified pairwise with the three terms in the Pauli--Jordan function:
\begin{equation}\label{1111}
\begin{aligned}
\!\!\!\!\bigg{(} & \int_{\mathbbm{R}^3} \frac{\diff^3 q \,  e^{\frac{3\omega^-(\vec q)}{\kappa}}}{2 \sqrt{m^2 + |\vec q|^2}} [ \ahat ({\vec p}),\dhat ({\vec q })] \,  e^{i {\vec p } \cdot  {\vec z} + i {\vec q} \cdot  {\vec y}}  e^{i \omega^+(\vec p) z^0 + i \omega^-(\vec q) y^0} + 
\\
& \int_{|\vec q | < \kappa} \frac{\diff^3 q \,  e^{\frac{3\omega^+(\vec q)}{\kappa}}}{2 \sqrt{m^2 + |\vec q|^2}} [ \ahat ({\vec p}),\ehat ({\vec q })] \,  e^{i {\vec p } \cdot  {\vec z} + i {\vec q} \cdot  {\vec y}}  e^{i \omega^+(\vec p) z^0 + i \omega^+(\vec q) y^0} +
\\
& \int_{|\vec q | > \kappa} \frac{\diff^3 q \,  e^{\frac{3\omega^+(\vec q)}{\kappa}}}{2 \sqrt{m^2 + |\vec q|^2}} [ \ahat ({\vec p}),\fhat ({\vec q })] \,  e^{i {\vec p } \cdot  {\vec z} + i {\vec q} \cdot  {\vec y}}  e^{i \omega^+(\vec p) z^0 + i \omega^+ (\vec q) y^0}  \bigg{)} =    e^{i   {\vec p} \cdot {\vec z} - i e^{\frac{\omega^+(\vec p)}\kappa} {\vec p} \cdot {\vec y}  }  e^{i \omega^+(\vec p) (z^0 - y^0) }   \,,
\end{aligned} 
\end{equation}
\begin{equation}\label{2222}
\begin{aligned}
\!\!\!\!\bigg{(} & \int_{\mathbbm{R}^3} \frac{\diff^3 q \,  e^{\frac{3\omega^-(\vec q)}{\kappa}}}{2 \sqrt{m^2 + |\vec q|^2}} [ \bhat^\dagger ({\vec p}),\dhat ({\vec q })] \,  e^{i {\vec p } \cdot  {\vec z} + i {\vec q} \cdot  {\vec y}}  e^{i \omega^-(\vec p) z^0 + i \omega^-(\vec q) y^0} + 
\\
 & \int_{|\vec q | < \kappa} \frac{\diff^3 q \,  e^{\frac{3\omega^+(\vec q)}{\kappa}}}{2 \sqrt{m^2 + |\vec q|^2}} [ \bhat^\dagger ({\vec p}),\ehat ({\vec q })] \,  e^{i {\vec p } \cdot  {\vec z} + i {\vec q} \cdot  {\vec y}}  e^{i \omega^-(\vec p) z^0 + i \omega^+(\vec q) y^0} +
\\
& \int_{|\vec q | > \kappa} \frac{\diff^3 q \,  e^{\frac{3\omega^+(\vec q)}{\kappa}}}{2 \sqrt{m^2 + |\vec q|^2}} [ \bhat^\dagger ({\vec p}),\fhat ({\vec q })] \,  e^{i {\vec p } \cdot  {\vec z} + i {\vec q} \cdot  {\vec y}}  e^{i \omega^-(\vec p) z^0 + i \omega^+(\vec q) y^0}  \bigg{)} = - e^{i  {\vec p} \cdot {\vec z}} e^{- i  e^{\frac{\omega^-(\vec p)}\kappa}{\vec p} \cdot {\vec y}}  e^{i \omega^-(\vec p)(z^0 - y^0) }   \,,
\end{aligned} 
\end{equation}
\begin{equation}\label{3333}
\begin{aligned}
\!\!\!\!\bigg{(} & \int_{\mathbbm{R}^3} \frac{\diff^3 q \,  e^{\frac{3\omega^-(\vec q)}{\kappa}}}{2 \sqrt{m^2 + |\vec q|^2}} [ \chat^\dagger ({\vec p}),\dhat ({\vec q })] \,  e^{i {\vec p } \cdot  {\vec z} + i {\vec q} \cdot  {\vec y}}  e^{i \omega^+(\vec p) z^0 + i \omega^-(\vec q) y^0} + 
\\
& \int_{|\vec q | < \kappa} \frac{\diff^3 q \,  e^{\frac{3\omega^+(\vec q)}{\kappa}}}{2 \sqrt{m^2 + |\vec q|^2}} [ \chat^\dagger ({\vec p}),\ehat ({\vec q })] \,  e^{i {\vec p } \cdot  {\vec z} + i {\vec q} \cdot  {\vec y}}  e^{i \omega^+(\vec p) z^0 + i \omega^+(\vec q) y^0} +
\\
 & \int_{|\vec q | > \kappa} \frac{\diff^3 q \,  e^{\frac{3\omega^+(\vec q)}{\kappa}}}{2 \sqrt{m^2 + |\vec q|^2}} [ \chat^\dagger ({\vec p}),\fhat ({\vec q })] \,  e^{i {\vec p } \cdot  {\vec z} + i {\vec q} \cdot  {\vec y}}  e^{i \omega^+(\vec p) z^0 + i \omega^+(\vec q) y^0}  \bigg{)} =   
-  e^{i  {\vec p} \cdot {\vec z}} e^{- i  e^{\frac{\omega^+(\vec p) }\kappa} {\vec p} \cdot {\vec y}}  e^{i \omega^+(\vec p) (z^0 - y^0) } \,.
\end{aligned} 
\end{equation}
Let us introduce the following ansatz:
\begin{equation}
\begin{aligned}
&[\ahat (\vec{p}) , \dhat(\vec{q}) ] = F_{aa} \, \delta^{(3)}[\vec{q} - {\vec S}_+(\vec{p}) ] \,,&
&[\ahat (\vec{p}) , \ehat(\vec{q}) ] = F_{ab} \, \delta^{(3)}[\vec{q} - {\vec S}_+(\vec{p}) ] \,,
&
&[\ahat (\vec{p}) , \fhat(\vec{q}) ] = F_{ac} \, \delta^{(3)}[\vec{q} - {\vec S}_+(\vec{p}) ] \,, 
\\
&[\bhat^\dagger (\vec{p}) , \dhat(\vec{q}) ] = F_{ba} \, \delta^{(3)}[\vec{q} - {\vec S}_-(\vec{p}) ]\,,
& 
&[\bhat^\dagger(\vec{p}) ,\ehat(\vec{q}) ] = F_{bb} \, \delta^{(3)}[\vec{q} - {\vec S}_-(\vec{p}) ] \,,
&
&[\bhat^\dagger (\vec{p}) ,\fhat(\vec{q}) ] = F_{bc} \,\delta^{(3)}[\vec{q} - {\vec S}_-(\vec{p}) ] \,,
\\
&[\chat^\dagger (\vec{p}) ,\dhat(\vec{q}) ] = F_{ca} \, \delta^{(3)}[\vec{q} - {\vec S}_+(\vec{p}) ] \,,
&
&[\chat^\dagger (\vec{p}) , \ehat(\vec{q}) ] = F_{cb} \, \delta^{(3)}[\vec{q} - {\vec S}_+(\vec{p}) ] \,,
&  
&[\chat^\dagger (\vec{p}) ,\fhat(\vec{q}) ] = F_{cc}\, \delta^{(3)}[\vec{q} - {\vec S}_+(\vec{p}) ] \,,
\end{aligned}
\end{equation}
the first equation, Eq.~\eqref{1111} turns into:
\begin{equation}
\begin{aligned}
 \frac{ e^{i {\vec p } \cdot  {\vec z} + i {\vec S}_+(\vec p) \cdot  {\vec y}} e^{i \omega^+(\vec p) z^0 }}{2 \sqrt{m^2 + |{\vec S}_+(\vec p)|^2}} \bigg{(} & e^{\frac{3\omega^-({\vec S}_+(\vec p))}{\kappa}} \, F_{aa} \,  e^{i \omega^-({\vec S}_+(\vec p)) y^0} + 
 \Theta\left(|{\vec S}_+(\vec p) | < \kappa \right)  e^{\frac{3\omega^+({\vec S}_+(\vec p))}{\kappa}} \, F_{ab} \,  e^{i \omega^+({\vec S}_+(\vec p)) y^0} +
\\
& \Theta\left(|{\vec S}_+(\vec p) | > \kappa \right)  e^{\frac{3\omega^+({\vec S}_+(\vec p))}{\kappa}} \, F_{ac} \,  e^{i \omega^+ ({\vec S}_+(\vec p)) y^0}  \bigg{)} \bigg{|}_{|\vec p|<\kappa} 
 \!\!\!\!\!\!\!\!\! =    e^{i   {\vec p} \cdot {\vec z} - i e^{\frac{\omega^+(\vec p)}\kappa} {\vec p} \cdot {\vec y}  }  e^{i \omega^+(\vec p) (z^0 - y^0) }   \,,
\end{aligned} 
\end{equation}
using the relations~(\ref{Omega_S_relations}), in particular
$\omega^- ({\vec S}_+(\vec{p}))=-\omega^+(\vec{p})$ if $|\vec{p}| < \kappa$, we see that 
$$
e^{i \omega^-({\vec S}_+(\vec p)) y^0}  \bigg{|}_{|\vec p|<\kappa} =  e^{-i \omega^+ (\vec p) y^0}  \,,
\qquad  
e^{i \omega^+({\vec S}_+(\vec p)) y^0}  \bigg{|}_{|\vec p|<\kappa} \neq e^{-i \omega^+ (\vec p) y^0}  \,,
$$
and so only the term multiplying $F_{aa}$ is identical to the right-hand side. The second equation, Eq.~\eqref{2222} is:
\begin{equation}
\begin{aligned}
 \frac{ e^{i {\vec p } \cdot  {\vec z} + i {\vec S}_-(\vec p) \cdot  {\vec y}} e^{i \omega^-(\vec p) z^0 }}{2 \sqrt{m^2 + |{\vec S}_-(\vec p)|^2}} \bigg{(} & e^{\frac{3\omega^-({\vec S}_-(\vec p))}{\kappa}} \, F_{ba} \,  e^{i \omega^-({\vec S}_-(\vec p)) y^0} + 
 \Theta\left(|{\vec S}_-(\vec p) | < \kappa \right)  e^{\frac{3\omega^+({\vec S}_-(\vec p))}{\kappa}} \, F_{bb} \,  e^{i \omega^+({\vec S}_-(\vec p)) y^0} +
\\
& \cancel{\Theta\left(|{\vec S}_-(\vec p) | > \kappa \right)  e^{\frac{3\omega^+({\vec S}_-(\vec p))}{\kappa}} \, F_{bc} \,  e^{i \omega^+ ({\vec S}_-(\vec p)) y^0}}  \bigg{)} \bigg{|}_{\vec p \in \mathbbm{R}^3} 
 \!\!\!\!\!\!\!\!\! =    e^{i   {\vec p} \cdot {\vec z} - i e^{\frac{\omega^-(\vec p)}\kappa} {\vec p} \cdot {\vec y}  }  e^{i \omega^-(\vec p) (z^0 - y^0) }   \,,
\end{aligned} 
\end{equation}
where the third term is crossed because $\Theta\left(|{\vec S}_-(\vec p) | > \kappa \right) \bigg{|}_{\vec p \in \mathbbm{R}^3}  = 0 $. Another look at relations~(\ref{Omega_S_relations}), in particular $\omega^+ ({\vec S}_-(\vec{p}))= -\omega^-(\vec{p})$ if $\vec{p} \in \mathbb{R}^3$, reveals that:
$$
e^{i \omega^-({\vec S}_-(\vec p)) y^0}  \bigg{|}_{\vec p \in \mathbbm{R}^3} \neq  e^{-i \omega^- (\vec p) y^0}  \,,
\qquad  
e^{i \omega^+({\vec S}_-(\vec p)) y^0}  \bigg{|}_{\vec p \in \mathbbm{R}^3} = e^{-i \omega^- (\vec p) y^0}  \,,
$$
so only the $F_{bb}$ term survives. Finally consider Eq.~\eqref{3333}:
\begin{equation}
\begin{aligned}
 \frac{ e^{i {\vec p } \cdot  {\vec z} + i {\vec S}_+(\vec p) \cdot  {\vec y}} e^{i \omega^+(\vec p) z^0 }}{2 \sqrt{m^2 + |{\vec S}_+(\vec p)|^2}} \bigg{(} & e^{\frac{3\omega^-({\vec S}_+(\vec p))}{\kappa}} \, F_{ca} \,  e^{i \omega^-({\vec S}_+(\vec p)) y^0} + 
\cancel{ \Theta\left(|{\vec S}_+(\vec p) | < \kappa \right)  e^{\frac{3\omega^+({\vec S}_+(\vec p))}{\kappa}} \, F_{cb} \,  e^{i \omega^+({\vec S}_+(\vec p)) y^0} } +
\\
& \Theta\left(|{\vec S}_+(\vec p) | > \kappa \right)  e^{\frac{3\omega^+({\vec S}_+(\vec p))}{\kappa}} \, F_{cc} \,  e^{i \omega^+ ({\vec S}_+(\vec p)) y^0}  \bigg{)} \bigg{|}_{|\vec p|>\kappa} 
 \!\!\!\!\!\!\!\!\! =    e^{i   {\vec p} \cdot {\vec z} - i e^{\frac{\omega^+(\vec p)}\kappa} {\vec p} \cdot {\vec y}  }  e^{i \omega^+(\vec p) (z^0 - y^0) }   \,,
\end{aligned} 
\end{equation}
with relations~(\ref{Omega_S_relations}), in particular $\omega^+ ({\vec S}_+(\vec{p}))=-\omega^+(\vec{p})$ if $|\vec{p}|>\kappa$:
$$
e^{i \omega^-({\vec S}_+(\vec p)) y^0}  \bigg{|}_{|\vec p| > \kappa} \neq  e^{-i \omega^+ (\vec p) y^0}  \,,
\qquad  
e^{i \omega^+({\vec S}_+(\vec p)) y^0}  \bigg{|}_{|\vec p|>\kappa} = e^{-i \omega^+ (\vec p) y^0}  \,,
$$
we see that only the $F_{cc}$ term survives. The three surviving coefficients then are constrained to be:
\begin{equation}
F_{aa} = 2 \sqrt{m^2 + |{\vec S}_+(\vec p)|^2} e^{\frac{3\omega^+(\vec p)}{\kappa}} \,,
~~
F_{bb} = - 2 \sqrt{m^2 + |{\vec S}_-(\vec p)|^2} e^{\frac{3\omega^-(\vec p)}{\kappa}} \,,
~~
F_{cc} = - 2 \sqrt{m^2 + |{\vec S}_+(\vec p)|^2} e^{\frac{3\omega^+(\vec p)}{\kappa}} \,,
\end{equation}
and the commutation relation are therefore:
\begin{equation}
\begin{aligned}
&[\ahat (\vec{p}) , \dhat(\vec{q}) ] = 2 \sqrt{m^2 + |{\vec S}_+(\vec p)|^2} e^{\frac{3\omega^+(\vec p)}{\kappa}}\, \delta^{(3)}[\vec{q} - {\vec S}_+(\vec{p}) ] \,,
\\
&[\bhat^\dagger (\vec{p}) ,\ehat(\vec{q}) ] = - 2 \sqrt{m^2 + |{\vec S}_-(\vec p)|^2} e^{\frac{3\omega^-(\vec p)}{\kappa}}  \, \delta^{(3)}[\vec{q} - {\vec S}_-(\vec{p}) ] \,,
\\
&[\chat^\dagger (\vec{p}) ,\fhat(\vec{q}) ] = - 2 \sqrt{m^2 + |{\vec S}_+(\vec p)|^2} e^{\frac{3\omega^+(\vec p)}{\kappa}} \, \delta^{(3)}[\vec{q} - {\vec S}_+(\vec{p}) ] \,.
\end{aligned}
\end{equation}
We can change variable in the delta functions on the right-hand sides, and isolate $\vec p$. Then the delta functions produce the determinant of a Jacobian that cancels with the exponential terms multiplying them:
\begin{equation}
\begin{aligned}
\text{if }~~ | \vec p | < \kappa \qquad e^{\frac{3\omega^+(\vec p)}{\kappa}}\, \delta^{(3)}[\vec{q} - {\vec S}_+(\vec{p}) ] 
&=
\frac{e^{\frac{3\omega^+(\vec p)}{\kappa}}}{\left| \text{det} \frac{\partial {\vec S}_+(\vec p)}{\partial \vec p} \right|}\, \delta^{(3)}[\vec{p} - {\vec S}_-(\vec{q}) ] = \delta^{(3)}[\vec{p} - {\vec S}_-(\vec{q}) ] \,, 
\\
\forall ~~  \vec p  \qquad
e^{\frac{3\omega^-(\vec p)}{\kappa}}\, \delta^{(3)}[\vec{q} - {\vec S}_-(\vec{p}) ] 
&=
\frac{e^{\frac{3\omega^-(\vec p)}{\kappa}}}{\left| \text{det} \frac{\partial {\vec S}_- (\vec p)}{\partial \vec p} \right|}\, \delta^{(3)}[\vec{p} - {\vec S}_+(\vec{q}) ]
=\delta^{(3)}[\vec{p} - {\vec S}_+(\vec{q}) ]  \,,
\\
\text{if }~~ | \vec p | > \kappa \qquad
e^{\frac{3\omega^+(\vec p)}{\kappa}}\, \delta^{(3)}[\vec{q} - {\vec S}_+(\vec{p}) ] 
&=
\frac{e^{\frac{3\omega^+(\vec p)}{\kappa}}}{\left| \text{det} \frac{\partial {\vec S}_+(\vec p)}{\partial \vec p} \right|}\, \delta^{(3)}[\vec{p} - {\vec S}_+(\vec{q}) ] = \delta^{(3)}[\vec{p} - {\vec S}_+(\vec{q}) ] \,,
\end{aligned}
\end{equation}
then the commutation relations take a particularly simple form:
\begin{equation}
\begin{aligned}
&[\ahat (\vec{p}) , \dhat(\vec{q}) ] = 2 \sqrt{m^2 + |\vec q|^2} \, \delta^{(3)}[\vec{p} - {\vec S}_-(\vec{q}) ] \,,
\\
&[\bhat^\dagger (\vec{p}) ,\ehat(\vec{q}) ] = -  2 \sqrt{m^2 + |\vec q|^2}   \, \delta^{(3)}[\vec{p} - {\vec S}_+(\vec{q}) ] \,,
\\
&[\chat^\dagger (\vec{p}) ,\fhat(\vec{q}) ] = - 2 \sqrt{m^2 + |\vec q|^2}  \, \delta^{(3)}[\vec{p} - {\vec S}_+(\vec{q}) ] \,.\end{aligned}
\end{equation}
Now use the definition~\eqref{StarredOperatorDef}
$$
\begin{aligned}
&[\ahat(\vec{p}) ,  \ahat^\dagger ({\vec S}_-({\vec q})) ] = 2 e^{\frac{3\omega^-(|\vec q |)}{\kappa}} \sqrt{m^2 + |\vec q|^2} \, \delta^{(3)}[\vec{p} - {\vec S}_-(\vec{q}) ] \,,
\\
&[\bhat^\dagger(\vec{p}) ,  \bhat({\vec S}_+({\vec q})) ] = -  2 e^{\frac{3\omega^+(|\vec q |)}{\kappa}} \sqrt{m^2 + |\vec q|^2}   \, \delta^{(3)}[\vec{p} - {\vec S}_+(\vec{q}) ] \,,
\\
&[\chat^\dagger(\vec{p}) ,  \chat ({\vec S}_+({\vec q})) ] = - 2 e^{\frac{3\omega^+(|\vec q |)}{\kappa}} \sqrt{m^2 + |\vec q|^2}  \, \delta^{(3)}[\vec{p} - {\vec S}_+(\vec{q}) ] \,,\end{aligned}
$$
and recall Eq.~\eqref{SofSrelations}:
$$
{\vec S}_+\left({\vec S}_+({\vec p }) \right)= {\vec p }  \qquad \text{if} ~~ |\vec{p}| > \kappa\,,
\qquad
{\vec S}_-\left({\vec S}_+({\vec p }) \right)= {\vec p }  \qquad \text{if} ~~ |\vec{p}| < \kappa\,,
\qquad
{\vec S}_+\left({\vec S}_-({\vec p }) \right)= {\vec p }  \qquad  \forall ~~ \vec{p} \in \mathbbm{R}^3 \,,
$$
we get:
$$
\begin{aligned}
&[\ahat(\vec{p}) ,  \ahat^\dagger (\vec{k}) ] = 2 e^{\frac{3\omega^-({\vec S}_+({\vec k}))}{\kappa}} \sqrt{m^2 + |{\vec S}_+({\vec k})|^2} \, \delta^{(3)}[\vec{p} - \vec{k}  ] \,,
\\
&[\bhat^\dagger(\vec{p}) ,  \bhat (\vec{k}) ] = -  2 e^{\frac{3\omega^+({\vec S}_-({\vec k}))}{\kappa}} \sqrt{m^2 + |{\vec S}_-({\vec k})|^2}   \, \delta^{(3)}[\vec{p} - \vec{k} ] \,,
\\
&[\chat^\dagger(\vec{p}) ,  \chat  (\vec{k}) ] = - 2 e^{\frac{3\omega^+({\vec S}_+({\vec k}))}{\kappa}} \sqrt{m^2 + |{\vec S}_+({\vec k})|^2}  \, \delta^{(3)}[\vec{p} - \vec{k} ] \,,\end{aligned}
$$
and finally, using one last time Eq.~\eqref{Omega_S_relations}:
$$
\begin{aligned}
 &\omega^+ ({\vec S}_-(\vec{q}))= -\omega^-(\vec{q}), \qquad \text{if} ~~~ \vec{q} \in \mathbb{R}^3 \,,
 \\
 &\omega^- ({\vec S}_+(\vec{q}))=-\omega^+(\vec{q}), \qquad \text{if} ~~~  |\vec{q}| < \kappa  \,,
 \\
 & \omega^+ ({\vec S}_+(\vec{q}))=-\omega^+(\vec{q}), \qquad \text{if} ~~~   |\vec{q}|>\kappa \,.
\end{aligned}
$$
we get the final form of the commutation relations:
\begin{equation}
\begin{aligned}
&[\ahat(\vec{p}) ,  \ahat^\dagger(\vec{k}) ] = 2 e^{-\frac{3\omega^+({\vec k})}{\kappa}} \sqrt{m^2 + |{\vec S}_+({\vec k})|^2} \, \delta^{(3)}[\vec{p} - \vec{k}  ] \,,
\\
&[\bhat^\dagger(\vec{p}) ,  \bhat (\vec{k}) ] = -  2 e^{-\frac{3\omega^-({\vec k})}{\kappa}} \sqrt{m^2 + |{\vec S}_-({\vec k})|^2}   \, \delta^{(3)}[\vec{p} - \vec{k} ] \,,
\\
&[\chat^\dagger(\vec{p}) ,  \chat  (\vec{k}) ] = - 2 e^{-\frac{3\omega^+({\vec k})}{\kappa}} \sqrt{m^2 + |{\vec S}_+({\vec k})|^2}  \, \delta^{(3)}[\vec{p} - \vec{k} ] \,.\end{aligned}
\end{equation}

\end{document}